\documentclass[11pt]{article}
\usepackage{CJKutf8} %chinese
\usepackage{graphicx} 
\usepackage{graphbox}
\usepackage[usenames, dvipsnames]{xcolor}
\usepackage{cancel}
\usepackage{tikz}
\usepackage{tikz-cd}
\usepackage{amsthm}
\usepackage{dsfont}
\usepackage{mathtools}
\usepackage{cancel}
\usepackage{jheppub}
\usepackage{orcidlink}
\usetikzlibrary{shapes.geometric, arrows}

\usepackage{hyperref}
\hypersetup{
	colorlinks = True,
   	citecolor = magenta,
   	linkcolor = cyan,
	urlcolor = cyan,
	bookmarks=true
}
\usepackage[export]{adjustbox}
\usepackage{float}

\newcommand{\vth}{\vartheta_1}
\renewcommand{\th}{\th}

\newcommand{\vphi}{\varphi}
\renewcommand{\d}{\text{d}}

\newcommand{\nn}{\nonumber}
\renewcommand{\c}[1]{\check{#1}}
\newcommand{\kn}{\text{KN}^\tau}
\newcommand{\la}{\langle}
\newcommand{\ra}{\rangle}
\newcommand{\pa}{\partial}
\newcommand{\gk}[1]{g^{(#1)}}
\newcommand{\im}{\mathrm{Im}\,}

\newcommand{\reg}{\mathrm{Reg}}
\renewcommand{\L}{\mathcal{L}}
\newcommand{\M}{\mathcal{M}}
\newcommand{\LS}{ {\mathcal{L}_{\omega,\eta}} }
\newcommand{\LSdual}{{\c{\mathcal{L}}_{\omega,\eta}}}
\newcommand{\mat}[1]{\underline{\boldsymbol{#1}}}

\newcommand{\bs}[1]{\boldsymbol{#1}}
\newcommand{\etaDe}{\eta_{\text{\tiny De}}}
\newcommand{\Gt}[1]{\tilde{\Gamma}\left(\begin{smallmatrix}1\\0\end{smallmatrix};#1|\tau\right)}
\newcommand{\Gtreg}[1]{\tilde{\Gamma}_\text{reg}\left(\begin{smallmatrix}1\\0\end{smallmatrix};#1|\tau\right)}
\newcommand{\re}{\mathrm{Re}}

\newcommand{\A}{\mathcal{A}}
\newcommand{\open}{\mathrm{open}}
\newcommand{\closed}{\mathrm{closed}}

\newcommand{\new}[1]{#1}

\preprint{UUITP--37/23}
\title{A double copy from twisted (co)homology at genus one}

\author[a,\orcidlink{0000-0001-7624-4421}]{Rishabh Bhardwaj,}
\emailAdd{rishabh\_bhardwaj@brown.edu}

\author[a,\orcidlink{0000-0003-1186-4624}]{Andrzej Pokraka,}
\emailAdd{andrzej\_pokraka@brown.edu}

\author[a,\orcidlink{0000-0002-0846-8017}]{Lecheng Ren,}
\emailAdd{lecheng\_ren@brown.edu}

\author[b,\orcidlink{0000-0001-6523-0841}]{Carlos Rodriguez}
\emailAdd{carlos.rodriguez@physics.uu.se}

\affiliation[a]{
	Department of Physics, 	
	Brown University, 	
	Providence, 	
	RI 02912, 
	USA
}

\affiliation[b]{
    Department of Physics and Astronomy, 
    Uppsala University, 
    Box 516, 
    75120 Uppsala, 
    Sweden
}

\abstract{
We study the twisted (co)homology of a family of genus-one integrals --- the so called Riemann-Wirtinger integrals. 
These integrals are closely related to one-loop string amplitudes in chiral splitting where one leaves the loop-momentum, modulus and all but one puncture un-integrated. 
While not actual one-loop string integrals, they share many properties and are simple enough that the associated twisted (co)homologies have been completely characterized \cite{Goto2022}. 
Using intersection numbers --- an inner product on the vector space of allowed differential forms --- we derive the Gauss-Manin connection for two bases of the twisted cohomology providing an independent check of \cite{Mano2012}. 
\new{We also use the intersection index --- an inner product on the vector space of allowed contours --- to derive a double-copy formula for the closed-string analogues of Riemann-Wirtinger integrals (one-dimensional integrals over the torus).} Similar to the celebrated KLT formula between open- and closed-string tree-level amplitudes, these intersection indices form a genus-one KLT-like kernel defining  bilinears in meromorphic Riemann-Wirtinger integrals that are equal to their complex counterparts. 
}

\begin{document}

\maketitle

%%%%%%%%%%%%%%%%%%%%%%%%%%%%%%%%%%%%%%%%%%%%%%%%%
%%%%%%%%%%%%%%%%%%%%%%%%%%%%%%%%%%%%%%%%%%%%%%%%%
%%%%%%%%%%%%%%%%%%%%%%%%%%%%%%%%%%%%%%%%%%%%%%%%%
\section{Introduction}

The double copy is a framework for computing closed-string/gravitational amplitudes using simpler open-string/gauge theory amplitudes as input. 
This framework has revealed important mathematical structures in field theory and string theory as well as providing an efficient method for computing scattering amplitudes that would otherwise be intractable. 
In particular, the double copy is a bilinear map on open-string/gauge theory amplitudes that makes the slogan $\text{gravity}=(\text{gauge theory})^2$ precise. 
This is extremely surprising and useful because gravitational amplitudes are orders of magnitude harder to compute than their gauge theory analogues.

The first instance of a double copy in the physics literature was discovered by Kawai-Lewellen-Tye (KLT) \cite{Kawai:1985xq}.  
These authors  factorized tree-level closed string scattering amplitudes into bilinears of tree-level open string amplitudes, their complex conjugates and the so-called KLT kernel. 
In parallel, similar relations appeared in the mathematics literature, more specifically in Aomoto's work on complex Selberg integrals \cite{Aomoto87}. 
In the decades since the seminal work \cite{Kawai:1985xq}, there has been a search for the analogous statement at genus-one 
and higher
\cite{Tourkine:2016bak, He:2016mzd, Ochirov:2017jby, Hohenegger:2017kqy, Casali:2019ihm, Casali:2020knc, Stieberger:2021daa, Edison:2021ebi, Stieberger:2022lss, Stieberger:2023nol}.

While its origin may be in string theory, the double copy is perhaps best understood in the field theory limit where there are several formulations \cite{Adamo:2022dcm}. 
% The field theory limit of the KLT double copy can be understood purely in terms of field theory objects. 
In the field theory limit, the matrix elements of the KLT kernel can be understood as color ordered amplitudes of bi-adjoint scalar theory \cite{Cachazo:2013iea} or generalizations thereof \cite{Mizera:2016jhj,Mizera:2017cqs, Chi:2021mio, Chen:2022shl, Chen:2023dcx}.

In the scattering equation approach by Cachazo, He and Yuan (CHY) \cite{Cachazo:2013hca, Cachazo:2013gna, Cachazo:2013iea}, with a loop-level extension in \cite{Geyer:2015jch, Geyer:2015bja, Geyer:2016wjx, Geyer:2018xwu, Geyer:2021oox, Geyer:2022cey}, $n$-point color-ordered amplitudes at $L$ loops are represented by an integral over the $(n+2L)$-punctured Riemann sphere. 
Here, the integrand is a numerator specific to the scattering process divided by a product of $n$ differences between the punctures.
In particular, the integrand only ever has simple poles when two punctures collide.  
Moreover, this integral representation of the amplitude localizes onto the locus of the scattering equations turning the integral into a discrete sum. 
To obtain gravity amplitudes in this formalism, one simply replaces the denominator of a color-ordered CHY integrand with the numerator factor corresponding to another color-ordered CHY integrand.

Additionally, the color-kinematics duality pioneered by Bern, Carrasco, and Johansson (BCJ) \cite{Bern:2008qj, Bern:2019prr, Bern:2022wqg, Adamo:2022dcm} has been tested and used at loop-level in a wide range of theories and up to four-loops in $\mathcal{N}\geq 4$ supergravities \cite{Bern:2011qn,Bern:2013uka,Bern:2014sna}. 
Schematically, the numerators of tree-level amplitudes or loop-level integrands are organized into the product of a color factor and a kinematic factor such that both the color and kinematic factors obey the Jacobi identity satisfied by the color algebra. 
Then, the corresponding gravity amplitude/integrand is obtained by replacing all color factors in the color-ordered amplitude/integrand with another copy of the kinematic factors, which can even belong to a different theory. 

In the CHY and string theory formulation of the double copy one inevitably encounters integrals of the generalized Euler-Mellin type. 
Twisted (co)homology provides a general mathematical framework for studying such integrals. 
In particular, it provides the necessary machinery to construct the double copy for any family of Euler-Mellin integrals \cite{berkesch2014euler,Matsubara-Heo:2020lzo}. 
In this language, the double copy of gravitational CHY and tree-level closed string amplitudes follows from a generalization of the well-known twisted Riemann bilinear relations \cite{hanamura1999hodge} (reviewed in section \ref{sec:Fund}). 

The integrand of a one-loop string amplitude also takes the form of a generalized Euler-Mellin integral on the torus. 
Thus, it is tempting to try using twisted (co)homology to build the corresponding double copy. 
In fact, this line of research was initiated in \cite{Casali:2019ihm} and \cite{Casali:2020knc}. 
Using twisted (co)homology, the authors derive the monodromy relations of \cite{Tourkine:2016bak,Hohenegger:2017kqy} for string integrals at genus-one and higher. 
While the authors of \cite{Casali:2019ihm} comment on the prospect of deriving the one-loop KLT kernel, the presence of unexpected, unphysical cycles in the basis of homology and $B$-cycle monodromies that depend on the other punctures complicate the analysis. The genus-one KLT formulae of \cite{Stieberger:2022lss,Stieberger:2023nol} derived from contour deformations call for an understanding in terms of twisted (co)homology.

Generalized Euler-Mellin integrals also make a prominent appearance in the computation of Feynman integrals/field theory amplitudes. 
In particular, twisted (co)homology has found interesting applications to the study of CHY amplitudes  \cite{delaCruz:2017zqr,Mizera:2017rqa, Mizera:2019gea, Mazloumi:2022nvi}, cosmological correlators \cite{De:2023xue} and Feynman integrals (in particular the coaction \cite{Abreu:2019xep, Britto:2021prf} and integral reduction  \cite{Mastrolia:2018uzb, Mizera:2019ose, Frellesvig:2019kgj, Frellesvig:2019uqt, Mizera:2019vvs, Frellesvig:2020qot, Chen:2020uyk, Caron-Huot:2021xqj, Caron-Huot:2021iev, Chen:2022lzr, Chen:2023kgw}).
 Moreover, the study of elliptic Feynman integrals has boomed over the last decade (see the review \cite{Bourjaily:2022bwx} and \cite{ Giroux:2022wav, Frellesvig:2023iwr, McLeod:2023qdf, He:2023qld, Gorges:2023zgv} for some recent works). 
Since both the $\alpha'$-expansion of one-loop string integrands (pre $\tau$ and loop-momentum integration) and elliptic Feynman integrals evaluate to the same class of functions --- elliptic multiple polylogarithms (eMPLs) \cite{Broedel:2014vla,Broedel:2017jdo} --- it is interesting to explore one-loop string integrands and their connection to field theory objects using tools from twisted (co)homology.

Motivated by the possibility of a KLT kernel for one-loop string integrands from twisted (co)homology and aiming for a better understanding the function space of genus-one integrals, we study the family of so-called Riemann-Wirtinger integrals.
This family of integrals is closely related to one-loop string integrands and serves as a toy model for their twisted (co)homology,  
which has been fully characterized \cite{mano2009, Mano2012, ghazouani2016moduli, Goto2022}. 
This allows us to present an explicit double copy formula for Riemann-Wirtinger \new{(RW)} integrals with a \new{(RW-)} KLT kernel built out of intersection indices an inner product on the space of homology!

\paragraph{Double copy teaser:} The simplest example of this double copy or genus-one \new{RW-}KLT formula is for $n=3$ punctures where the twisted (co)homology is 2-dimensional. 
Explicitly, 
\begin{align}\label{eq:z1integral}
    &\int_M \d^2 z_1\ 
    e^{2\pi i s_{1A} (z_1-\bar{z}_1) }
    \left\vert
        \frac{\vth(z_{12}\vert \tau)}{\vth(z_{13}\vert \tau)}
    \right\vert^{2s_{12}} \
    \overline{F(z_{1a},\eta|\tau)} \
    F(z_{1b},\eta|\tau)
    \\& \qquad
    = \frac{i}{2} 
    \frac{\sin \pi s_{12}}{\sin \pi s_{1A}}
    \begin{pmatrix}
        [\gamma_{2A}\vert\vphi_b\ra
        &
        [\gamma_{23}\vert\vphi_b\ra
    \end{pmatrix}
    \cdot
    \begin{pmatrix}
        0 
        & 
        e^{-i \pi (s_{1A}-s_{12})}
        \\
        -e^{i \pi (s_{1A}-s_{12})}
        & 2 i \sin \pi (s_{1A}-s_{12})
    \end{pmatrix}
    \cdot
      \begin{pmatrix}
        \overline{[\gamma_{2A}\vert\vphi_a\ra}
        \\
        \overline{[\gamma_{23}\vert\vphi_a\ra}
    \end{pmatrix}
    \nn
\end{align}
where
\begin{align}
    [\gamma\vert\vphi_a\ra = 
    \int_{\gamma} \d z_1\ 
    e^{2\pi i s_{1A} z_1 }
    \left(
        \frac{\vth(z_{12}\vert \tau)}{\vth(z_{13}\vert \tau)}
    \right)^{s_{12}}
    F(z_{1a},\eta|\tau) \, ,
\end{align}
$a,b\in \{2,3\}$ and the integration contours are $\gamma_{2A}=\{z_1\in[z_2,z_2+1]\}$ and $\gamma_{23}=\{z_1\in[z_2,z_3]\}$. 
We also set $z_{ij}=z_i-z_j$ and often fix $z_2$ as the origin: $z_2=0$. 
In the complex integral of \eqref{eq:z1integral}, $M$ is the fundamental parallelogram with the punctures removed, and the measure is chosen such that $\int_P \d^2 z_1 = 1$. 
Here, $\vth(z\vert \tau)$ is the odd Jacobi-theta function, $F(z,\eta\vert \tau)$ is the Kronecker-Eisenstein series and $\{s_{12},s_{1A}\}$ are real numbers to be thought of as Mandelstam invariants and $\eta\in\mathbb{C}$. 
The genus-one \new{RW-}KLT formula \eqref{eq:z1integral} is contingent on the combination\footnote{\new{Mathematically, this condition ensures that the integrand in the LHS of \eqref{eq:z1integral} is be doubly periodic in $z_1$. We give a physical interpretation to this condition in connection to the integration of loop-momentum in string theory around \eqref{eq:realityOfSB}.}}
\begin{align}
    -\eta + s_{1A} \tau -s_{12} z_3 = s_{1B}  \, .
\end{align}
conspiring to a real constant $s_{1B}$ (we also generalize to complex Mandelstams $s_\bullet\in\mathbb{C}$).
\new{To our knowledge, this is the first genus-one KLT-like formula where the Mandelstam variables are generic and the modulus can have arbitrary real part: $\re(\tau)\neq0$. The closest examples to this formula in the literature are a pair of double-copy formulas exhibited in \cite{Stieberger:2023nol} for 2-point, 1-loop string amplitudes with integer Mandelstam variable $s_{12}$ and arbitrary real part of the modulus $\tau$.}

A similar double copy \new{formula} will be given for the generalization of the complex $z_1$-integral \eqref{eq:z1integral} with additional unintegrated punctures $z_4, z_5, \cdots, z_n$. 
At $n$-points, the \new{RW-}KLT kernel is an $(n-1)\times(n-1)$ matrix whose entries are known rational functions of the monodromies $e^{2\pi i s_\bullet}$. \new{Such matrices have been constructed and studied before in \cite{ghazouani2016moduli,Goto2022}}.

\paragraph{Outline of paper:}
In section \ref{sec:Fund}, we review the formalism of twisted (co)homology and the associated \new{construction of a double copy.} 
% in the context of tree-level string amplitudes. 
Then, in section \ref{sec:1L+RW}, we review one-loop string integrands and introduce the family of Riemann-Wirtinger integrals. 
The twisted cohomology of Riemann-Wirtinger integrals is \new{reviewed} in section \ref{sec:RWco-eta}.
\new{In particular, we provide a basis of differential forms compatible with the twisted (co)homology double copy in section \ref{sec:RWhom-basis}.}
\new{While not needed for the practical construction of the twisted (co)homology double copy, 
a well-defined intersection number --- an inner product on cohomology --- is essential for the existence of a double copy from twisted (co)homology. 
Therefore, we demonstrate how the intersection number is computed at genus-one in section \ref{sec:RWhom-int}
but relegate most of the details and results to appendix \ref{sec:RWco-app} to not distract from the flow of the RW double copy narrative.}
We review the twisted homology of Riemann-Wirtinger integrals in section \ref{sec:RWhom} and the intersection indices --- an inner product on homology --- for a basis of homology in section \ref{sec:RWhom-int}. 
These intersection indices make up the double copy kernel. 
The double copy of Riemann-Wirtinger integrals is constructed in section \ref{sec:DC-RW} and verified numerically for both real (section \ref{sec:real_s}) and complex (section \ref{sec:complex_s}) Mandelstams. 
In \ref{sec:modular} we write a version of the double copy of Riemann-Wirtinger integrals that has nice modular properties, which looks like a closed-integral after integrating out the loop momentum.
\new{Currently, it is not known how to define Riemann-Wirtinger integrals where two (or more) punctures are integrated.
Since understanding this is an important step towards making the twisted (co)homology double copy applicable to actual string integrands, we provide a discussion of the current problems preventing the definition of multi-puncture Riemann-Wirtinger integrals and conjecture how one might overcome these problems in section \ref{sec:doubleint}.}
We conclude in section \ref{sec:conclusion} with a discussion and future directions.

\new{For the interested reader, we have collected additional novel results on Riemann-Wirtinger integrals that would benefit any future work aimed at an analytic double copy in appendix \ref{sec:RWco-app}. 
} 
\new{In appendix \ref{sec:RWco-app-deqs}, we verify the Gauss-Manin connection satisfied by the RW family of integrals that was derived through independent means \cite{Mano2012}. 
While not directly used to verify the numeric double copy in the main text, the RW-DEQs are essential for obtaining analytic expressions for the RW integrals and their double copy in terms of eMPLs. 
In particular, we obtain the leading order solution in the $\alpha'$ expansion to the RW-DEQs in terms of eMPLs in appendix \ref{sec:RWco-app-alphaprime}.
We also provide boundary values for the RW-DEQs in appendix \ref{sec:RWco-app-bdval}.
Moreover, since only genus-zero intersection numbers have appeared in the physics literature, these genus-one intersection number calculations are interesting in their own right and have potential application to the integral reduction of hyperelliptic Feynman integrals \cite{Giroux:2022wav}.
}
% Using the intersection number --- an inner product on cohomology --- we verify the Gauss-Manin connection satisfied by this family of integrals that was derived through independent means in \cite{Mano2012} (section \ref{sec:RWco-app-alphaprime}).

%%%%%%%%%%%%%%%%%%%%%%%%%%%%%%%%%%%%%%%%%%%%%%%%%
%%%%%%%%%%%%%%%%%%%%%%%%%%%%%%%%%%%%%%%%%%%%%%%%%
%%%%%%%%%%%%%%%%%%%%%%%%%%%%%%%%%%%%%%%%%%%%%%%%%
\section{Fundamentals of twisted (co)homology and the double copy\label{sec:Fund}}

In this section, we give an overview of twisted (co)homology in the context of a simple example: the tree-level four-point open string amplitude (or the Euler beta function).
We slowly build up the twisted (co)homology formalism ending with the twisted Riemann bilinear relations. 
Then, we describe how these relations can be generalized to produce the double copy and express the tree-level four-point closed string amplitude as a square of the open string amplitudes.

Historically, twisted (co)homology has been primarily developed to study properties of hypergeometric functions; see the textbooks \cite{aomoto2011theory, yoshida2013hypergeometric} or 
\cite{deligne1986monodromy,
Matsumoto1994,
cho1995,
kita_matsumoto_1997,
Matsumoto1998,
matsumoto1998kforms,
hanamura1999hodge,
mimachi2002intersection,
Mimachi2003,
mimachi2004intersection,
Majima2000,
Goto2015,
Matsumoto:2018aa,
goto2020homology,
Matsubara-Heo:2020lzo,
MatsubaraHeo2023a}
for a very incomplete list of research papers. 
Fortunately for physicists, these authors are often concerned with explicit calculations and many worked examples can be found in their writtings.
Interestingly, the twisted (co)homology version of the double copy was first discovered by mathematicians studying Selberg integrals that appear in conformal field theory correlation functions \cite{hanamura1999hodge, mimachi2002intersection, Mimachi2003, mimachi2004intersection, Matsubara-Heo:2020lzo}. 
More recently, the twisted (co)homology double copy  was proved from a motivic perspective \cite{Brown:2018omk,Brown:2019jng}.
In particular, we want to test whether the double copy prescription of \cite{hanamura1999hodge, mimachi2002intersection, Mimachi2003, mimachi2004intersection, Brown:2018omk,Brown:2019jng, Matsubara-Heo:2020lzo} holds at higher genus and access the possible application to one-loop closed string amplitudes.

%%%%%%%%%%%%%%%%%%%%%%%%%%%%%%%%%%%%%%%%%%%%%%%%%
%%%%%%%%%%%%%%%%%%%%%%%%%%%%%%%%%%%%%%%%%%%%%%%%%
\subsection{A simple example}

To understand the construction of twisted (co)homology and its relation to the double copy, it is best to have a simple example in mind.
To this end, consider the tree-level massless four-point open string integral (Euler beta function) 
\begin{align} \label{eq:beta}
    \A^\open(s,t) \propto \int_0^1 \d z\ z^{s-1} (1-z)^{t-1}
\end{align}
where we have used the $\text{SL}(2,\mathbb{C})$ gauge symmetry to fix three of the punctures to $0,1$ and $\infty$ and are not concerned with the overall constant kinematic factors and the momentum conserving delta function. 
It satisfies interesting linear relations such as 
\begin{align} \label{eq:A4_open_contiguity}
\begin{aligned}
    \A^\open(s+1,t) = \frac{s}{s+t} \A^\open(s,t)
    \,, 
    \qquad\text{and}\qquad
    \A^\open(s,t+1) = \frac{t}{s+t} \A^\open(s,t)
    \,, 
\end{aligned}
\end{align}
as well as, the quadratic identity 
\begin{align} \label{eq:A4_open_quad}
    \A^\open(s,t) \A^\open(-s,-t) 
    &= - \pi \left(\frac{1}{s}+\frac{1}{t}\right)
        \left(\frac{1}{\tan(\pi s)}+\frac{1}{\tan(\pi t)}\right)
        . 
\end{align}
The analogue of \eqref{eq:A4_open_contiguity} in quantum field theory corresponds to the dimension shifting identities and differential equations satisfied by dimensionally regulated Feynman integrals. 
While Feynman integrals also satisfy quadratic identities like \eqref{eq:A4_open_contiguity} they are not particularly well studied in the literature. 
The corresponding closed string amplitude is 
\cite{Mizera:2017cqs}
\begin{align} \label{eq:A4_closed}
    \A^\closed (s,t) \propto \int_{\mathbb{C}} 
    |z|^{2s} |1-z|^{2t}
    \frac{\d z \wedge \d \bar{z}}{|z|^2 |1-z|^2}
    =
    \frac{
        - \pi (s+t)^2
        \Gamma(s) \Gamma(t) \Gamma(-s-t)
    }{
        \Gamma(1-s) \Gamma(1-t) \Gamma(1+s+t)
    }
    \,.
\end{align}
Again, we have dropped the proportionality factors and gauge fixed three of the punctures. 
Of course it is well known that the integral \eqref{eq:A4_closed} admits a double copy
\begin{align} \label{eq:A4_closed_DC}
    \A^\closed(s,t) = \left(\frac{1}{\tan(\pi s)}+\frac{1}{\tan(\pi t)}\right)^{-1}
    \left( \A^\open(s,t) \right)^2
    \,.
\end{align}
That is, gravity (closed string) amplitudes are the square of gauge (open string) amplitudes.

%%%%%%%%%%%%%%%%%%%%%%%%%%%%%%%%%%%%%%%%%%%%%%%%%
%%%%%%%%%%%%%%%%%%%%%%%%%%%%%%%%%%%%%%%%%%%%%%%%%
\subsection{Lightning review of twisted (co)homology}

Twisted (co)homology provides a systematic formalism for understanding relations such as \eqref{eq:A4_open_contiguity}, \eqref{eq:A4_open_quad}, \eqref{eq:A4_closed_DC} and more!
Roughly speaking, twisted (co)homology is the (co)homology theory for integrals whose integrands have  prescribed multiplicative monodromies.
More explicitly, the theory of twisted (co)homology applies to integrals of the form 
\begin{align} \label{eq:twisted period}
    \int_\gamma u\ \vphi
\end{align}
where $u$ is a universal multi-valued function called the twist, $\vphi$ is a single-valued differential form on the space $M=E\setminus\{u=0,\infty\}$ and $\gamma$ is a contour on $M$.
In this language, equation \eqref{eq:beta} becomes 
\begin{align}
     \A^\open(s,t) = \int_{0}^{1} u\ \vphi
     \,,
\end{align}
with 
\begin{align}
     u = z^s(1-z)^t,
     \quad
     \vphi = \d\log\frac{z}{1-z}
     \quad\text{and}\quad
     M = \mathbb{CP} \setminus \{0,1,\infty\}
     \,.
\end{align}
Note that the monodromies of the twist in this example are $e^{2\pi i s}$ and $e^{2\pi i t}$ as $z$ winds around the origin and $z=1$ (there is also the monodromy $e^{-2\pi i(s+t)}$ as $z$ winds about the point $\infty$). 
The so-called \emph{dual}\footnote{The adjective ``dual'' is a matter of convention and its significance will become clear once we have introduced the intersection pairings.}
local system,
\begin{align}
    \c{\mathcal{L}}(s,t) = \mathbb{C}u,
\end{align}
keeps track of the branch choices of $u$.
For this reason, twisted (co)homology is also known as (co)homology with coefficients in a local system. 

Importantly, the integral \eqref{eq:twisted period} is not well defined until a branch choice for $u$ is made. 
Thus, twisted contours must store information associated to the topological integration contour $\gamma$ as well as a branch choice $u_\gamma$ on this contour. 
Explicitly, a twisted contour is the tensor product $\gamma \otimes u_\gamma$ where $\otimes u_\gamma$ is the aforementioned coefficient in the dual local system. 
Returning to our example, we can now write the 
open-string four-point amplitude 
as an integral over a twisted contour:
\begin{align} \label{eq:beta_twisted_period}
    % B(a,b) = 
    \A^\open(s,t) =
    \int_{(0,1) \otimes u} \vphi
    \,. 
\end{align}
While the differential form $\vphi$ is an ordinary differential form on $M$, we will also call it a twisted differential form.\footnote{Technically, one should work with $\mathcal{L}$-valued differential forms. However, since the local system is a trivial line bundle, there exists a global trivialization and we can work with ordinary complex differential forms without loss of information.}

Even in the twisted setting, there exists a version of Stokes' theorem \cite{aomoto2011theory, yoshida2013hypergeometric}
\begin{align}
    \int_{\gamma \otimes u_\gamma} \nabla_\omega \vphi
    = \int_\gamma u_\gamma \nabla_\omega \vphi
    = \int_\gamma \d(u_\gamma\ \vphi) 
    = \int_{\partial\gamma} u_\gamma\ \vphi
    = \int_{\partial_\omega(\gamma \otimes u_\gamma)} \vphi
    \;. 
\end{align}
Here, we have introduced the covariant derivative $\nabla_\omega := \d + \omega \wedge$ where $\omega := \d\log(u)$ is a flat connection $\omega\wedge\omega=0$ (this guarantees that the covariant derivative is nilpotent: $\nabla_\omega \circ \nabla_\omega = 0$).
% , as well as, implicitly defined the action of the boundary operator on twisted cycles. 
% Importantly, the ordinary exterior differential is replaced by the covariant derivative $\nabla$ in the twisted setting.
Since the total derivative of the multi-valued integrand $\d(u\ \vphi)$ is equivalent to the covariant derivative of the single-valued differential form $\nabla_\omega \vphi$, we can effectively ``forget'' about $u$ and work only with single-valued differential forms by replacing the normal exterior derivative with the covariant derivative.

The boundary of a twisted contour is defined to be the topological boundary of the contour with each boundary component tensored with the corresponding branch induced from the original branch choice.
This is most easily illustrated on a one-dimensional contour that has been smoothly triangulated. 
Using the standard notation for $m$-simplicies, the (triangulated) path from point $a$ to $b$ is denoted by $\la ab \ra$ and has boundary $\partial \la ab \ra = \la b \ra - \la a \ra$. 
Then, the boundary of the corresponding twisted contour $\la ab \ra \otimes u_{\la ab \ra}$ is 
\begin{align}
    \partial_\omega \left( \la ab \ra \otimes u_{\la ab \ra}(z) \right)
    := \la b \ra \otimes u_{\la b \ra}(b) 
    - \la a \ra \otimes u_{\la a \ra}(a)
    \,,
\end{align} 
where the branch choices $u_{\la b \ra}$ and $u_{\la b \ra}$ are induced from $u_{\la ab \ra}$ by restriction.
While the generalization to higher dimensional simplices is straightforward, it will not be needed in this work. 
The interested reader is encouraged to consult \cite{aomoto2011theory, yoshida2013hypergeometric} as well as \cite{mimachi2003intersection, mimachi2004intersection, Mizera:2017cqs, Casali:2019ihm, Casali:2020knc, Duhr:2023bku,Baune:2023uut}. 

In our running example, both the twisted contour and twisted differential form are \emph{closed} (i.e., their respective image under $\partial_\omega$ or $\nabla_\omega$ vanishes). 
The twisted contour in the definition of the beta function is closed because $u$ vanishes at the end points
\begin{align}
    \partial_\omega \left[ (0,1) \otimes u_{(0,1)}(z) \right]
    = \partial_\omega \left[ \la 01 \ra \otimes u_{\la01\ra}(z) \right]
    = \la1\ra \otimes u_{\la1\ra}(1) - \la0\ra \otimes u_{\la0\ra}(0)
    = 0
    \,.
\end{align}
On the other hand, the twisted differential form in \eqref{eq:beta_twisted_period} is closed because it is closed in the usual sense (has no image under $\d$) and has vanishing wedge product with $\omega$:
\begin{align}
    \nabla_\omega \left( \d\log\frac{z}{1-z} \right)
    = \d^2 \log\frac{z}{1-z} + \omega \wedge \d \log\frac{z}{1-z} 
    = 0
    \,.
\end{align}
Closed twisted contours and differential forms are called twisted \emph{cycles} and \emph{cocycles} respectively. 
Intuitively, cycles are contours with no boundary.
Unfortunately, cocycles do not have a simple intuitive interpretation --- they are simply differential forms that vanish when (covariantly) differentiated.

While twisted cycles and cocycles are the ``interesting'' twisted contours/differential forms on $M$, they are not uniquely defined. 
The integral of a twisted cocycle ($\vphi$) over a twisted cycle ($\gamma \otimes u_\gamma$) does not change when the twisted cycle (cocycle) is shifted by an \emph{exact} twisted contour (differential form).
Here, an exact twisted contour is the image of the boundary operator ($\partial_\omega(\delta \otimes u_\delta)$) and an exact twisted differential form is the image of the covariant derivative ($\nabla_\omega \psi$).
Using Stokes theorem, it is easy to verify that integrals over twisted (co)cycles are indeed invariant under such shifts
\begin{align}
    \int_{\gamma \otimes u_\gamma} \left( \vphi + \nabla_\omega \psi \right) 
   & = \int_{\gamma \otimes u_\gamma} \vphi + \int_{\partial_\omega(\gamma \otimes u_\gamma)} \psi 
    = \int_{\gamma \otimes u_\gamma} \vphi 
    \,,
\\
    \int_{\gamma \otimes u_\gamma + \partial_\omega(\delta \otimes u_\delta)} \vphi
    &= \int_{\gamma \otimes u_\gamma} \vphi
    + \int_{\delta \otimes u_\delta} \nabla_\omega \vphi
    = \int_{\gamma \otimes u_\gamma} \vphi
    \,,
\end{align}
since $\partial_\omega (\gamma\otimes u_\gamma) = 0 = \nabla_\omega \vphi$ by definition.
To remove the above redundancy in defining twisted (co)cycles, it is useful to define the twisted (co)homology and work with equivalence classes of (co)cycles. 

The $p$-th twisted homology $H_p$ group is the space of closed twisted $p$-contours modulo exact twisted $(p+1)$-contours
\begin{align}
    H_p(M,\check{\mathcal{L}}_\omega)
    &= \frac{
        \ker\left(
            \partial_\omega: 
            C_{p}(M,\check{\mathcal{L}}_\omega) 
            \to 
            C_{p-1}(M,\check{\mathcal{L}}_\omega)
        \right)
    }{
        \text{im}\left(
            \partial_\omega: 
            C_{p+1}(M,\check{\mathcal{L}}_\omega)
            \to 
            C_p(M,\check{\mathcal{L}}_\omega)
        \right)
    }
    \,,
\end{align}
where $C_p(M,\c{\L}_\omega)$ is the space of twisted $p$-contours. 
Its elements (twisted homology classes) are equivalence classes of twisted cycles and are denoted by a square bra.
Explicitly, given a twisted $p$-cycle $\gamma \otimes u_\gamma\in C_p(M,\c{\L}_\omega)$, its twisted homology class is
\begin{align}
    [\gamma \otimes u_\gamma \vert 
    = \{
        \gamma \otimes u_\gamma + \partial_{\omega} (\beta \otimes u_\beta) \,
        \vert \,
        \forall\, (\beta \otimes u_\beta) \in C_{p+1}(M,\c{\L}_\omega)
    \}
    \,.
\end{align}
When two twisted $p$-cycles, $\gamma \otimes u_\gamma$ and $\gamma^\prime \otimes u_{\gamma^\prime}$, differ by the boundary of a twisted $(p+1)$-contour $\beta \otimes u_\beta$, 
\begin{align}
    \gamma \otimes u_\gamma \simeq \gamma^\prime \otimes u_{\gamma^\prime}
    \iff
    \gamma \otimes u_\gamma - \gamma^\prime \otimes u_{\gamma^\prime}
    = \partial_\omega(\beta \otimes u_\beta) 
    \,,
\end{align}
we say that they are homologous or equivalent in homology.

Similarly, the $p$-th twisted cohomology $H^p$ group is the space of closed twisted $p$-forms modulo exact twisted $(p-1)$-forms
\begin{align}
    H^p(M,\nabla_\omega)
    &= \frac{
        \ker\left(
            \nabla_\omega: 
            \Omega^{p}(M,\nabla_\omega) 
            \to 
            \Omega^{p+1}(M,\nabla_\omega)
        \right)
    }{
        \text{im}\left(
            \nabla_\omega: 
            \Omega^{p-1}(M,\nabla_\omega)
            \to 
            \Omega^p(M,\nabla_\omega)
        \right)
    }
    \,,
\end{align}
where $\Omega^p(M,\nabla_\omega)$ is the space of twisted $p$-forms.
Its elements (twisted cohomology classes) are equivalence classes of twisted cocycles and are denoted by an angle ket.
Explicitly, given a twisted $p$-cocycle $\vphi \in \ker( \nabla_\omega: \Omega^p(M,\nabla_\omega) \to \Omega^{p+1}(M,\nabla_\omega))$, its twisted cohomology class is
\begin{align}
    \vert \vphi \ra
    = \{
        \vphi + \nabla_{\omega} \psi \,
        \vert \,
        \forall\, \psi \in \Omega^{p-1}(M,\nabla_\omega)
    \}
    \,.
\end{align}
When two twisted $p$-cocycles, $\vphi$ and $\vphi^\prime$, differ by the covariant derivative of a twisted $(p-1)$-form $\nabla_\omega \psi$, 
\begin{align}
    \vphi \simeq \vphi^\prime 
    \iff
    \vphi - \vphi^\prime 
    = \nabla_\omega \psi
    \,,
\end{align}
we say that $\vphi$ and $\vphi^\prime$ are cohomologous or equivalent in cohomology.
In the physics literature, shifting a differential form by a total derivative is commonly known as integration-by-parts and is the standard method for reducing integrals that appear in the calculation of multi-loop perturbative amplitudes to a minimal basis of so-called master integrals.
Developing fast integration-by-parts computer algorithms and researching alternative integral reduction methods is currently an active area of research 
\cite{Klappert:2020nbg, Barakat:2022qlc, Smirnov:2023yhb, Wu:2023upw, Heinrich:2023til} since integral reduction is essential to precision phenomenology.
The intersection number, introduced below, is one of the most promising alternatives to traditional integration-by-parts algorithms being developed
\cite{Mastrolia:2018uzb, Mizera:2019ose, Frellesvig:2019kgj, Frellesvig:2019uqt, Mizera:2019vvs, Chen:2020uyk, Frellesvig:2020qot, Caron-Huot:2021xqj, Caron-Huot:2021iev, Chen:2022lzr, Chen:2023kgw}.

An important fact about twisted (co)homology is that only the middle degree (co)homology is non-trivial
\begin{align} \label{eq:dimH}
    H_p(M,\c{\L}_\omega) = 0
    = H^p(M,\nabla_\omega)  \quad\text{if}\quad p\neq n
    \,,
\end{align}
where $n$ is the complex dimension of $M$ ($n=\dim_{\mathbb{C}} M $) or equivalently half the real dimension of $M$ ($ n= \frac{1}{2}\dim_{\mathbb{R}} M$).
\new{As a consequence, the dimension} of the middle (co)homology groups is given by the Euler characteristic of $M$ \cite{aomoto2011theory}\footnote{In the context of Feynman integrals, it is well known that the Euler characteristic gives the number of independent integrals (dimension of the cohomology) \cite{Bitoun:2018afx}.}
\begin{align}
    \dim H_n(M,\c{\L}_\omega) 
    = \dim H^n(M,\nabla_\omega)
    = \vert \chi(M) \vert
    \,.
\end{align}
Another important fact is that twisted (co)homology groups are actually vector spaces (existence of an inner product structure)!

%%%%%%%%%%%%%%%%%%%%%%%%%%%%%%%%%%%%%%%%%%%%%%%%%
%%%%%%%%%%%%%%%%%%%%%%%%%%%%%%%%%%%%%%%%%%%%%%%%%
\subsection{Pairings \label{sec:pairings}}

\begin{figure}
    \centering
    \includegraphics[width=.5\textwidth]{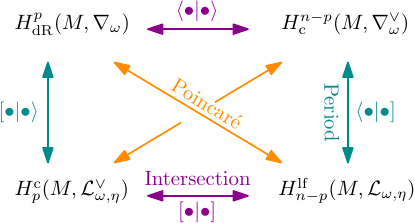}
    \caption{Diagram detailing the connections between twisted (co)homology groups and their duals.
    The {\color{BlueGreen!70!black} period pairings} (integration) are denoted by the vertical arrows, while the {\color{Plum} intersection pairings} are denoted by the horizontal arrows. 
    {\color{YellowOrange}Poincar\'e duality} is indicated by the diagonal arrows. 
    We have also included some subscripts and superscripts for the experts and can be safely ignored by readers unfamiliar with twisted (co)homology.
    }
    \label{fig:dualities}
\end{figure}

Now that we have defined the twisted (co)homology, we can interpret integration as a pairing between homology and cohomology 
\begin{align}
    [\bullet \vert \bullet \ra: 
    H_p(M,\c{\L}_\omega) \times H^p(M,\nabla_\omega) 
    \to \mathbb{C}
    \qquad 
    \text{where}
    \qquad
    [\gamma \otimes u_\gamma \vert \vphi \ra
    := \int_{\gamma\otimes u_\gamma} \vphi
    \,.
\end{align}
Formally, this pairing is know as the \emph{period} pairing. 
In particular, we can now think of the four-point open-string amplitude \eqref{eq:beta} as a twisted period
\begin{align} \label{eq:beta_P}
    \A^\open(s,t) =
    \bigg[(0,1) \otimes z^a (1-z)^b
    \ \bigg\vert \
    \d\log\frac{z}{1-z} \bigg\ra
    \,.
\end{align}
However, it turns out that we can construct other pairings that involve only homology or only cohomology: 
the \emph{intersection index}\footnote{Here, we are following the naming convention of Pham \cite{hwa1966homology, pham2011singularities}. Often the intersection index is called the homological intersection number. We prefer intersection index because it is more succinct.}
and \emph{intersection number}. 
While it may seem unnatural to try and pair cycles with cycles and cocycles with cocycles, these pairings define inner products on the vector space of homology and cohomology.
Using the intersection pairings we can project any twisted (co)cycle onto a basis of (co)homology. 
In particular, this provides a systematic way of discovering linear relations between integrals.
Moreover, both intersection pairings are essential for constructing a double copy. 

The intersection pairing pairs the twisted (co)homology with an associated dual twisted (co)homology. 
The dual twisted (co)homology is simply defined to be the twisted (co)homology associated to the inverse twist: $\c{u}:=u^{-1}$:
\begin{align}
    [H_p(M,\c{\L}_\omega) ]^\vee 
    := H_p(M,\L_\omega)
    \qquad \text{and} \qquad 
    [H^p(M,\nabla_\omega) ]^\vee 
    := H^p(M,\c{\nabla}_\omega)
\end{align}
where $\L_\omega = \mathbb{C}u^{-1}$ is the local system and $\c{\nabla}_\omega = \nabla_{-\omega} = \d - \omega\wedge$ is the dual covariant derivative.
Dual integrals or dual periods correspond to the following pairing 
\begin{align}
    \la\bullet \vert \bullet ]: 
    H^p(M,\c{\nabla}_\omega) \times H_p(M,\L_\omega)
    \to \mathbb{C}
    \qquad 
    \text{where}
    \qquad
    \la \c{\vphi} \vert \c{\gamma} \otimes \c{u} ] :=
    \int_{\c{\gamma}\otimes \c{u}_{\c{\gamma}}} \c{\vphi}
    \,,
\end{align}
where $\vert\c{\gamma}\otimes\c{u}_{\c{\gamma}}]$ and $\la \c{\vphi} \vert$ denote dual homology and cohomology classes.
For example, the integral $\A^\open(-s,-t)$ corresponds to the dual period
\begin{align} \label{eq:beta_Pdual}
    \A^\open(-s,-t) = \bigg\la \d\log\frac{z}{1-z} \bigg\vert (0,1) \otimes z^{-s} (1-z)^{-t} \Bigg]
    \,.
\end{align}
From Poincar\'e duality \cite{aomoto2011theory, pham2011singularities} (see figure \ref{fig:dualities}), the (normal) twisted cohomology is isomorphic to the dual homology $H^p(M,\nabla_\omega) \cong H_{n-p}(M,\mathcal{L}_\omega)$ and the (normal) twisted homology is isomorphic to the dual cohomology $H_p(M,\c{\L}_\omega) \cong H^{n-p}(M,\c{\nabla}_\omega)$. 
Moreover, the \emph{dual} local system $\c{\L}_\omega$ corresponds to the functions that are annihilated by the \emph{dual} covariant derivative ($\c{\nabla} (\mathbb{C} u)=0$) while the local system $\L_\omega$ corresponds to the functions that are annihilated by the covariant derivative ($\nabla (\mathbb{C} u^{-1})=0$). 
This is why the dual local system appears in the (normal) twisted homology.

Now that the dual (co)homology has been defined, we can introduce the intersection pairings 
\begin{align} \label{eq:Hint}
\begin{aligned}
    % [\gamma \otimes u_\gamma \vert 
    % \c{\gamma} \otimes \c{u}_{\c{\gamma}} ]:
    [ \bullet \vert \bullet ]:
    &\
    H_p(M,\c{\L}_\omega) \times 
    H_{n-p}(M,\L_\omega) \to \mathbb{C}
    \\&
    \quad\text{by}\quad
    [\gamma \otimes u_\gamma |\c{\gamma} \otimes \c{u}_{\c{\gamma}}] 
    := \sum_{x \in \gamma \cap \c{\gamma}} 
    \left(u_\gamma \c{u}_{\c{\gamma}}\vert_{x}\right) 
    \big[\reg[\gamma] \vert \c{\gamma}\big]^{\text{top}}_x
    \,,
\end{aligned}
\end{align}
and
\begin{align} \label{eq:Cint}
    % \la \c{\vphi} \vert \vphi \ra: 
    \la \bullet \vert \bullet \ra: 
    &\
    H^{n-p}(M,\c{\nabla}_\omega) \times 
    H^p(M,\nabla_\omega) \to \mathbb{C}
    \quad\text{by}\quad
    \la\c{\vphi}\vert\vphi\ra 
    := \int_{M} \reg[\c{\vphi}]\wedge\vphi
    \,.
\end{align}
In \eqref{eq:Hint} and \eqref{eq:Cint}, the dual cycle or dual cocycle must be regularized.
In particular, they must be replaced by a (co)homologous (co)cycle that has compact support. 
For cycles, this regularization procedure introduces rational functions of the monodromies (i.e., $e^{2\pi i s}$ and $e^{2\pi i t}$). 
On the other hand, the regularization of dual cocycles introduces rational functions of the exponents in the twist (i.e., $s$ and $t$).
We will postpone the explicit recipe for the regularization procedure to sections \ref{sec:RWco-eta} and \ref{sec:RWhom}.

In definition \eqref{eq:Hint}, the set $\gamma\cap\c{\gamma}$ corresponds to the set of all points where $\gamma$ and $\c{\gamma}$ intersect. 
Here, $[\bullet \vert \bullet ]^{\text{top}}_x$ is the topological intersection number at $x$ that evaluates to $\pm1$ depending on the relative orientation of $\gamma$ and $\c{\gamma}$. 
Additionally, the factor $(u_\gamma \c{u}_{\c{\gamma}}\vert_x)$ simply evaluates to a phase since $\c{u}=u^{-1}$.
Thus, we see that the intersection index \eqref{eq:Hint} counts the number of times a (regularized) twisted cycle and dual cycle intersects (with sign) weighted by rational functions of the monodromies. 
Similarly, the intersection number \eqref{eq:Cint} counts the number of overlapping singularities between twisted forms and dual forms weighted by rational functions of the exponents in the twist.
While not obvious without knowing the regularization procedure, the integral in  \eqref{eq:Cint} is simple to compute and always evaluates to a sequence of residues!

Using the intersection pairings as inner products on (co)homology, we have the following decompositions of identity
\begin{align} \label{eq:decomp_id}
    \mathds{1} = 
    \left\vert 
        \c{\gamma}_i \otimes \c{u}_{\c{\gamma}_i} 
    \right]
    H_{ij}^{-1}
    \left[ 
        \gamma_j \otimes u_{\gamma_j} 
    \right\vert
    \qquad
    \text{and}
    \qquad
    \mathds{1} = 
    \left\vert 
        \vphi_i  
    \right\ra
    C_{ij}^{-1}
    \left\la
        \c{\vphi}_j
    \right\vert
    \,.
\end{align}
Here, $H_{ij} = [\gamma_i \otimes u_{\gamma_i} \vert \c{\gamma}_j \otimes u_{\c{\gamma}_j} ]$ is the homology intersection matrix corresponding to the bases $\{ [ \gamma_i \otimes u_{\gamma_i} \vert \}$ and $\{ \vert \c{\gamma}_i \otimes u_{\c{\gamma}_i} ] \}$. 
Similarly, $C_{ij} = \la \c{\vphi}_i  \vert \vphi_j \ra$ is the cohomology intersection matrix corresponding to the bases $\{ \vert \vphi_i \ra \}$ and $\{ \la \c{\vphi}_i \vert \}$. 
Using these formula, any (co)cycle can be projected onto a  basis.

In our example of the beta function, only the first (co)homology is non-trivial. Moreover, it is a one dimensional vector space since $|\chi(M)|=1$. 
Therefore, we can choose the bases of (co)homology and dual (co)homology to be 
\begin{align}
    &H_1(M,\c{\L}_\omega) {=} \text{Span}\left\{
        \left[ (0,1) \otimes z^s(1-z)^t \right\vert
    \right\}
    \,,
    \quad
    H^1(M,\nabla_\omega) {=} \text{Span}\left\{
        \left\vert \d\log\frac{z}{1-z} \right\ra
    \right\}
    \,,
    \\
    &H_1(M,\L_\omega) {=} \text{Span}\left\{
        \left\vert (0,1) \otimes z^{-s}(1-z)^{-t} \right]
    \right\}
    \,,
    \quad
    H^1(M,\c{\nabla}_\omega) {=} \text{Span}\left\{
        \left\la \d\log\frac{z}{1-z} \right\vert
    \right\}
    \,.
\end{align}
Note that one can choose the same topological cycles and differential forms for the dual bases since the underlying topology of $M$ is the same.
With these choices, equation \eqref{eq:decomp_id} becomes
\begin{align}
    \mathds{1} = 
    \left\vert 
        (0,1) \otimes z^{-s}(1-z)^{-t} 
    \right]
    H^{-1}
    \left[ 
        (0,1) \otimes z^s(1-z)^t
    \right\vert
\end{align}
and
\begin{align}
    \mathds{1} = 
    \left\vert 
        \d\log\frac{z}{1-z} 
    \right\ra
    C^{-1}
    \left\la
        \d\log\frac{z}{1-z} 
    \right\vert
    \,,
\end{align}
where 
\begin{align} \label{eq:beta_HC}
    H = \frac{i}{2} \left(\frac{1}{\tan(\pi s)} + \frac{1}{\tan(\pi t)}\right)
    \,,
    \qquad\text{and}\qquad
    C = (2\pi i) \frac{s+t}{st}
    \,.
\end{align}
Now, we can understand the origin of the first equation of \eqref{eq:A4_open_contiguity} by acting with the decomposition of identity on $\vert z\ \d\log\frac{z}{1-z}\ra$:
\begin{align}
    \bigg\vert 
        z\ \d\log\frac{z}{1-z}
    \bigg\ra
    = 
    \frac{s}{s+t}
    \bigg\vert 
        \d\log\frac{z}{1-z}
    \bigg\ra
    \,,
\end{align}
where $\la z\ \d\log\frac{z}{1-z} \vert \d\log\frac{z}{1-z} \ra = \frac{s}{s+t} C $.
We also remark that the leading term in the $\alpha'$ expansion of $H$ corresponds to a four-point bi-adjoint scalar amplitude ($s,t \propto \alpha')$.

%%%%%%%%%%%%%%%%%%%%%%%%%%%%%%%%%%%%%%%%%%%%%%%%%
%%%%%%%%%%%%%%%%%%%%%%%%%%%%%%%%%%%%%%%%%%%%%%%%%
\subsection{Quadratic identities and the double copy}

Having introduced all the pieces featured in
\eqref{eq:A4_open_contiguity} (equations \eqref{eq:beta_P}, \eqref{eq:beta_Pdual} and \eqref{eq:beta_HC}), we turn our attention to the quadratic identity \eqref{eq:A4_open_quad}.
This identity follows from the twisted Riemann bilinear relations (also known as twisted period relations)
\begin{align} \label{eq:Rbilin}
    \mat{C} 
    =  \left(
        \c{\mat{P}} \cdot \mat{H}^{-1} \cdot \mat{P}
    \right)^\top
    \qquad \text{or} \qquad
    \mat{H} 
    =  \mat{P} \cdot \mat{C}^{-1} \cdot \c{\mat{P}}
    \,.
\end{align}
Here, $\mat{P}$ and $\c{\mat{P}}$ are the twisted period matrices for a given basis choice of the twisted (co)homology and dual (co)homology,  
\begin{align} \label{eq:Pmats}
    P_{ij} = \big[ \gamma_i \otimes u_{\gamma_i} \big\vert \vphi_j \big\ra
    \qquad\text{and}\qquad
    \c{P}_{ij} = \big\la \c{\vphi}_i \big\vert \c{\gamma}_j \otimes u_{\c{\gamma}_j} \big]
    \,.
\end{align}
Now, while \eqref{eq:A4_open_quad} and \eqref{eq:Rbilin} are not quite what we mean by a double copy they are only one step away. 
In the remainder of this section we will describe how to generalize equation \eqref{eq:Rbilin} and obtain the double copy \eqref{eq:A4_closed_DC}. 

The closed string amplitude \eqref{eq:A4_closed} already looks like a double copy since its integrand is simply two copies of the same thing (with one complex conjugated)
\begin{align}
    \A^\closed(s,t) 
    = \int_\mathbb{C} \overline{(u \ \vphi)} \wedge (u\ \vphi)
    = \int_\mathbb{C} \vert u \vert^2 \
    \bar{\vphi} \wedge \vphi
\end{align}
where $\vert u \vert^2 = |z|^{2s}|1-z|^{2t}$ and $\vphi = \d\log\frac{z}{1-z}$.
It also looks like the intersection number \eqref{eq:Cint} where we use the complex conjugate of $\vphi$ instead of the regulated dual form. 
That is, we can interpret $\A^\closed$ as the pairing 
\begin{align}
    \la\bar{\bullet}\vert\bullet\ra: 
    H^n(\bar{M},\nabla_{\bar\omega}) \times H^n(M,\nabla_\omega)
    \to \mathbb{C} 
    \quad \text{by} \quad 
    \la\bar{\vphi}_a\vert\vphi_b\ra :=
    \int_M \vert u \vert^2 \
    \bar{\vphi}_a \wedge \vphi_b
    \,.
\end{align}
Note that the complex conjugated dual local system has the same monodromies as the local system for real exponents ($s,t \in \mathbb{R}$).
For example,  $\bar{u} \to \overline{e^{2\pi i s}} \bar{u} = e^{-2\pi i s} \bar{u}$ and  $u^{-1} \to e^{-2\pi i s} u^{-1}$ as $z$ winds around the origin.
Thus, the local system is isomorphic to the complex conjugate of the dual local system: 
\begin{align}
    \mathbb{C} u^{-1} =: \L_\omega(s,t) \cong \overline{\c{\L}_\omega(s,t)} := \mathbb{C} \bar{u}
\end{align}
for real exponents $s$ and $t$.
Replacing dual (co)cycles by complex conjugated (co)cycles in the Riemann bilinear relations yields the \emph{double copy}
\begin{align} \label{eq:DC}
    \boxed{
    \la\bar{\vphi}_a\vert\vphi_b\ra 
    := \int_M \vert u \vert^2 \
    \bar{\vphi}_a \wedge \vphi_b
    =  \left(\
        \overline{\mat{P}} \cdot \mat{H}^{-1} \cdot \mat{P} \
    \right)_{ba}
    }
    \,.
\end{align}
Note that since the $\overline{\c\L_\omega} \cong \L_\omega$, we do not need to recompute the intersection matrix $\mat{H}$. 
Moreover, since the intersection matrix $\mat{H}$ is meromorphic in the exponents of the twist, we can take these exponents to be complex.
We can check that the above prescription indeed reproduces equation \eqref{eq:A4_closed_DC}
\begin{align} \label{eq:A4_closed_DC_example}
    \bar{P} H^{-1} P 
    &= \left(\frac{1}{\tan(\pi s)}+\frac{1}{\tan(\pi t)}\right)^{-1}
    \left( \A^\open(s,t) \right)^2
    \,,
    \nn\\
    &=\frac{
        - \pi (s+t)^2
        \Gamma(s) \Gamma(t) \Gamma(-s-t)
    }{
        \Gamma(1-s) \Gamma(1-t) \Gamma(1+s+t)
    }
    \,,
    \nn\\
    &=\int_{\mathbb{C}} 
    |z|^{2s} |1-z|^{2t}
    \frac{\d z \wedge \d \bar{z}}{|z|^2 |1-z|^2}
    =\A^\closed(s,t) 
    \,.
\end{align}
Note that the double copy \eqref{eq:DC} (and \eqref{eq:A4_closed_DC_example}) are manifestly single-valued. 
Thus, the double copy procedure is a mechanism to generate interesting single-valued integrals.

The main goal of this work is to explore the structure of twisted cohomology at higher genus and test if the double copy prescription of \cite{Brown:2019jng} holds. 
For one-loop string amplitudes, the KLT kernel has been long sought after
\cite{Stieberger:2014hba, Tourkine:2016bak, Stieberger:2016xhs, Ochirov:2017jby, Hohenegger:2017kqy, Casali:2019ihm, Casali:2020knc, Stieberger:2021daa, Stieberger:2022lss} culminating in \cite{Stieberger:2022lss, Stieberger:2023nol}. 
Our goal is to asses whether twisted (co)homology is indeed the correct formalism to compute KLT kernels for higher genus string amplitudes. 
Our hope is that twisted (co)homology could simplify the procedure of determining the KLT kernel and turn the complicated contour deformation arguments in \cite{Stieberger:2022lss, Stieberger:2023nol} into more linear algebra like statements.

%%%%%%%%%%%%%%%%%%%%%%%%%%%%%%%%%%%%%%%%%%%%%%%%%
%%%%%%%%%%%%%%%%%%%%%%%%%%%%%%%%%%%%%%%%%%%%%%%%%
%%%%%%%%%%%%%%%%%%%%%%%%%%%%%%%%%%%%%%%%%%%%%%%%%
\section{One-loop string integrals and the Riemann-Wirtinger integral \label{sec:1L+RW}}

In this section, we introduce one-loop string integrals after chiral splitting (section \ref{sec:1-loop_string}) and explain how the Riemann-Wirtinger family of integrals (the main focus of this work) is related (section \ref{sec:RW_integral_foundations}).

%%%%%%%%%%%%%%%%%%%%%%%%%%%%%%%%%%%%%%%%%%%%%%%%%
%%%%%%%%%%%%%%%%%%%%%%%%%%%%%%%%%%%%%%%%%%%%%%%%%
\subsection{One-loop string integrals \label{sec:1-loop_string}}

In the  chiral splitting formalism \cite{DHoker:1988pdl}, $n$-point one-loop open-string integrals take the form 
\begin{align} \label{eq:1LSI}
    &\int \d^{D}\ell \int \d\tau\
    e^{\pi i \alpha^\prime \tau \ell^2 }
   \int \d^n z\
    e^{2\pi i \alpha^\prime \ell\cdot \sum_{j=1}^n k_j z_j}
    \prod_{1 \leq j < k < n}
    e^{\alpha^\prime k_j \cdot k_k \log \frac{\vth(z_j-z_k)}{\vth^\prime(0)}}
    f_\text{string}(\ell^\mu, z_j, \tau)
    \nn\\
    &\propto 
    \int \d^D\ell
    \int \d\tau\ 
    e^{\pi i \alpha^\prime \tau \ell^2 }
    \int\d^n z\
    \kn_n\
    f_\text{string}(\ell^\mu, z_j, \tau)
    \, 
\end{align}
where $D$ is the space-time dimension.
Here, $\kn_n$ is the Koba-Nielsen for $n$-punctures is a multi-valued function universal to all one-loop string integrals
\begin{align} \label{eq:KN}
    \kn_n =  \left(
        \prod_{i=1}^n
        e^{2\pi i s_{iA} z_i}
    \right) 
    \left(
        \prod_{1 \leq j < k \leq n}
        \vth^{s_{jk}}(z_j-z_k)
    \right)
\end{align}
where $\vth$ is the Jacobi theta function and we have hidden most of the loop-momentum dependence in the $s_i$'s
\begin{align}
    s_{iA} = \ell \cdot k_i\,. 
\end{align}
The $s_{ij}$'s are the familiar planar Mandelstam invariants $s_{ij} = s_{ji} = \frac{1}{2} (k_i + k_j)^2 = k_i \cdot k_j$ where $k_i^\mu$ is the momentum associated to the $i^\text{th}$ particle. 
Moreover, these variables satisfy the momentum conservation constraint $\sum_{j\neq i} s_{ij} = 0$.

One-loop string integrals are integrals over the moduli space of the $n$-punctured torus: $\mathcal{M}_{1,n}$.
The modulus $\tau$ (ratio of the $A$- and $B$-cycle periods) controls the shape of the torus, while the $z_i$ are the position of punctures on the torus. 
Meromorphic functions on the torus are called elliptic functions and must be doubly periodic under integer shifts of the $A$- and $B$-cycle periods. 
Another way of saying this is that elliptic functions transform covariantly under modular transformations ($\text{SL}(2,\mathbb{Z})$ transformation of the periods). 
In particular, the modulus can always be mapped to the upper half plane using a modular transformation. 
Thus, without loss of generality, we normalize the $A$-cycle period to one and the $B$-cycle period to $\tau \in \mathbb{H}$. 

These periods define a lattice $\Lambda_{\tau} = \mathbb{Z} + \tau \mathbb{Z}$ that tiles $\mathbb{C}$ and the torus is the corresponding quotient $E = \mathbb{C}/\Lambda_\tau$. 
We define the fundamental parallelogram to be $P=\{z = a + b \tau \vert a,b \in [0,1] \}$ and assume all $z_i \in P$ (one can always preform a modular transformation to move a $z_i$ outside of the fundamental parallelogram into $P$).
While functions on the torus should be doubly periodic it is often convenient to work with quasi-periodic functions like the odd Jacobi-theta function $\vth$. 
We also use conventions for the Jacobi-theta function such that 
\begin{align}
    \vth(z+1\vert\tau) = - \vth(z\vert\tau),
    \qquad
    \vth(z+\tau\vert\tau) 
    = - e^{-i \pi (\tau+2z)} \vth(z\vert\tau).
\end{align}
In later sections, we will often omit the explicit dependence on $\tau$ since we always work at fixed $\tau$.

In this work, we concentrate on the $\d\tau\ \d^D\ell$ integrands of string integrals, which are themselves integrals over the configuration space of the $n$-punctured torus. 
Under $A$- and $B$-cycle shifts of the punctures, the Koba-Nielsen factor becomes 
\begin{align}
    \label{eq:KN-Amon}
    A\text{-cycle}:&\qquad 
    \frac{\kn_n\vert_{z_i\to z_i+1}}{\kn_n}
    = \exp\left[2\pi i s_{iA} \right]
    ,
    \\
    \label{eq:KN-Bmon}
    B\text{-cycle}:&\qquad 
    \frac{\kn_n\vert_{z_i\to z_i+\tau}}{\kn_n}
    = \exp\left[2\pi i \left( 
        s_{iA} \tau 
        + \underset{j\neq i}{\sum_{j=1}^n} s_{ij} z_j
    \right)\right], 
\end{align}
where we have used momentum conservation $\sum_{j \neq i} s_{ij}$ = 0. 

The function $f_\text{string}$ in \eqref{eq:1LSI} is a meromorphic and almost periodic function of punctures $z_{i=1,\dots,n}$, modulus $\tau$ and the loop momentum $\ell^\mu$. 
At genus-one, products of the Kronecker-Eisenstein function,
\begin{align}
    F(z,\eta\vert\tau)
    &= \frac{
        \vth^\prime(0\vert\tau)
        \vth(z+\eta\vert\tau)
    }{
        \vth(z\vert\tau)
        \vth(\eta\vert\tau)
    },\label{eq:def_F}
\end{align}
at $z\to z_i-z_j$ form generating series for one-loop integrands $f_\text{string}$. 
Here, $\eta$ is treated as a formal parameter and the coefficients of the $\eta$-expansion of $F$ appear in the one-loop integrands:
\begin{align}
    F(z,\eta\vert\tau) 
    = \sum_{k\geq0} g^{(k)}(z\vert\tau)\  
    \eta^{k-1}. 
\end{align}
With the exception of $\gk{0}(z\vert\tau) = 1$, all other Kronecker-Eisenstein coefficients $\gk{k\geq1}(z\vert\tau)$ are meromorphic functions of $z$ with at most simple poles. 
In particular, $\gk{k\geq1}(z) \d z$ should be thought of as the  genus-one analogue of $\d\log$-forms. 
These functions are the integration kernels that define elliptic multiple polylogarithms (eMPLs), which are becoming increasingly well studied in the physics literature (see 
\cite{Duhr:2020gdd,
DHoker:2020hlp,
Bourjaily:2021vyj,
Kristensson:2021ani,
Duhr:2021fhk,
Hidding:2022vjf,
Giroux:2022wav,
Frellesvig:2023iwr,
McLeod:2023qdf,
He:2023qld,
Gorges:2023zgv}
for an incomplete survey of the last few years).

Elliptic functions cannot have only a simple pole.\footnote{Double periodicity forces elliptic functions to have a double pole or more than one simple pole.} Therefore, the $\gk{k\geq1}$ are not elliptic functions and have non-trivial $B$-cycle transformations
\begin{align}
    \label{eq:g-Amon}
    A\text{-cycle}:&\qquad 
    \gk{k}(z+1\vert\tau) 
    = \gk{k}(z\vert\tau) 
    ,
    \\
    \label{eq:g-Bmon}
    B\text{-cycle}:&\qquad 
    \gk{k}(z+\tau\vert\tau) 
    = \sum_{j=0}^k \frac{(-2\pi i)^j}{j!} \gk{k-j}(z,\tau)
    .
\end{align}
While the $\gk{k\geq1}$'s are not elliptic, one can construct doubly periodic linear combinations. 
On the other hand, the generating series $F$ has comparatively simple $B$-cycle monodromies
\begin{align}
    \label{eq:F-Amon}
    A\text{-cycle}:&\qquad 
    F(z+1,\eta\vert\tau) 
    = F(z\vert\tau)
    ,
    \\
    \label{eq:F-Bmon}
    B\text{-cycle}:&\qquad 
    F(z+\tau,\eta\vert\tau) 
    = F(z,\eta\vert\tau) e^{-2 \pi i \eta}
    .
\end{align}
Thus, it is often more convenient to consider a generating series of one-loop string integrands built from Kronecker-Eisenstein functions. 

Moreover, the Kronecker-Eisenstein functions satisfy a genus-one analogue of partial fractions called the Fay identity
\begin{align} \label{eq:fay}
    F(z_{ik},\zeta_i) F(z_{jk},\zeta_j) 
    &= F(z_{ik},\zeta_{ij}) F(z_{ji},\zeta_j) 
    + F(z_{jk},\zeta_{ij}) F(z_{ij},\zeta_i)
    \\
    z_{ij} &= z_i - z_j 
    \\
    \zeta_{ij} &= \zeta_i + \zeta_j
    \,.
\end{align}
Since the Fay identity holds order by order in $\eta$, the coefficient functions $\gk{k}$ also satisfy partial fraction like identities. 
However, these $\gk{k}$ identities are much more complicated --- another reason to prefer working with the generating functions $F$ over the $\gk{k}$'s.

In section \ref{sec:modular}, we will also use the non-meromorphic but doubly-periodic version of the Kronecker Eisenstein series:
\begin{align}
\Omega(z,\eta\vert\tau)=\exp \big(2 \pi i \eta \frac{\im z}{\im \tau} \big)F(z,\eta \vert\tau) \, .
\end{align}
Like its meromorphic cousin, the doubly-periodic Kronecker-Eisenstein series can be $\eta$-expanded 
\begin{align}
    \Omega(z,\eta\vert\tau) 
    = \sum_{k\geq0} f^{(k)}(z\vert\tau)\  
    \eta^{k-1}
    \,, 
\end{align}
to obtain doubly-periodic coefficients $f^{(k)}$.

%%%%%%%%%%%%%%%%%%%%%%%%%%%%%%%%%%%%%%%%%%%%%%%%%
%%%%%%%%%%%%%%%%%%%%%%%%%%%%%%%%%%%%%%%%%%%%%%%%%
\subsection{The Riemann-Wirtinger integral}\label{sec:RW_integral_foundations}

The one-loop string integrals \eqref{eq:1LSI} are closely related to a family of genus-one integrals --- the so-called Riemann-Wirtinger integral ---  recently studied in the mathematics literature \cite{Goto2022, ghazouani2017moduli, ghazouani2016moduli, Mano2012, mano2009}. The integrals are essentially the $\d z_1$-part of \eqref{eq:1LSI}
\begin{align} \label{eq:RW-def}
   \int_{\gamma_i} u(z_1) \
   F(z_1-z_j,\eta\vert\tau)\ \d z_1
   ,
   \quad
   u(z_1) = e^{2\pi i s_{1A} z_1 } \prod_{i=2}^n \vth^{s_{1i}}(z_1-z_i),
\end{align}
with $n\geq 3$ punctures and an extra condition  
\begin{align}
    \label{eq:eta-def}
     \boxed{
     s_{1B} 
     = s_{1A} \tau + \sum_{j=2}^n s_{1j} z_j - \eta 
     = \mathrm{const.}
     \,,
     }
\end{align} 
where $s_{1B} \in \mathbb{C}$ is a generic complex number.\footnote{To convert from the notation of \cite{Goto2022} to our notation one makes the replacement $c_\infty \to -s_{1B}$, $c_{i>0} \to s_{1i}$, $ c_{0} \to s_{1A}$.}
Crucially, this condition ensures that both the $A$- and $B$-cycle monodromies of the integrand in \eqref{eq:RW-def} are treated symmetrically:
\begin{align}
     \label{eq:RW-Amon}
    A\text{-cycle}:&\qquad 
    \frac{\kn_n\ F(z,\eta\vert\tau)\vert_{z_1 \to z_1+1}}{\kn_n\ F(z\vert\tau)}
    = e^{2 \pi i s_{1A}}
    ,
    \\
    \label{eq:RW-Bmon}
    B\text{-cycle}:&\qquad 
    \frac{\kn_n\ F(z,\eta\vert\tau)\vert_{z_1 \to z_1+\tau}}{\kn_n\ F(z\vert\tau)}
    = e^{2 \pi i s_{1B}}
    .
\end{align}
The twist also has monodromies as $z_1$ loops around another puncture $z_j$
\begin{align}
    \label{eq:RW-jmon}
    z_1\text{ loop around }z_j
    :&\qquad 
    \frac{\kn_n\vert_{\circlearrowleft_{1j}}}{\kn_n}
    = e^{2 \pi i s_{1j}}
    .
\end{align} 
To understand the linear and quadratic relations among Riemann-Wirtinger integrals, we construct the corresponding twisted (co)homology.
In particular, we need to understand how the local system is modified at genus-one.

In the following, we review the construction of the local system since there are new features at genus-one. 
Specifically, the ``total'' local system is the tensor product of two local systems: one from the multi-valuedness of the twist and one from the quasi-periodicity of the Kronecker-Eisenstein function.
While the local system will be important, much of the underlying mathematical formalism (see appendix \ref{app:locSys} and references therein) can be ignored by first-time readers. 

We begin with the local system and dual local system associated to the multi-valuedness of the twist
\begin{align} 
    \label{eq:KN-LS}
    \mathcal{L}_\omega 
    &:= \mathcal{L}_\omega(s_{1A},s_{12},\dots,s_{1n}) = \mathbb{C} u^{-1}
    \,,
    \\
    \label{eq:KN-LSdual}
    \c{\mathcal{L}}_\omega 
    &:= \mathcal{L}_{-\omega}(-s_{1A},-s_{12},\dots,-s_{1n}) = \mathbb{C} u
    \,,
    \\
    \omega &:= \d\log(u)
    \,.
\end{align}  
These local systems are line bundles that keep track of the branch choices of the twist and dual (inverse) twist.
More precisely, the universal covering $\tilde{M}$ of $M$ is a principle $\pi_1(M)$ bundle where  $\pi_1(M)$ is the first homotopy group. 
By exponentiating the monodromies of $u$ we define a one-dimensional representation of $\pi_1(M)$. 
Then, the local system is the associated line bundle. 
More succinctly, the (dual) local system is the locally constant sheaf of solutions to the equation $\nabla f = 0$ ($\c{\nabla} f = 0$) with 
the flat\footnote{By flat, we mean that $\omega\wedge\omega=0 \implies  \nabla^2=0$.}
(dual) covariant derivative  $\nabla = \d + \omega\wedge$ ($\c\nabla = \d - \omega\wedge$). 
In other words, the (dual) local system is the set of all possible branch choices for the dual twist $\c{u}=u^{-1}$ (twist $u$). 
These covariant derivatives replace the normal exterior derivative in the twisted setting.

So far the genus-one construction mirrors the genus-zero construction in section \ref{sec:Fund}.
However, at genus-one, we have an additional local system associated to the multi-valued-ness of the Kronecker-Eisensten functions \eqref{eq:F-Bmon}.
Following \cite{Mano2012, Goto2022}, we define a 1-dimensional representation of the fundamental group $\pi_1(E)$ for $\eta\in\mathbb{C}$. \
We denote this by $e_\eta: \pi_1(E) \simeq \Lambda_{\tau} \ni \gamma \mapsto e_\eta(\gamma) \in \mathbb{C}^*$ such that $e_\eta(1) = 1$ and $e_\eta(\tau) = e^{- 2\pi i \eta}$. 
This representation corresponds to the allowed phases of our integrand modulo the twist (recall equation \eqref{eq:F-Amon} and \eqref{eq:F-Bmon}). 
We denote the rank-1 local system on $E$ determined by the above representation of $\pi_1(E)$ by $\mathcal{L}_\eta$. 

Thus, the combined local system of the integrand of \eqref{eq:RW-def} and its dual are
\begin{align} \label{eq:LS}
    &\mathcal{L}_{\omega,\eta} (s_{1A}, s_{12}, \dots, s_{1n}, s_{1B})
    = \mathcal{L}_\omega 
        \otimes
        \mathcal{L}_\eta
    = \mathbb{C}u^{-1} \otimes \mathcal{L}_\eta
    \,,
    \\
    \label{eq:dualLS}
    &\c{\mathcal{L}}_{\omega,\eta}
    = \mathcal{L}_{\omega}(-s_{1A},-s_{12},\dots,-s_{1n},-s_{1B}) 
    \otimes \L_{-\eta}
    = \mathbb{C}u \otimes \mathcal{L}_{-\eta}
    \,.
\end{align} 
As usual, these local systems are connected by the involution $\vee: s_\bullet \mapsto -s_\bullet$.
From this, we construct the twisted cohomology and homology of the Riemann-Wirtinger integral ($H^p(M,\L_{\omega,\eta})$ and $H_p(M,\c{\L}_{\omega,\eta})$) as well as their duals ($H^p(M,\c{\L}_{\omega,\eta})$ and $H_p(M,\L_{\omega,\eta})$) in sections \ref{sec:RWco-eta} and \ref{sec:RWhom}.

%%%%%%%%%%%%%%%%%%%%%%%%%%%%%%%%%%%%%%%%%%%%%%%%%
%%%%%%%%%%%%%%%%%%%%%%%%%%%%%%%%%%%%%%%%%%%%%%%%%
%%%%%%%%%%%%%%%%%%%%%%%%%%%%%%%%%%%%%%%%%%%%%%%%%
\section{Twisted cohomology of the Riemann-Wirtinger integral \label{sec:RWco-eta}}

\new{Since one cannot put any differential form into the double copy formula, the space of allowed differential forms needs to be understood.}
\new{To this end,} \new{we review the construction of} the twisted cohomology of the Riemann-Wirtinger integral family. 
We define the twisted cohomology for $\eta\neq0$ in section \ref{sec:RWco-basis} and give two different bases one of which has a smooth $\eta=0$ limit (see appendix \ref{sec:RWco-app-0} and \cite{Goto2022} for more). 
Then, in section \ref{sec:RWco-int}, we define the associated dual twisted cohomology and an inner product called the intersection number. 
\new{While only the existence of a well-defined intersection number is needed to have a well-defined double copy, we illustrate how to compute intersection numbers at genus-one in order to provide a complete presentation of the twisted (co)homology of Riemann-Wirtinger integrals. 
As a simple application of the intersection number, we verify the over-completeness of the second basis of cohomology by deriving the linear relation satisfied by its elements.}

\new{The interested reader can find a derivation of the DEQs satisfied by the RW family of integrals for both $\eta\neq0$ and $\eta=0$ via intersection numbers in appendices \ref{sec:RWco-app-deqs} and \ref{sec:RWco-app-0}.
% Moreover, we expect genus-one and higher intersection numbers to have future application to the integral reduction of hyperelliptic Feynman integrals \cite{Giroux:2022wav}.
}

% Using the intersection number, we compute the differential equations satisfied by the Riemann-Wirtinger integrals in section \ref{sec:RWco-app-alphaprime}.
% In section \ref{sec:alphaprime}, we solve the differential equations to order $\mathcal{O}(\alpha')$ in terms of eMPLs for generic boundary values.
% We postpone a more detailed discussion of boundary values to section \ref{sec:RWhom} where we introduce a basis of cycles. 
% Lastly, in section \ref{sec:RWco-0}, we describe the cohomology in the $\eta\to0$ limit and discuss the corresponding differential equations.

%%%%%%%%%%%%%%%%%%%%%%%%%%%%%%%%%%%%%%%%%%%%%%%%%
%%%%%%%%%%%%%%%%%%%%%%%%%%%%%%%%%%%%%%%%%%%%%%%%%
\subsection{Basis of cohomology \label{sec:RWco-basis}}

Recall that the twisted cohomology is essentially the cohomology with respect to the covariant derivative $\nabla = \d + \omega\wedge$ where $\omega=\d\log u$ and $\nabla^2=0$. 
This covariant derivative keeps track of the monodromies generated by the twist $u$. It also means that we can trade the multi-valued integrand $u\vphi$ for simpler integrands ($\mathcal{L}_\eta$-valued differential forms) since $\d(u \vphi) = u (\nabla \vphi)$. 
We denote the space of twisted differential $p$-forms on $M$ by $\Omega^p_\eta(M) = \Omega^p(M) \otimes \L_\eta$.

Because the coefficients $s_\bullet$ are generic, the total derivative always vanishes
\begin{align}
	\int \d(u\ \vphi) = \int u (\nabla \vphi) = 0.
\end{align}
This means that the Riemann-Wirtinger integrand is not unique --- we can always add a total covariant derivative to $\vphi$ without changing the value of the integral. 
For this reason, we would like to work with equivalence classes of forms that are unique and independent of such shifts. 
We would also like any representative of such a class to be closed: $\nabla\vphi=0$. 
This ensures that there are no boundary terms in the twisted version of Stokes' theorem. 
Thus, we are lead to define the twisted cohomology 
\begin{align}
	H^{p}(M,\L_{\omega,\eta}) 
	= \frac{
		\text{ker}\left( 
			\nabla: \Omega^{p}_\eta(M) \to \Omega^{p+1}_\eta(M)
		\right) 
	}{
		\text{im}\left( 
			\nabla: \Omega^{p-1}_\eta(M) \to \Omega^{p}_\eta(M)
		\right) 
	}
\end{align}
Explicitly, the twisted cohomology class represented by $\vphi$ is 
\begin{align}
	\vert\vphi\ra = \{
		\vphi + \nabla \psi \vert 
		\psi \in \Omega^{\text{deg}\vphi -1}_\eta(M)
		\quad\text{and}\quad
		\nabla \vphi = 0
	\}
    \,.
\end{align}
Here, the local system appears in the second argument of $H^p$ instead of $\nabla$ because $\nabla$ only knows about the $\L_\omega$ half of the local system $\LS$. 

For each $p=0,\dots,\text{dim}_\mathbb{R}M$, the twisted cohomology is a finite dimensional vector space.  
\new{Moreover, from equations \eqref{eq:dimH}, only the middle-dimensional cohomology $p=\frac{1}{2}\text{dim}_\mathbb{R}M$ is non-trivial.
Thus, there are no non-trivial closed twisted 0- or 2-forms}
\begin{align}
	\text{dim} H^{p=0,2}(M,\L_{\omega,\eta})  = 0.
\end{align}
By a theorem from \cite{Mano2012, Goto2022}, we also know that the dimension of the twisted cohomology of 1-forms is 
\begin{align}
    \text{dim}H^1(M,\L_{\omega,\eta})
    = n-1
\end{align}
where $n$ is the number of punctures $z_i$. 
\new{Note that in \cite{Mano2012, Goto2022} they effectively have one additional puncture (the variable $u$ in \cite{Mano2012, Goto2022} should be thought of as the puncture $z_{n+1}$) and therefore, the dimension of their cohomology is $n$ instead of $n-1$.}

We close this section by introducing two useful bases for the twisted cohomology. 
The first basis contains forms that have at most a simple poles at one of the punctures $z_{i\geq2}$
\begin{align}
    \vphi_{a}  &= F(z_1-z_{a+1}, \eta)\, \d z_1 
    \quad\text{for}\quad 
    a= 1,2,\cdots, n-1
    . 
\end{align}
The second is a spanning set of forms
\begin{align}
    \xi^{(p)}_1 &= \eta F(z_1-z_p,\eta)\, \d z_1 
    \nn\\ \label{eq:xi-basis}
    \xi^{(p)}_p &= \partial_1 F(z_1-z_p,\eta)\, \d z_1     
    \\
    \xi^{(p)}_{a \geq 2} &= 
    \left[
        F(z_1-z_a,\eta) - F(z_1-z_p,\eta)
    \right] \d z_1 
    \qquad\text{for}\qquad a \neq p
    \nn
\end{align}
 subject to a single relation
\begin{align}
    \label{eq:xirel}
    0 \simeq \left[
        2 \pi i s_{1A}
        + s_{1p}\, \gk{1}(\eta)
        + \underset{j\neq p}{\sum_{j=2}^n} s_{1j} 
        \left( 
            F(z_j-z_p,\eta)
            - \gk{1}(z_j-z_p)
        \right)
    \right]
    \vert \xi^{(p)}_1\ra
    \nn\\
    - (s_{1,p} - 1) \eta
    \vert \xi^{(p)}_p\ra
    + \eta\, \underset{j\neq p}{\sum_{j=2}^n} s_{1j}\, F(z_j-z_p,\eta)
    \ \vert \xi^{(p)}_j\ra.
\end{align}
This identity follows from the fact that $\nabla \xi^{(p)}_p \simeq 0$ in cohomology (and can be shown using the Fay identity or intersection numbers).
This basis has a well defined $\eta\to0$ limit and spans the twisted cohomology in this limit \cite{Goto2022} (see appendix \ref{sec:RWco-app-0}).
In the next section, we will verify that the intersection matrix of these sets has rank $n-1$ confirming that the set $\{\vphi_a\}$ is indeed a basis and that the set $\{\xi_a^{(p)}\}$ is over-complete.

%%%%%%%%%%%%%%%%%%%%%%%%%%%%%%%%%%%%%%%%%%%%%%%%%
%%%%%%%%%%%%%%%%%%%%%%%%%%%%%%%%%%%%%%%%%%%%%%%%%
\subsection{Intersection numbers: an inner product on cohomology \label{sec:RWco-int}}

Recall from section \ref{sec:pairings} that the dual cohomology isomorphic to the twisted cohomology $H^p(M,\L_{\omega,\eta})$ is defined by changing the sign of all $s_{\bullet}$ and consequently $\eta$
\begin{align}
	\left[ H^p(M,\L_{\omega,\eta}) \right]^\vee
	:= H^p(M,\c{\L}_{\omega,\eta}).
\end{align}
The dual covariant derivative is $\c{\nabla} = \d + \c{\omega}\wedge$ where $\c{\omega} = \d\log{u^{-1}} = -\omega$. 
As a consequence of changing the signs of the $s_\bullet$'s, the dual Riemann-Wirtinger integral comes with a dual twist $\c{u}=u^{-1}$.
We can also think of $\vee$ as a map from $H^p(M,\L_{\omega,\eta})$ to the dual cohomology: $\c{\vphi} = [\vphi]^\vee := \vphi\vert_{s_\bullet \to -s_\bullet, \eta \to -\eta}$.
In particular, the set $\{\c{\vphi}_a\}$ is basis for the dual cohomology and $\{\c{\xi}^{(p)}_a\}$ spans the dual cohomology subject to the dualized version of \eqref{eq:xirel}.

The intersection number pairs the cohomology and its dual to form an inner product
\begin{align}
\la\bullet\vert\bullet\ra: H^p(M,\c{\L}_{\omega,\eta}) \times H^p(M,\L_{\omega,\eta}) \to \mathbb{C}( s_\bullet)
\end{align}
by 
\begin{align} \label{eq:intNumDef}
    \la\c{\vphi}\vert\vphi\ra 
    = 
    \int_M \text{Reg}[\c{\vphi}] \wedge \vphi 
\end{align}
where $\text{Reg}$ maps the representative $\c{\vphi}$ to a new representative that has no support in the neighbourhood of each puncture $z_{i\geq2}$.\footnote{Formally, $\reg$ is a map from the dual twisted cohomology to the compactly supported dual twisted cohomology $\reg: H^p(M,\c{\nabla}) \to H^p_c(M,\c{\nabla})$.}
Note that our definition for the intersection number differs from \cite{Goto2022} by a sign because we choose to put the dual forms on the left hand side because it is more natural in the bra-ket notation of quantum mechanics.

Since the Riemann-Wirtinger integral is one-dimensional we only need to know the map $\reg$ in this case. 
Explicitly, for any 1-form $\c{\vphi}$, its image under $\reg$ is 
\begin{align}
    \reg[\c{\vphi}]
    &:= \c{\vphi} 
    - \sum_{i=2}^n \nabla \bigg( 
        \Theta(|z_{1i}|<\epsilon)\  
        \c{\psi}_i 
    \bigg)
    \nn\\&
    = \bigg( 1 {-} \sum_{i=2}^n \Theta(|z_{1i}|<\epsilon) \bigg) \c{\vphi}
    - \sum_{i=2}^n \c{\psi}_i\ \d\Theta(|z_{1i}|<\epsilon),
\end{align}
where $\epsilon$ is arbitrarily small enough such that the Heaviside-$\Theta$ functions do not overlap. 
Here, the 0-form $\c{\psi}_i$ is a local primitive for $\c{\vphi}$ near $z_1=z_i$: $\c{\nabla}\c{\psi}_i = \c{\vphi}$ for $z_1\sim z_i$.  
While any global primitive is indeed multi-valued, the local primitive $\c{\psi}_i$ is single-valued and can be expressed as a Laurent polynomial in $z_1-z_i$. 
Clearly, $\reg[\c{\vphi}]$ has no support for $z_1 \sim z_i$. 
Moreover, it differs from $\c{\vphi}$ by a total derivative and is thus equivalent in cohomology.\footnote{Forms differing by a covariant derivative are in the same cohomology class and are said to be cohomologous. The replacement of $\c{\vphi}$ with $\text{Reg}[\c{\vphi}]$ does not change the intersection number because it only depends on the class not the representative.}
It is essential that we use the compactly supported version of $\c{\vphi}$ otherwise the intersection number would not be finite.\footnote{More precisely, for generic $s_\bullet$, $\text{Reg}$ is an isomorphism between $H^p(M,\c{\L}_{\omega,\eta})$ and its compactly supported version $H^p_c(M,\c{\L}_{\omega,\eta})$. Alternatively, one could also regulate Riemann-Wirtinger cohomology instead of the dual cohomology.} 
Moreover, since only the anti-holomorphic part of $\reg[\c{\vphi}]$ survives the wedge product in \eqref{eq:intNumDef}, the intersection number of Riemann-Wirtinger 1-forms is given by the following simple residue formula
\begin{align} \label{eq:intNum}
    \la\c{\vphi}\vert\vphi\ra
    = - \sum_{i=2}^n \int_M 
        \c{\psi}_i\ \d\Theta(|z_{1i}|<\epsilon)
        \wedge \vphi
    = - \sum_{i=2}^n \text{Res}_{z_1=z_i}[ \c{\psi}_i \vphi ].
\end{align}
One important feature of this formula is that the intersection number vanishes whenever the forms $\c{\vphi}$ and $\vphi$ do not have overlapping singularities. 
Moreover, the intersection numbers satisfy the following relation
\begin{align}
    [\la \c\vphi \vert \phi \ra]^\vee
    = - \la \c\phi \vert \vphi \ra
    \,.
\end{align}
Thus, one only has to compute the upper/lower triangular part of the intersection matrix.

In particular, we can already predict that the intersection matrix $\la\c{\vphi}_a\vert\vphi_b\ra$ is diagonal. 
To compute the proportionality constant, we need to construct the primitives of $\c{\vphi}_a$
\begin{align}
    \label{eq:phidualprim-diag}
    \c{\psi}_{a,i=a} &= - \frac{1}{s_{1a}} + 
    \left[
        \frac{2 \pi i s_{1A}}{ (s_{1a}{-}1) s_{1a} }
        - \frac{\gk{1}(-\eta)}{s_{1a}{-}1}
        + \frac{
            \sum_{i \neq a} s_{1i}\ \gk{1}(z_{1a})
        }{ (s_{1a}{-}1) s_{1a} }
    \right] z_{1a}
    + \mathcal{O}\left(z_{1a}^2\right),
    \\
    \label{eq:phidualprim-off-diag}
    \c{\psi}_{a,i\neq a} 
    &= -\frac{F(z_{ia},-\eta)}{s_{1i}{-}1} z_{1i}
        + \mathcal{O}\left(z_{1i}^2\right),
\end{align}
where $\c{\nabla}\c{\psi}_{a,i} = \c{\vphi}_a$ near $z_1=z_i$. 
Using \eqref{eq:intNum}, we obtain
\begin{align}
    (C_\vphi)_{ab} 
    = \la\c{\vphi}_a\vert\vphi_b\ra 
    = 2\pi i \frac{\delta_{ab}}{s_{1a}}.
\end{align}
In particular, this intersection matrix has full rank verifying that the set $\{\vphi_a\}$ forms a basis.

Equations \eqref{eq:phidualprim-diag} and \eqref{eq:phidualprim-off-diag} are also enough to determine almost all of the intersection numbers for the set $\{\xi^{(p)}_a\}$:
\begin{align} \label{eq:xi-int1}
     \frac{
        \la \c{\xi}^{(p)}_1 \vert \xi^{(p)}_a \ra
    }{-2\pi i}
    &= 
    \begin{cases}
        \frac{\eta^2}{s_{1p}} 
        & b=1, 
        \\
        -\frac{\eta}{s_{1p}} 
        & a=1, b \geq 2 \text{ and } b \neq p, 
        \\
        \frac{
            2 i \pi s_{1A}
            - s_{1p}\ \gk{1}(\eta)
            + \sum_{i=2;i \neq p}^n s_{1i}\
                \gk{1}(z_{pi})
        }{(s_{1p}-1) s_{1p}}
        & b=p,
    \end{cases}
    \\ \label{eq:xi-int2}
    \frac{
        \la \c{\xi}^{(p)}_{a\geq2, a\neq p} 
        \vert \xi^{(p)}_{b \geq 2}
    \ra}{-2\pi i}
     &= 
    \begin{cases}
        -\frac{s_{1a}+s_{1p}}{s_{1a}s_{1p}} 
        & b = a 
        \\
        -\frac{1}{s_{1p}} 
        & b \neq a, p
        \\
        \frac{
            - s_{1p}\ F(z_{ap},\eta)
            - s_{1p}\ \gk{1}(\eta)
            + \sum_{i=2; i\neq p}^n s_{1i}\ \gk{1}(z_{pi})
            + 2 i \pi  s_{1A}
        }{(s_{1p}-1) s_{1p}}
        & b=p.
    \end{cases}
\end{align}
For $a=b=p$, one needs to go beyond the local primitives \eqref{eq:phidualprim-diag} and \eqref{eq:phidualprim-off-diag}
\begin{align} \label{eq:xi-int3}
    \frac{
        \la \c{\xi}^{(p)}_p \vert \xi^{(p)}_p \ra
    }{-2\pi i}
    &=
        (2 \pi s_{1A})^2
        + 2 s_{1p}^2\ \gk{2}(\eta)
        + \sum_{i=2,i\neq p}^n s_{1p} s_{1i}\ \gk{1}(z_{pi})
        - \left(
            \sum_{i=2,i\neq p}^n s_{1i}\ \gk{1}(z_{pi})
        \right)^2
    \nn\\&
    - 4 i \pi  s_{1A} 
    \sum_{i=2,i\neq p}^n s_{1i}\ \gk{1}(z_{pi})
    - 2 s_{1p} \sum_{i=2,i\neq p}^n s_{1i}\  \wp(z_{pi})
    .
\end{align}
All remaining $\la \xi^{(p)}_a \vert \xi^{(p)}_b \ra$ intersection numbers can be obtained from the above using the relation $[\la \c\vphi \vert \phi \ra]^\vee = - \la \c\phi \vert \vphi \ra$.
One can then verify that the $n \times n$ intersection matrix $\la \xi^{(p)}_{a=1,\dots,n} \vert \xi^{(p)}_{b=1,\dots,n} \ra$ has rank $n-1$ confirming that the set $\{\xi^{(p)}\}_{a=1,\dots,n}$ is over-complete.
In particular, the identity \eqref{eq:xirel} follows from a  straightforward application of the resolution of identity \eqref{eq:decomp_id}. 
Choosing $\{\xi_a^{(p)}\}_{a \neq p}$ and $\{\c\xi_a^{(p)}\}_{a \neq p}$ for our basis of cohomology and dual cohomology, we set $C_{ab} = \la \c\xi_a^{(p)} \vert \xi_b^{(p)} \ra$ and apply \eqref{eq:decomp_id} to $\vert \xi_p^{(p)}\ra$:
\begin{align}
    \vert\xi_p^{(p)}\ra
    \simeq \sum_{a,b\neq p} \vert\xi_a^{(p)}\ra\ 
        C_{ab}^{-1}\  
        \la\c\xi_{b}^{(p)}\vert\xi_p^{(p)}\ra
    = \frac{\mathrm{RHS}\eqref{eq:xirel}}{\eta (s_{12}-1)}
    \,.
\end{align}
While more complicated than the forms $\vphi_a$, the spanning set $\xi^{(p)}_a$ is still useful because it spans the cohomology in the $\eta\to0$ limit (see appendix \eqref{sec:RWco-app-0}).

%%%%%%%%%%%%%%%%%%%%%%%%%%%%%%%%%%%%%%%%%%%%%%%%%
%%%%%%%%%%%%%%%%%%%%%%%%%%%%%%%%%%%%%%%%%%%%%%%%%
%%%%%%%%%%%%%%%%%%%%%%%%%%%%%%%%%%%%%%%%%%%%%%%%%
\section{Twisted homology of the Riemann-Wirtinger integral \label{sec:RWhom}}

After reviewing the construction of the twisted homology for Riemann-Wirtinger integrals, we give a basis for homology (section \ref{sec:RWhom-basis}). 
Then, in section \ref{sec:RWhom-int},  we compute the intersection matrix that becomes the double copy kernel in section \ref{sec:DC-RW}. 

\new{For the interested reader, we describe how to take the $\tau \to i \infty$ limit of the Riemann Wirtinger integral and obtain boundary values for the twisted contours in section \ref{sec:RWco-app-bdval}. }

%%%%%%%%%%%%%%%%%%%%%%%%%%%%%%%%%%%%%%%%%%%%%%%%%
%%%%%%%%%%%%%%%%%%%%%%%%%%%%%%%%%%%%%%%%%%%%%%%%%
\subsection{Basis of homology \label{sec:RWhom-basis}}

As reviewed in section \ref{sec:Fund}, twisted or loaded contours are represented by the tensor product of a ``normal''  topological contour with a branch choice. 
That is, given a topological contour $\tilde{\gamma}$ on the manifold $M$, we ``load'' it with a branch choice of the twist $u_{\tilde{\gamma}}$ (section of $\LSdual$) to form a twisted contour $\gamma$
\begin{align}
     \gamma = \tilde{\gamma} \otimes u_{\tilde{\gamma}} 
     \,.
     % (u_{\tilde{\gamma}} e_{\tilde{\gamma}}) .
\end{align}
Technically, the contour should also be loaded with a branch choice $e_{\tilde{\gamma}}$ on $E$ (section of $\c{\L}_\eta$). 
However, since we can regard the Kronecker-Eisenstein function $F(z,\eta)$ as a global meromorphic section of $\c{\L}_\eta$, such a loading will not play an important role in the computation of intersection indices.

The twisted homology is defined in the usual way
\begin{equation} \label{eq:twisted_homology_def}
    H_m(M,\LSdual) = \frac{\text{ker}\left(\partial_{\omega,\eta}: C_m(M,\LSdual) \to  C_{m-1}(M,\LSdual)  \right)}{\text{im}\left(\partial_{\omega,\eta}: C_{m+1}(M,\LSdual) \to  C_{m}(M,\LSdual)  \right)} 
\end{equation}
where $C_{m}(M,\LSdual)$ is the space of twisted $m$-contours. 
Elements of the twisted homology are equivalence classes of closed twisted cycles (twisted contours with no boundary) modulo exact cycles (twisted contours that are boundaries) and denoted by the square bra. 
    Explicitly, the homology class of an $m$-cycle $\gamma \otimes u_\gamma \in \ker(\partial_{\omega,\eta}: C_{m}(,\c{\L}_\omega) \to C_{m-1}(,\c{\L}_\omega) )$ is 
\begin{align}
    [\gamma \otimes u_\gamma \vert 
    = \{
        \gamma \otimes u_\gamma + \partial_{\omega} (\beta \otimes u_\beta) \,
        \vert \,
        (\beta \otimes u_\beta) \in C_{m+1}(,\c{\L}_\omega)
    \}
    \,.
\end{align}
That is, a representative of a twisted homology class, $\gamma \otimes u_\gamma$, is a twisted contour that has no boundary and is not a boundary itself.
We also define the dual twisted homology by exchanging the dual local system $\c{\L}_{\omega,\eta}$ in \eqref{eq:twisted_homology_def} for the local system $\L_{\omega,\eta}$
\begin{equation} 
    [H_m(M,\LSdual)]^\vee
    := H_m(M,\LS) 
    = \frac{
        \text{ker}\left(
            \c{\partial}_{\omega,\eta}: 
            C_m(M,\LS) \to  C_{m-1}(M,\LS) 
        \right)
    }{
        \text{im}\left(
            \c{\partial}_{\omega,\eta}: 
            C_{m+1}(M,\LS) \to  C_{m}(M,\LS)  
        \right)
    }
\end{equation}
where $\LS = [\LSdual]^\vee = \LSdual(-s_{1A}, -s_{12}, \dots, -s_{1n}, -s_{1B}) = \mathbb{C} u^{-1}$, $C_{m}(M,\LS)$ is the space of dual twisted $m$-contours and $\c{\partial}_{\omega,\eta} = \partial_{-\omega,-\eta}$ is the dual boundary operator. 
Representatives of dual homology classes are topological cycles loaded with a branch choice of the inverse twist: $\c{\gamma} = \c{\gamma}^\prime \otimes u_{\c{\gamma}^\prime}^{-1}$.
Given a representative of the dual homology, its class is denoted by a square ket $\vert\c{\gamma}^\prime \otimes u_{\c{\gamma}^\prime}^{-1}]$.

Since only the middle dimensional twisted cohomology is non-trivial, $\dim H_0 = \dim H_2 = 0$. 
We also know from \cite{Mano2012, Goto2022} that $\dim H_1 = n-1$ where $n$ is the number of punctures.\footnote{\new{Note that in \cite{Mano2012, Goto2022}, the variable $u$ should be thought of as an additional puncture $z_{n+1}$ and therefore, their homology has $n$ elements.}}
Of course, this counting is the same for the cohomology (section \ref{sec:RWco-eta}).

\begin{figure}
    \centering
    \includegraphics[align=c, scale=.8]{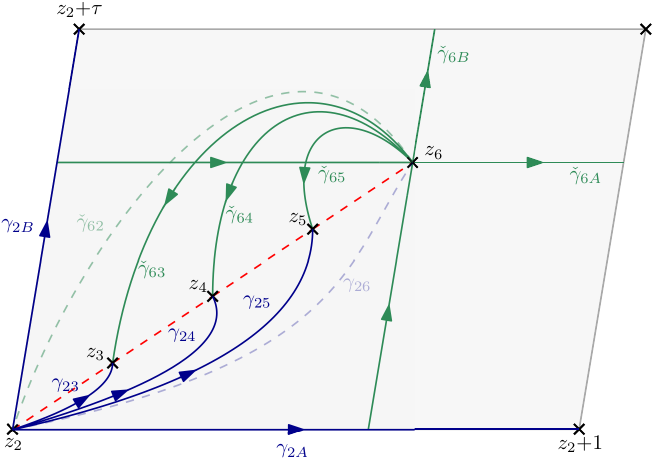}
    \caption{
    An over-complete set of {\color{blue!80!black}cycles} in {\color{blue!80!black}blue} and an over-complete set of {\color{Green!80!black}dual cycles} in {\color{Green!80!black} green} for $n=6$ punctures.
    The (dual) cycles depicted by solid lines form a basis
    and the dashed {\color{Red!80!black} red} line denotes our choice of {\color{Red!80!black} branch cut}. 
    All of the cycles start at the point $z_2$ while all of the dual cycles start at the point $z_n=z_6$.
    The solid $n-1=5$ (dual) cycles define a basis for the (dual) twisted homology. 
    The dashed cycles are naturally defined but linearly depend on the cycles denoted by solid lines.
    The contours drawn here are technically cycles because the end points do not belong to $M$. Such cycles are known as locally finite or Borel-Moore cycles in the mathematics literature. 
    }
    \label{fig:hom_basis}
\end{figure}

The remaining task is to provide a basis for the twisted homology and its dual.
A particularly convenient choice of topological cycles is depicted in figure \ref{fig:hom_basis} with {\color{blue!80!black}cycles} in {\color{blue!80!black}blue} and {\color{Green!80!Black}dual cycles} in {\color{Green!80!black} green}. 
All of the cycles start at the puncture $z_2$ while the dual cycles start at $z_n$ (where $n=6$ in figure \ref{fig:hom_basis}).
These topological cycles are then loaded with a branch of $u$ ($\c{u}$) to form twisted (dual) cycles where the {\color{Red!80!black}branch cut} is denoted by the dashed {\color{Red!80!black}red} line in figure \ref{fig:hom_basis}. 
We will always choose branches such that on the interval $\Delta_k = [z_k,z_{k+1}]_{k=2,\dots,n-1}$
\begin{align} \label{eq:branch}
    \arg\left\lbrack \frac{z_1 - z_j}{z_1 - z_i} \right\rbrack = \pi \quad \text{ for } 2 \leq i \leq k < j \leq n \,.
\end{align}
Furthermore, note that the cycles in figure \ref{fig:hom_basis} intersect only one dual cycle. 
This is enough to guarantee that the corresponding intersection matrix is diagonal. 

We have actually drawn an over complete set of $n$ (dual) cycles in figure \ref{fig:hom_basis} since we know that $\dim H_1=n-1$.
To find a linear relation amongst these cycles, start with a closed circular contour that does not encircle any singularities. 
Then, expand this contour without crossing the branch cuts until the contour runs along the edge of the fundamental parallelogram and the branch cut. 
Rewriting the deformed contour in terms of the set $\{\gamma_{2A}, \gamma_{2B}\} \cup \{ \gamma_{2j}\}_{j=3,\dots,n}$ yields the relation \cite{Goto2022}
\begin{align} \label{eq:homology_cycle_overcomplete_basis}
    (e^{2 \pi i s_{1A}} - 1) [\gamma_{2B}\vert
    + (1 - e^{2 \pi i s_{1B}}) [\gamma_{2A}\vert
    - \sum_{j=3}^n 
        e^{- 2 \pi i (s_{12} + \cdots + s_{1j})}
        (1 - e^{2 \pi i s_{1j}})
        [\gamma_{2j}\vert
    \simeq 0, 
\end{align}
(a similar identity exists for the set of dual cycles $\{ \c\gamma_{nA}, \c\gamma_{nB} \} \cup \{ \c{\gamma}_{nj}\}_{j=2,\dots,n-1}$).
We will often choose the cycles drawn in solid lines as our basis for the (dual) homology. 
Moreover, we will also rediscover the relation \eqref{eq:homology_cycle_overcomplete_basis} using the intersection indices in the next section.

In string theory context, relations like \eqref{eq:homology_cycle_overcomplete_basis} are often called monodromy relations because one projects a contour with one color ordering onto another with different color ordering, which corresponds to the winding or exchanging of punctures.
Such identities relate various color orderings for open string amplitudes.
The tree-level gauge theory equivalent of \eqref{eq:homology_cycle_overcomplete_basis} are the so-called \textit{Kleiss-Kuijf} and \emph{BCJ} relations. 
The Kleiss-Kuijf follow from the leading $\alpha'$ expansion of the string monodromy relations while the BCJ relations follow from the subleading order in $\alpha'$.\footnote{Note that the BCJ relations are \emph{linear} relations among color-ordered amplitudes and should not be confused with color-kinematics duality.}

%%%%%%%%%%%%%%%%%%%%%%%%%%%%%%%%%%%%%%%%%%%%%%%%%
%%%%%%%%%%%%%%%%%%%%%%%%%%%%%%%%%%%%%%%%%%%%%%%%%
\subsection{The intersection index: an inner product on homology \label{sec:RWhom-int}}

In this section, we regularize the cycles defined in the previous section in order to compute the intersection matrix that becomes the double copy kernel in section \ref{sec:DC-RW}. 
The intersection indices for Riemann-Wirtinger integrals were first derived in \cite{ghazouani2016moduli} for $\eta=0$ and more recently revisited in \cite{Goto2022} for $\eta\neq0$. 

As reviewed in section \ref{sec:Fund},  the intersection index \eqref{eq:Hint} is a pairing between the twisted homology with local system $\LSdual$ and the twisted homology with local system $\LS$
\begin{align}
    [ \bullet \vert \bullet ]:
    &\
    H_p(M,\c{\L}_\omega) \times 
    H_{n-p}(M,\L_\omega) \to \mathbb{C}
    \,,
\end{align}
given by  
\begin{align}  \label{eq:Hint2}
     [\gamma \otimes u_\gamma |\c{\gamma} \otimes \c{u}_{\c{\gamma}}] 
    := \sum_{x \in \gamma \cap \c{\gamma}} 
    \left(u_\gamma \c{u}_{\c{\gamma}}\vert_{x}\right) 
    \big[\reg[\gamma] \vert \c{\gamma}\big]^{\text{top}}_x
    \,,
\end{align}
where the set $\gamma\cap\c{\gamma}$ corresponds to the set of all points where $\gamma$ and $\c{\gamma}$ intersect and the factor $(u_\gamma \c{u}_{\c{\gamma}}) \vert_x$, evaluates to a phase. 
Here, $[\bullet \vert \bullet ]^{\text{top}}_x$ is the topological intersection index at $x$ that evaluates to $\pm1$ depending on the relative orientation of $\gamma$ and $\c{\gamma}$.
Following the conventions of  \cite{Mizera:2017cqs, Mimachi_2003}, we define
\begin{align}\label{fig:topological_intersection_numbers}
\begin{gathered}
    \includegraphics[scale=0.5]{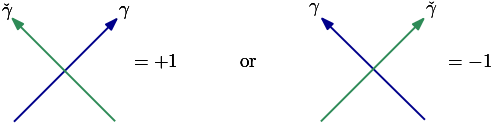}
\end{gathered}\,.     
\end{align}
The only remaining piece is to specify the map $\reg$. 

\begin{figure}
    \centering
    \includegraphics[align=c, scale=.8]{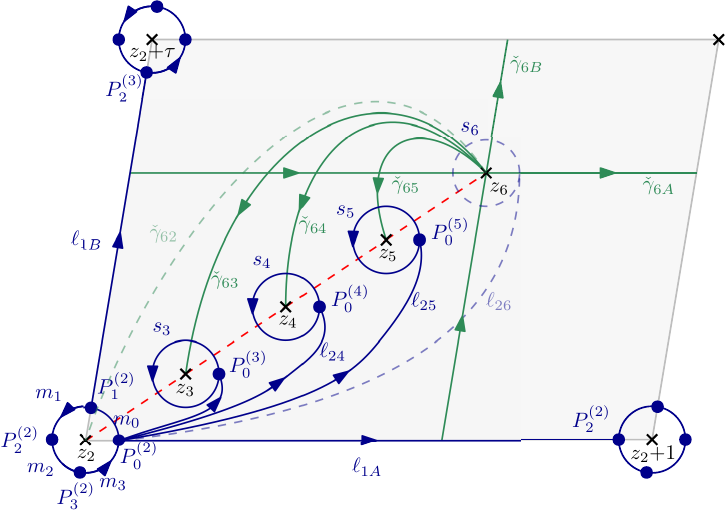}
    \caption{
    The over-complete set of {\color{blue!80!black}cycles} ({\color{Green!80!black}dual cycles}) subject to \eqref{eq:homology_cycle_overcomplete_basis} in the compact homology (dual homology). 
    Those depicted by solid lines form a basis. 
    % Note that all of the $l_i$ are below any of the {\color{Red!80!black}branch cuts}. 
    In particular, this choice has an orthogonal intersection matrix since each {\color{blue!80!black}cycle} only intersects one {\color{Green!80!black}dual cycles}! 
    }
    \label{fig:hom_basis_reg}
\end{figure}

In order for the definition \eqref{eq:Hint2} to make sense, at least one contour should have compact support. 
Thus, $\reg$ is a map from the normal\footnote{To be precise, the homology described in section \ref{sec:RWhom-basis} is called locally finite or Borel-Moore \cite{hwa1966homology}.}
 twisted homology described in section \ref{sec:RWhom-basis} to the compactly supported twisted homology 
 \begin{align}
    \reg: H_{m}(M,\LSdual) 
    \to H_{m}^c(M,\LSdual)
    \,.
\end{align}
When all of the exponents $s_\bullet$ are generic, this map is an isomorphism and we can use it to compute intersections involving non-compact cycles. 

Compact cycles cannot have endpoints like in figure \ref{fig:hom_basis} even if the endpoints do not belong to the space (as in our case).
Thus, compact cycles must wind around the punctures instead of ending at the punctures. 
This necessarily means that the compact cycles will cross a branch cut.
To be a twisted cycle with compact support, it must have no boundary. 
Because such a cycle crosses branch cuts there will be relative phases that prevent $\partial_{\omega,\eta}$ from vanishing. 
Only by breaking up the compact cycle into pieces and combining these pieces with particular phases do we obtain a genuine compact twisted cycle.

This procedure is best illustrated through examples. 
Consider the cycles of figure \ref{fig:hom_basis_reg} in {\color{blue!80!black}blue}.
Clearly, these are compact.
The compact cycles corresponding to the non-compact $\gamma_{2j=3\dots,n, A, B}$ cycles have been broken up into three pieces: one encircling the puncture $z_2$, one encircling the puncture $z_j$ and one connecting each of these circles below the branch cut. 
When $j=A$ or $j=B$, the compact cycle encircles $z_2{+}1$ or $z_2{+}\tau$.
To understand how to combine the contours $m_j, S_j$ and $\ell_{2j}$ into a genuine compact twisted cycle, we enumerate their twisted boundaries:
\begin{equation} \label{eq:action_bdry_op_chains}
\begin{aligned}
    \partial_{\omega,\eta} m_j
    &= \begin{cases}
    	 P^{(2)}_{j+1}-P^{(2)}_{j} 
	 	& \text{for}\ i \in \{1,2,3\}
		\,,
	 \\
	 e^{2\pi i s_{12}} P^{(2)}_1-P^{(2)}_0 
	 	&\text{for} \ j= 0 
		\,,
    \end{cases}
\\
     \partial_{\omega,\eta} S_{j} 
   	&= (e^{2\pi i s_{1j}}-1)P^{(j)}_0 \quad \text{for}\ j \in \{3,\dots,n\},
\\
    \partial_{\omega,\eta} \ell_{2j} 
    &=\begin{cases}
    	P_0^{(j)}-P^{(2)}_0 
		& \text{for}\ i \in \{1,2,3\}
		\,,
	\\
	e^{2\pi i s_A}P^{(2)}_2-P^{(2)}_0 
		& \text{for}\ j=A
		\,,
	\\
	e^{-2\pi i s_B}P^{(2)}_3-P^{(2)}_1
		& \text{for}\ j=B
		\,.
    \end{cases}
 \end{aligned}
\end{equation}
Using \eqref{eq:action_bdry_op_chains}, the reader can verify that the regulated contours 
\begin{align}
    \reg[\gamma_{2A}]  &:= 
        \frac{
            m_0 
            + e^{2\pi i s_{12}}
            (m_1+m_2+m_3)
        }{
            e^{2\pi i s_{12}} -1
        }
        {+} \ell_{2A}
        {-} e^{ 2 \pi i s_{1A} }
            \frac{
                (m_2+m_3+m_0) 
                + e^{2 \pi i s_{1A} } m_1 
            }{  
                e^{2\pi i s_{12}} -1 
            }\label{eq:reg_gammaA}
    \\
    \reg[\gamma_{2B}] &:= 
        \frac{m_1+m_2+m_3+m_0}{ e^{2\pi i s_{12}} -1 }
        + \ell_{2B}
        - e^{-2 \pi i s_{1B}}
            \frac{
                (m_3+m_0)
                + (m_1+m_2) 
                    e^{2 \pi i s_{12}}
            }{
                 e^{2\pi i s_{12}} - 1
            }\label{eq:reg_gammaB}
    \\
    \reg[\gamma_{2j}] &:= 
        \frac{
            m_0 
            + e^{2\pi i s_{12}}
            (m_1+m_2+m_3)
        }{
            e^{2\pi i s_{12}} -1
        }
            + \ell_{2j}
            - \frac{S_j}{
            e^{ 2\pi i s_{1j}-1}
        }
    \qquad\text{for}\quad j=3, \dots, n \label{eq:reg_gamma2j}
\end{align}
have no twisted boundary.
In practice, this regularisation procedure is equivalent to an analytic continuation by using a Pochhammer contour \cite{Mizera:2017cqs}. 
The regularization of these cycles almost identical to the genus-zero case \new{and equivalent to \cite{goto2020homology}}.

Since the intersection matrix is a basis dependent object, there is no unique best choice.
Therefore, we will provide the intersection matrix for two choices of dual cycles in this section. 
Choosing the {\color{Green!80!black}dual basis} in figure \ref{fig:hom_basis} and \ref{fig:hom_basis_reg}, yields the following diagonal intersection matrix 
\begin{align} \label{eq:H2n}
	H^{(2n)}_{ab} 
	&:= [\gamma_{2a} \vert \c{\gamma}_{nb}]
	\\
	&= \mathrm{diag}\scalebox{.98}{$\left(
		\frac{
			e^{2\pi i (s_{12}+s_{13})}
		}{
			1-e^{2\pi i s_{13}}
		}, 
		\frac{
			e^{2\pi i (s_{12}+s_{13}+s_{14})}
		}{
			1-e^{2\pi i s_{14}}
		},
		\cdots,
		\frac{
			e^{2\pi i (s_{12}+s_{13}+ \dots + s_{1,n-1})}
		}{
			1-e^{2\pi i s_{1,n-1}}
		},
		e^{-2\pi i (s_{1n}+s_{1B})},
		e^{-2\pi i s_{1A}}
	\right)$},
	\nn
\end{align}
where $a=3,\dots, n-1, A, B$ and $b=3, \dots, n-1, B, A$. Note that we have to invert the ordering of the $A$- and $B$-cycles in the dual basis to make the intersection matrix diagonal.
This form of the intersection matrix is most useful in the decomposition of identity \eqref{eq:decomp_id} since $\mat{H}^{(2n)}$ is easy to invert.
For example, we can prove \eqref{eq:homology_cycle_overcomplete_basis} by applying \eqref{eq:decomp_id} to $[\gamma_{2n}\vert$
\begin{align}
	[\gamma_{2n}\vert
	&\simeq \sum_{a=3}^{n-1} \frac{
		[\gamma_{2n} \vert \c{\gamma}_{na} ]
	}{
		[\gamma_{2a} \vert \c{\gamma}_{na} ]
	} [\gamma_{2a} \vert 
	+ \frac{
		[\gamma_{2n} \vert \c{\gamma}_{nB} ]
	}{
		[\gamma_{2A} \vert \c{\gamma}_{nB} ]
	} [\gamma_{2A} \vert 
	+\frac{
		[\gamma_{2n} \vert \c{\gamma}_{nA} ]
	}{
		[\gamma_{2B} \vert \c{\gamma}_{nA} ]
	} [\gamma_{2B} \vert 
	=  \frac{- \mathrm{LHS} \eqref{eq:homology_cycle_overcomplete_basis} }{1-e^{2\pi i s_{1n}}} 
	\,. 
\end{align} 
Here, we have used \eqref{eq:H2n} and 
\begin{align}
	[\gamma_{2n} \vert \c{\gamma}_{nb} ]_{b=3, \dots, n-1, B, A}
	&= \frac{-1}{ 1-e^{2 i \pi  s_{1n}} }
	\left\{
		1, \dots, 1, 
		 e^{ 2\pi i s_{1n} } (1- e^{ -2\pi i s_{1B} } ) , 
		(1-e^{ 2\pi i s_{1A} } )
	\right\}
	\,. 
\end{align}

While the dual basis $\{\c{\gamma}_{nb}\}_{b=3,\dots,n-1,B,A}$ is convenient for many computations, it is not the best for the double copy because it requires evaluating two sets of twisted periods: $[\gamma_{2a} \vert \vphi_b\ra$ and $\la \c{\vphi}_a \vert \c{\gamma}_{nb}]$. 
On the other hand, useing the basis of topological cycles $\gamma_{2a}$ for the dual basis as well (i.e., use $\c{\gamma}_{2b} = [\gamma_{2b}]^\vee = \gamma_{2b} \vert_{s_\bullet\to-s_\bullet}$ as our dual basis), one easily obtains the dual periods from the ``known'' integrals% 
% $[\gamma_{2a} \vert \vphi_b\ra$
\new{: $\la \c{\vphi}_a \vert \c{\gamma}_{2b}] = [\gamma_{2b} \vert \vphi_a \ra\vert_{s_\bullet \to -s_\bullet}$}. 
The price one pays is that the intersection matrix becomes incredibly dense and hard to invert. 
\new{However, inverting this matrix is still much easier than evaluating two sets of periods. }
Explicitly, we set 
\begin{align} \label{eq:H22}
	H^{(22)}_{ab} := [\gamma_{2a} \vert \c{\gamma}_{2b} ] \,,
\end{align}
for $a=3,\dots, n-1, A, B $ and $b=3,\dots, n-1, A, B$ where 
\begin{align}
\begin{aligned}
    [\gamma_{2j}|\c{\gamma}_{2k}] 
    	&= \begin{cases}
       	 	\frac{e^{2\pi i s_{1j}}}{1-e^{2\pi i s_{1j}}}&\text{for}~j<k\\
	        \frac{1}{1-e^{2\pi i s_{1j}}}&~~~~ j>k,\\ 
	    \end{cases} 
    \qquad\qquad ~
     [\gamma_{2j}|\c{\gamma}_{2j}] 
     	= \frac{1-e^{2\pi i(s_{12}+s_{1j})}}{(1-e^{2\pi i s_{12}})(1-e^{2\pi i s_{1j}})},
    \\
    [\gamma_{2j}|\c{\gamma}_{2A}] 
   	 &= \frac{e^{2\pi i s_{12}}(1-e^{-2\pi i s_{1A}})}{1-e^{2\pi i s_{12}}},
    \qquad\qquad~~
    [\gamma_{2A}|\c{\gamma}_{2j}] = \frac{1-e^{2\pi i s_{1A}}}{1-e^{2\pi i s_{12}}},
    \\
    [\gamma_{2j}|\c{\gamma}_{2B}] 
	    &= \frac{e^{2\pi i s_{12}}(1-e^{2\pi i s_{1B}})}{1-e^{2\pi i s_{12}}},
    \qquad\qquad~~~~
    [\gamma_{2B}|\c{\gamma}_{2j}] = \frac{1-e^{-2\pi i s_{1B}}}{1-e^{2\pi i s_{12}}},
    \\
    [\gamma_{2A}|\c{\gamma}_{2A}] 
	    &= -\frac{(e^{2\pi i s_{1A}}-1)(e^{2\pi i s_{1A}}-e^{2\pi i s_{12}})}{e^{2\pi i s_{1A}}(1-e^{2\pi i s_{12}})},
    \\
    [\gamma_{2B}|\c{\gamma}_{2B}] 
	    &= -\frac{(e^{2\pi i s_{1B}}-1)(e^{2\pi i s_{1B}}-e^{2\pi i s_{12}})}{e^{2\pi i s_{1B}}(1-e^{2\pi i s_{12}})},
    \\
    [\gamma_{2A}|\c{\gamma}_{2B}] 
	    &=  \frac{e^{2\pi i s_{12}}-e^{2\pi i (s_{1A}+s_{12})}-e^{2\pi i (s_{1B}+s_{12})}+e^{2\pi i (s_{1B}+s_{1A})}}{1-e^{2\pi i s_{12}}},
    \\
    [\gamma_{2B}|\c{\gamma}_{2A}] 
	    &=  \frac{1-e^{-2\pi i s_{1B}}-e^{-2\pi i s_{1A}}+e^{2\pi i (s_{12}-s_{1B}-s_{1A})}}{1-e^{2\pi i s_{12}}}.
\end{aligned}
\label{eq:all_homology_intersection_number}
\end{align}
\new{Aside from a change in notation, the above intersection indices can be found in \cite{Goto2022} and are quoted here for the convenience of the reader. 
Pedagogical examples of selected intersection indices have also been provided in appendix \ref{app:intInd}.}

Even though it is not obvious that $\mat{H}^{(22)}$ is invertible, its determinant is easily computed \begin{equation}
   	 \det(\mat{H}^{(22)}) 
	= \frac{1-e^{-2\pi i s_{1n}}}{\prod_{m=2}^{n-1}(1-e^{2\pi i s_{1m}})} 
	\neq 0
	\,, 
\end{equation}
and non-vanishing. 
Similarly, the determinant of $\mat{H}^{(2n)}$ is also non-vanishing.  
However, notice that these determinants are finite only when $s_{12},\dots,s_{1n-1}\notin \mathbb{Z}$.
Thus, the bases of twisted cycles introduced in this section are only valid when the Mandelstams are generic: $s_{12},\dots,s_{1n} \notin \mathbb{Z}$. 
Also note that when $s_{1A} = s_{1B}$ 
\begin{align}
	\begin{pmatrix} 
		[\gamma_{2A} \vert \c\gamma_{2A} ]
		& [\gamma_{2A} \vert \c\gamma_{2B} ]
		\\
		[\gamma_{2B} \vert \c\gamma_{2A} ]
		& [\gamma_{2B} \vert \c\gamma_{2B} ]
	\end{pmatrix}
	= \begin{pmatrix} 
		\phantom{-}0 & 1
		\\
		-1 & 0
	\end{pmatrix}
\end{align}
and we can identify $\gamma_{2A}$ and $\gamma_{2B}$ with the usual basis of the non-twisted homology $H_1(E,\mathbb{Z})$. 

Before closing this section, we note that intersection indices can be used to compute the monodromy matrices associated to $z_j$ $A$- and $B$-cycles as well as encircling $z_{k\geq2}$ with $z_{j\geq2}$ (i.e., the circuit matrix $M_{jk}$) \cite{Goto2022}.
While we do not explicitly need these matrices in this work, it is worth stressing that their computation is rather straightforward from the intersection point of view. 
Moreover, these matrices may be the key to understanding Riemann-Wirtinger  integrals when more than one puncture is integrated (see section \ref{sec:doubleint} for some speculation on how this may work)!
While the derivation of the circuit matrices is almost identical to the analogous genus-zero case, something new and interesting happens when computing the $z_{j\geq2}$ $A$- and $B$-cycle monodromy matrices.
In addition to a phase, the $z_{j\geq2}$ $A$- and $B$-cycles induce a change in the local system where
\begin{align}
    z_{j\geq2}\ A\text{-cycle}:
    s_{1B} \mapsto s_{1B} - s_{1j}
    \,,
    \\ 
    z_{j\geq2}\ B\text{-cycle}:
    s_{1A} \mapsto s_{1A} + s_{1j}
    \,.
\end{align}
The corresponding monodromy matrices realize a linear map from the homology with $s_{1A}$ and $s_{1B}$ to the homology with $s_{1B}-s_{1j}$ or $s_{1A}+s_{1j}$. 
For more details, see \cite{Goto2022}.

%%%%%%%%%%%%%%%%%%%%%%%%%%%%%%%%%%%%%%%%%%%%%%%%%
%%%%%%%%%%%%%%%%%%%%%%%%%%%%%%%%%%%%%%%%%%%%%%%%%
%%%%%%%%%%%%%%%%%%%%%%%%%%%%%%%%%%%%%%%%%%%%%%%%%
\section{The double copy of Riemann-Wirtinger integrals \label{sec:DC-RW}}

In this section, we introduce the closed-string analogues of the Riemann-Wirtinger integrals, which we call complex or single-valued  Riemann-Wirtinger integrals (the first example of which was considered in section 4.4.4 of \cite{ghazouani2016moduli}).
% The complex Riemann-Wirtinger integrals are single-valued periods in the sense of \cite{Brown:2019jng} and admit a double copy as reviewed in section \ref{sec:Fund}.
We define the complex Riemann-Wirtinger integral and its double copy in section \ref{sec:DC-RW-eta}. 
Then, in sections \ref{sec:real_s} we study the double copy for real Mandelstam variables $\{s_{1j}\}_{j=2}^{n}, s_{1A},s_{1B} \in \mathbb{R}$. 
However, since double copy formulas are meromorphic functions of the Mandelstams, we can take them to be complex quantities in the end.
Two natural ways of relaxing the reality condition are explored in sections \ref{sec:complex_s} and \ref{sec:modular}.

%%%%%%%%%%%%%%%%%%%%%%%%%%%%%%%%%%%%%%%%%%%%%%%%%
%%%%%%%%%%%%%%%%%%%%%%%%%%%%%%%%%%%%%%%%%%%%%%%%%
\subsection{Complex Riemann-Wirtinger integrals \label{sec:DC-RW-eta}}

In section \ref{sec:Fund}, we defined the single-valued pairing between a twisted form and a complex-conjugated twisted form as the intersection pairing  
\begin{align}
    \la\overline{\bullet}\vert\bullet\ra:
    H^1(\overline{M},\overline{\LSdual})
    \times H^1(M,\LS)
    \to \mathbb{C}
    \,. 
\end{align}
This pairing is single valued in the sense that the integrand insensitive to all $z_1$-monodromies.
Explicitly, we define
\begin{align} \label{eq:complex-RW-def}
   \langle \bar{\vphi}_k \vert \vphi_l \rangle 
   := \int_{M} |u(z_1)|^2 \
   F(z_1-z_l,\eta\vert\tau)\ 
   \overline{F(z_1-z_k,\eta\vert\tau)}\ 
    \d^2 z_1
   ,
\end{align}
where $\d^2 z_1 =  \frac{i}{2} \frac{\d z_1 \wedge \d \overline{z_1}}{\im\tau}$ is the normalized volume form, 
\begin{align} \label{eq:complex-u-def}
     |u(z_1)|^2 
     := e^{2\pi i s_{1A} (z_1-\overline{z}_1) } \prod_{j=2}^n |\vth(z_1-z_j)|^{2 s_{1j}},
\end{align}
and $P$ is the fundamental parallelogram.
Note that equation \eqref{eq:complex-u-def} is a definition when the Mandelstams are complex since, in this case, $u \bar{u} \neq |u|^2$.

One can think of this pairing as an intersection number where the dual twist is $\c{u}=\bar{u}$. 
To see this, we note that there is secretly a hidden factor of $\c{u}\ u$ inside the integral defining the intersection number \eqref{eq:Cint}.
Usually (outside of this section) $\c{u}=u^{-1}$ and the product $\c{u}\ u=1$ drops out of the formula;
the resulting integrand is single-valued/doubly-periodic.
However, it is crucial to choose $\c{u}$ such that it cancels the multi-valuedness of the integrand when defining the intersection number. 
Equivalently, this conditions tells us what spaces we are allowed to pair in this manner.
For example, the choice $\c{u}=\bar{u}$ also renders the integrand single-valued/doubly-periodic. 
This suggests that we can take the replacement $\c{u}\to\bar{u}$ in the twisted Riemann-bilinear relations seriously and define the double copy
\begin{align}
    \label{eq:complex-RW-integral-klt}
    \langle \bar{\vphi}_a \vert \vphi_b \rangle
    = \frac{i}{2}\sum_{i,j = 1}^{n-1}
    \la \bar{\vphi}_a \vert \bar{\delta}_{i} ]\
    (H_{ij}^{-1})^{\top}\ 
    [ \gamma_{j} \vert \vphi_b\ra
    = \frac{i}{2}\sum_{i,j = 1}^{n-1}
    [ \gamma_{i} \vert \vphi_b\ra \
    H_{ij}^{-1}\ 
     \overline{[ \delta_{j} \vert \vphi_a\ra}
    \, ,
\end{align}
where $\mat{H}$ is the homology intersection matrix corresponding to a basis choice $\{\gamma_i\}$ and $\{\delta_i\}$.
The factor of $i/2$ in \eqref{eq:complex-RW-integral-klt} comes from our particular normalization of $\d^2z_1$. 

However, as written, one  should not expect \eqref{eq:complex-RW-def} to have any cohomological meaning for generic complex $\eta$ since the integrand is not necessarily doubly-periodic, even if $s_{1A}$ and the
$\{s_{1j}\}_{j=1}^{n}$ (all but $s_{1B}$) are real. 
In fact, the integrand of \eqref{eq:complex-RW-def} precisely fails to be doubly-periodic because it has a $B$-cycle monodromy:
\begin{align}
    \label{eq:complex-RW-Bmon}
    B\text{-cycle}:&\qquad 
    \frac{|u(z_1)|^2 \ F(z_1-z_j,\eta\vert\tau) \overline{F(z_1-z_k,\eta\vert\tau)}\vert_{z_1 \to z_1+\tau}}{|u(z_1)|^2 \ F(z_1-z_j\vert\tau)\overline{F(z_1-z_k,\eta\vert\tau)}}
    = e^{2 \pi i (s_{1B}-\overline{s}_{1B})}
    .
\end{align}
Thus, to ensure that the $z_1$-integral over the fundamental parallelogram is indeed the integral over the whole space, we further require the reality of the Mandelstam variable
\begin{align}
\label{eq:realityOfSB}
    \overline{s_{1B}} =s_{1B}
    \, .
\end{align}
Since $s_{1B}$ is defined to satisfy \eqref{eq:eta-def}, complex conjugation has a natural action on $s_{1B}$ via the complex conjugation on punctures $z_j$, $\eta$ and the modulus $\tau$ and thus \eqref{eq:realityOfSB} is non-trivial.\footnote{A version of \eqref{eq:realityOfSB} is found in equation (63) of \cite{ghazouani2016moduli}} One possible way of satisfying \eqref{eq:realityOfSB} numerically is by first choosing a value of $s_{1B}$ and then using the constraint \eqref{eq:eta-def} to find a value of $\eta$ given $\tau$ and the rest of the Mandelstam variables. 

Another important reason why we require the condition in \eqref{eq:realityOfSB} is because we will use the isomorphism of local systems $\mathcal{L}_{\omega,\eta} \cong \overline{\LSdual}$ where 
\begin{align}
    \overline{\LSdual}
    &= \c{\L}_{\omega,\bar{\eta}}(-\overline{s}_{1A},-\overline{s}_{12},\dots,-\overline{s}_{1n},-\overline{s}_{1B}) 
    = \mathbb{C} \overline{u} \otimes \L_{\bar{\eta}}
    \, ,
\end{align} 
is the complex conjugate of the dual local system (c.f., equation \eqref{eq:dualLS}). 
However, this coincides with \eqref{eq:LS} 
\begin{align} \label{eq:complex-conjugateLS}
    \overline{\LSdual}
    &\cong 
    \L_{\omega,\eta}(s_{1A},s_{12},\dots,s_{1n},s_{1B}) 
    \quad \text{if} \quad s_{\bullet}\in\mathbb{R}
    \,,
\end{align} 
only if all $s_\bullet$'s are real --- in this case, $u^{-1}$ and $\bar{u}$ have the same monodromies (we will show this more explicitly in section \ref{sec:real_s}).

Strictly speaking, the double copy formula \eqref{eq:complex-RW-integral-klt} should include the complex-conjugate twisted-cycles $\overline{\gamma_j} \in H_m(M,\overline{\LSdual})$.
However, the isomorphism $\mathcal{L}_{\omega,\eta} \cong \overline{\LSdual}$\footnote{Explicitly, given a section, $a \bar{v}$, of $\overline{\c\L_{\omega}}$ where $a \in \mathbb{C}$, this isomorphism is given by
\begin{align*}
    a \bar{v} 
    \mapsto 
    a \frac{\bar{v}}{\bar{u} u}
    \,.
\end{align*}
Since $\bar{v}/\bar{u} \in \mathbb{C}$ is a phase, $a \frac{\bar{v}}{\bar{u} u} \in \mathbb{C} u^{-1} = \L_\omega$. 
Furthermore, this induces an isomorphism between homologies via $\bar{\gamma} = \tilde{\gamma} \otimes \bar{v} \to \tilde{\gamma} \otimes \frac{\bar{v}}{\bar{u}} u^{-1}$.
} 
allows us to recycle the intersection indices of section \ref{sec:RWhom-int} without modification. 
In fact, we can use the same intersection indices even when the $s_{\bullet}$'s are complex since they enter the double copy in a meromorphic way (sections \ref{sec:complex_s} and \ref{sec:modular}).
Similarly, we should also have the conjugated period pairing $\la \overline{\vphi_a} \vert \overline{\gamma_{j} }]$ in \eqref{eq:complex-RW-integral-klt} where the conjugated cohomology comes with the connection $\bar{\nabla}$ or equivalently the local system $\overline{\LS}$. 

We start by building up our intuition about the double copy for real $s_\bullet$ in section \ref{sec:real_s} and describe how this condition can be relaxed later in sections \ref{sec:complex_s} and \ref{sec:modular}. 
In particular, we have checked the identity \eqref{eq:complex-RW-integral-klt} both for real and complex Mandelstams  $\{s_{1A}, s_{1B}, s_{1j=2,\dots,n}\}$ whenever the LHS of \eqref{eq:complex-RW-integral-klt} converges. 
Moreover, one can interpret \eqref{eq:complex-RW-integral-klt} as defining the analytic continuation of $\langle \bar{\vphi}_a \vert \vphi_b \rangle$ when \eqref{eq:complex-RW-def} does not converge.

%%%%%%%%%%%%%%%%%%%%%%%%%%%%%%%%%%%%%%%%%%%%%%%%%
%%%%%%%%%%%%%%%%%%%%%%%%%%%%%%%%%%%%%%%%%%%%%%%%%
\subsection{The double copy for real Mandelstam variables \label{sec:real_s}}

In this section, we test the double copy formula \eqref{eq:complex-RW-integral-klt} when all Mandelstams (including $s_{1B}$) are real. 

We start by understanding the local system $\overline{\LSdual}$. 
It is defined by the multi-valuedness of the integrands $\overline{u\ F(z,\eta\vert\tau)}$:
\begin{align}
     \label{eq:complex-conjugate-RW-Amon}
    A\text{-cycle}:&\qquad 
    \frac{\overline{u\ F(z,\eta\vert\tau)\vert_{z_1 \to z_1+1}}}{\overline{u\ F(z\vert\tau)}}
    = e^{-2 \pi i s_{1A}}
    ,
    \\
    \label{eq:complex-conjugate-RW-Bmon}
    B\text{-cycle}:&\qquad 
    \frac{\overline{u\ F(z,\eta\vert\tau)\vert_{z_1 \to z_1+\tau}}}{\overline{u\ F(z\vert\tau)}}
    = e^{-2 \pi i s_{1B}}
    ,
\end{align}
where \eqref{eq:eta-def} implies that we have
\begin{align}
    \label{eq:complex-conjugate-eta-def}
     \boxed{
     s_{1B} 
     = s_{1A} \overline{\tau} + \sum_{j=2}^n s_{1j} \overline{z_j} - \overline{\eta} 
     = \mathrm{const} \in \mathbb{R}.
     }
\end{align} 
Then, the twisted cycles $\overline{\gamma_j} \in H_1(\overline{M},\overline{\LSdual})$ can be paired up with twisted forms $\overline{\vphi_a}\in H^1(\overline{M},\overline{\L_{\omega,\eta}})$ to form complex-conjugated Riemann-Wirtinger integrals
\begin{align}
\label{eq:defn-complex-conjugated-RW}
\la \overline{\vphi_a} \vert \overline{\gamma_{j} }]
=
\overline{[ \gamma_{j} \vert \vphi_a\ra} \, .
\end{align}
These conjugated integrals also satisfy a version of the over-completeness relations \eqref{eq:homology_cycle_overcomplete_basis}:
 \begin{align}
    &
    (e^{-2 \pi i s_{1A}} - 1) \la \overline{\varphi} \vert \overline{\gamma_{2B}} ]
    + (1 - e^{2 \pi i s_{1B}}) \la \overline{\varphi} \vert \overline{\gamma_{2A}} ]
    \nn\\&\qquad
    - \sum_{j=3}^n 
        e^{-2 \pi i (s_{12} + \cdots + s_{1j})}
        (1 - e^{-2 \pi i s_{1j}})
        \la \overline{\varphi} \vert \overline{\gamma}_{2j} ]
    = 0,
    \label{eq:homology_cycle_overcomplete_basis-cc-real-rw-integral}
\end{align}
which is as a direct corollary of \eqref{eq:defn-complex-conjugated-RW} and complex-conjugating \eqref{eq:homology_cycle_overcomplete_basis}. 

In principle, as long as one uses the same topological cycles for a basis of both homologies and the same complex differential forms for the basis of both cohomologies, the conjugated integrals do not need to be evaluated from scratch.
However, for the purposes of numerics, the following fact\footnote{\new{This comes naturally from the contour integral as a Riemann integral, and linearity of complex integration (using a real parametrization of the contour), see equation (4) in Chapter 4 of  \cite{ahlfors1979complex} and the paragraphs below equation (6) in the same chapter for a similar expression.} } provides a practical definition of the conjugated Riemann-Wirtinger integral. 
Given a meromorphic function $f:\mathbb{C}\to\mathbb{C}$ and a real-parameterized path $\gamma:[0,1]\mapsto \mathbb{C}$ we have the fact 
\begin{align}\label{eq:cc-intf}
\overline{\int_\gamma f(z)\d z}= \int_{\overline{\gamma}}\overline{f(\overline{z})} \d z .
\end{align}
Thanks to \eqref{eq:cc-intf}, the conjugated integral can be rewritten as 
\begin{align} \label{eq:cc-RW-def-integ-real-mandelstams}
   \la \overline{\vphi_a} \vert \overline{\gamma_{j} } ]
   & =
   \int_{\overline{\gamma_j}} \overline{u(\overline{z_1})} \
   \overline{F(\overline{z_1}-z_a,\eta\vert\tau)}\ 
   \d z_1
    =
    \int_{\overline{\gamma_j}} \overline{u(\overline{z_1})} \
   F(z_1-\overline{z_a},\overline{\eta} \vert - \overline{\tau})\ \d z_1
   \,,
    \\
   \overline{u(\overline{z_1})} &= e^{-2\pi i s_{1A} z_1 } \prod_{i=2}^n \overline{\vth(\overline{z_1}-z_i\vert \tau)}^{s_{1i}}
   =
   e^{-2\pi i s_{1A} z_1 } \prod_{i=2}^n \vth^{s_{1i}}(z_1-\overline{z_i}\vert-\overline{\tau})
   \,,
\end{align}
which we contrast to \eqref{eq:RW-def}. Note that we have used the following identity for theta functions,
\begin{align}
\overline{\vth(z\vert \tau)}
=
 \vth(\overline{z}\vert -\overline{\tau}) \, ,
\end{align}
which serves to emphasize that the integrand is a meromorphic function of $z_1$, up to the expected branch cuts\footnote{This is because, if  $f(z)$ satisfies the Cauchy-Riemann relations, so does $\overline{f(\overline{z})}$.}.

For most numerical checks of \eqref{eq:complex-RW-integral-klt}, the regularized twisted cycles (\ref{eq:reg_gammaA}-\ref{eq:reg_gamma2j}) are overkill since there are many choices of $s_{1j}$, $z_j$ and $\tau$ for which the integrals over the un-regulated cycles converge. 
For example,
\begin{align}
\int_{0}^{1} u(z_1)\ F(z_1-z_a,\eta |\tau)\ \d z_1 \, 
\label{eq:unreg-RW-example}
\end{align}
is convergent for $\re (s_{12}) > 0$. 
As a function of $s_{12}$, the regulated cycle $\gamma_{2A}=\textrm{Reg}(0,1)$ (remember that we have gauged $z_2=0$) analytically continues this integral to negative $s_{12}$. 
A numeric implementation of these integrals and the resulting double copy is provided in the ancillary \texttt{Mathematica} notebook. 

\begin{figure}
    \centering
    \includegraphics[scale=0.6]{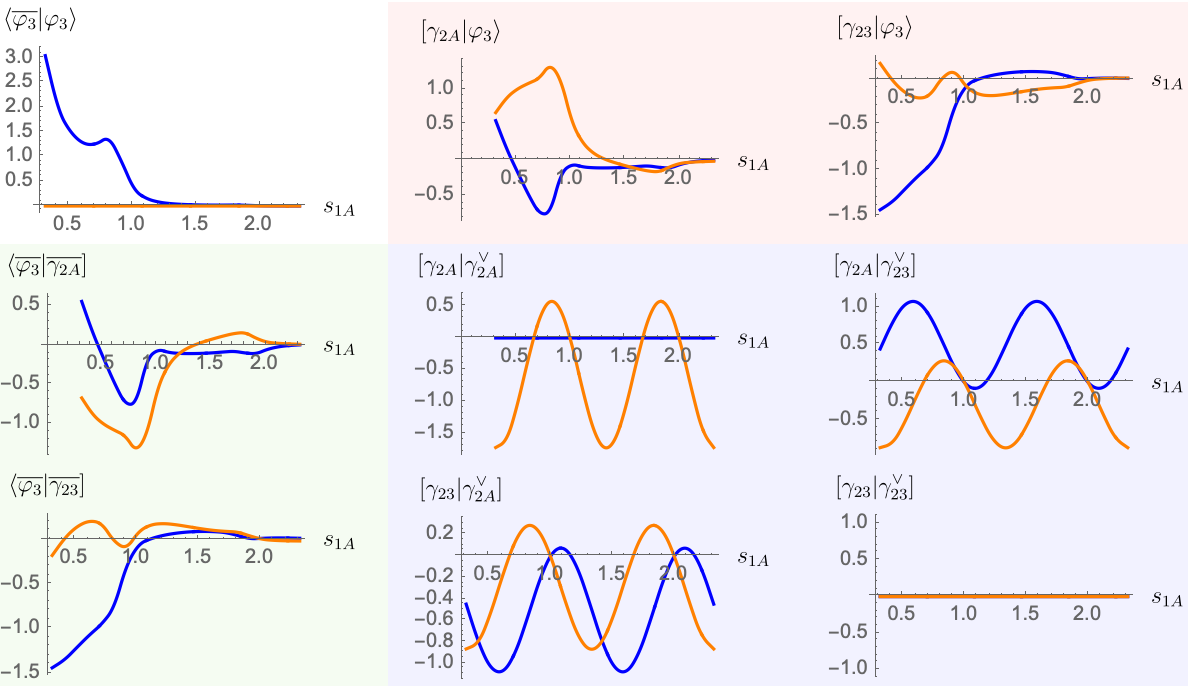}
    \caption{Plots of each component entering \eqref{eq:complex-RW-integral-klt} for $n=3$ and $(s_{12},s_{1B},z_3,\tau)=(-\frac{1}{3},-\frac{\pi}{3},\frac{1}{3}+\frac{i}{3},i+\frac{1}{11})$ as $s_{1A}$ varies.
    In each subplot, the real part of the quantity is in blue and its imaginary part is in orange. 
    Here, we choose a symmetric basis for homology and dual homology: $\{\gamma_{2A}, \gamma_{2B}\}$ and $\{\c\gamma_{2A}, \c\gamma_{2B}\}$. 
    In this case, several quantities plotted above are related by complex conjugation.
    }
    \label{fig:real-mans-symmetric}
\end{figure}

\begin{figure}[h]
    \centering
    \includegraphics[scale=0.6]{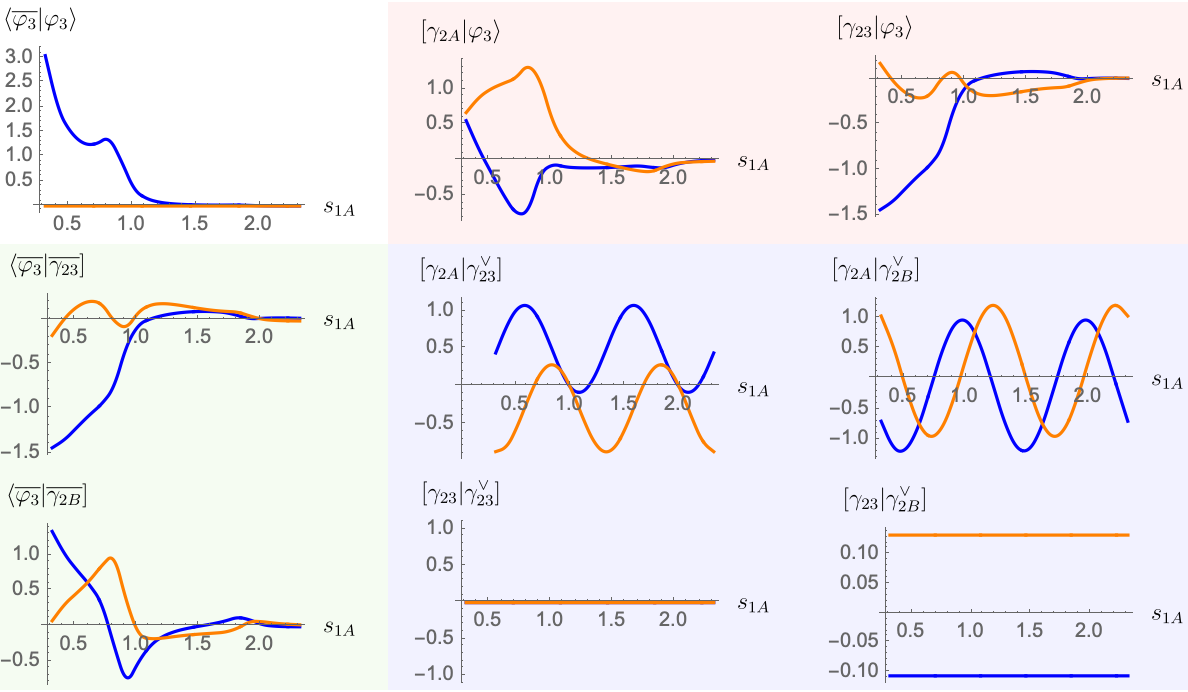}
    \caption{The analogous plot to figure \ref{fig:real-mans-symmetric} using an asymmetric choice for the basis of homology and dual cohomology: $\{\gamma_{2A}, \gamma_{23}\}$ and $\{\c\gamma_{23}, \c\gamma_{2B}\}$.
    }
    \label{fig:real-mans-asymmetric}
\end{figure}

In all cases, we will choose $\{\vphi_{a=2,\dots,n}\}$ as the basis for the cohomology and $\{\bar{\vphi}_{a=2,\dots,n}\}$ as the basis for the complex-conjugated cohomology.
In particular, the double copy of the ``diagonal'' pairings $\la \bar{\vphi}_a \vert \vphi_a \ra$ are real and positive.
Furthermore, we note that intersection matrix (and its inverse transpose) must be anti-Hermitian in order for the RHS of \eqref{eq:complex-RW-integral-klt} to be consistent with the reality of the LHS \eqref{eq:complex-RW-integral-klt}.
To see this, choose the same basis of topological cycles for the homology and its dual in \eqref{eq:complex-RW-integral-klt}. 
This is one way to rediscover the identity \cite{goto2020homology}
\begin{align}
[\gamma \vert \c\delta] = - [\delta \vert \c\gamma]^\vee, 
\end{align}
where we recall that the operator $\vee$ implements the involution  (see below \eqref{eq:dualLS})
\begin{align}
(s_{1A},s_{1B}, s_{12}, \ldots , s_{1n} ) \rightarrow 
(-s_{1A},-s_{1B}, -s_{12}, \ldots , -s_{1n}) 
\,.
\end{align}
Thus, the operator $\vee$ is equivalent to complex conjugation of the intersection number, $[\delta \vert \c\gamma]^\vee=\overline{[\delta\vert \c\gamma]}$ whenever the Mandelstam variables are real.

\begin{figure}[h]
    \centering
    \includegraphics[scale=0.55]{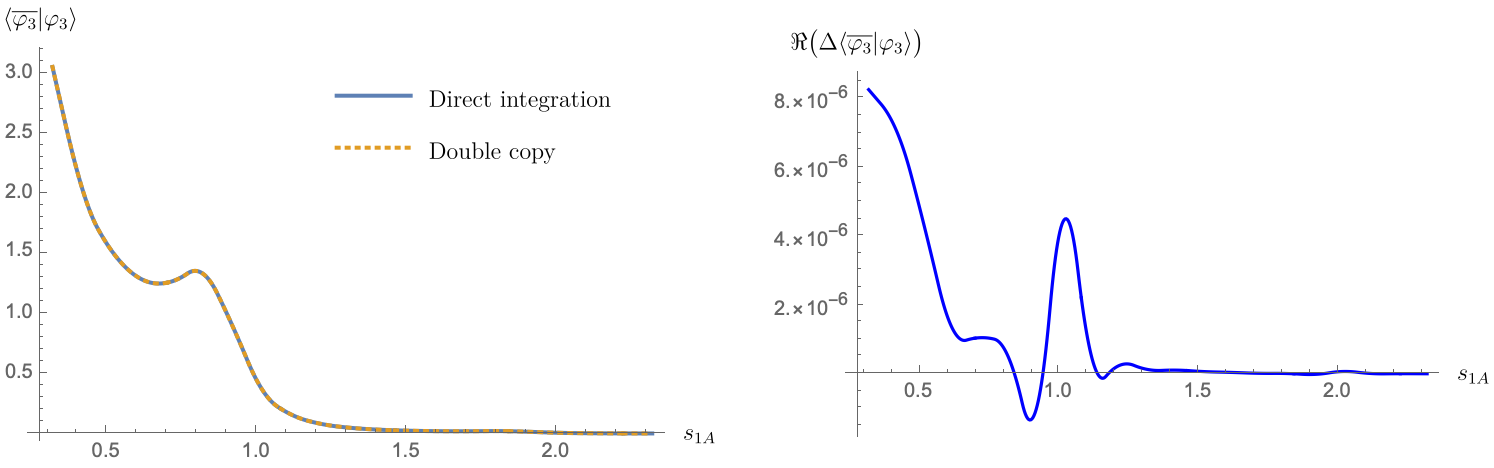}
    \caption{Numerical comparison of the (real part of the) complex Riemann-Wirtinger integral $\langle \overline{\varphi}_3 \vert \varphi_3 \rangle$ by numerical integration \eqref{eq:complex-RW-def} and
    by using the double copy formula \eqref{eq:complex-RW-integral-klt}.
    We use $(s_{12},s_{1B},z_3,\tau)=(-\frac{1}{3},-\frac{\pi}{3},\frac{1}{3}+\frac{i}{3},i+\frac{1}{11})$ and plot (a) $\langle \overline{\varphi}_3 \vert \varphi_3\rangle$ according to numerical integration of the complex integral \eqref{eq:complex-RW-def} and the double copy formula, and (b) the real  part of $\Delta \langle \overline{\varphi}_3 \vert \varphi_3\rangle$ for different values of  $s_{1A}$. Note that $ \langle \overline{\varphi}_3 \vert \varphi_3\rangle$ is a real quantity for these values of the Mandelstam variables, so the imaginary parts here are numerically zero.}
    \label{fig:real-mans-delta-comapared}
\end{figure}

In figures \ref{fig:real-mans-symmetric} and \ref{fig:real-mans-asymmetric} we plot the numerical values for every term that appears in the double copy formula \ref{eq:complex-RW-integral-klt} for $n=3$ as $s_{1A}$ varies. 
Figure \ref{fig:real-mans-symmetric} corresponds to choosing a symmetric basis for homology and dual homology: $\{\gamma_{2A}, \gamma_{2B}\}$ and $\{\c\gamma_{2A}, \c\gamma_{2B}\}$ and figure \ref{fig:real-mans-asymmetric} corresponds to choosing an asymmetric basis for homology and dual homology: $\{\gamma_{2A}, \gamma_{23}\}$ and $\{\c\gamma_{23}, \c\gamma_{2B}\}$.
While some of the individual building blocks look very different when comparing the two cases, they combine to form the same complex Riemann-Wirtinger integral!
We also compare the direct evaluation of the complex Riemann-Wirtinger integral \eqref{eq:complex-RW-def} to the one obtained by the double copy of \eqref{eq:complex-RW-integral-klt}. The difference of these quantities, 
\begin{align}
\Delta \langle \overline{\varphi}_3 \vert \varphi_3\rangle = \langle \overline{\varphi}_3 \vert \varphi_3\rangle\bigg|_{\text{\eqref{eq:complex-RW-def}}} -  \langle \overline{\varphi}_3 \vert \varphi_3\rangle\bigg|_{\text{\eqref{eq:complex-RW-integral-klt}}}
\end{align}
is shown in figure \ref{fig:real-mans-delta-comapared} and found to be numerically negligible.

\begin{figure}[h]
    \centering
    \includegraphics[scale=0.52]{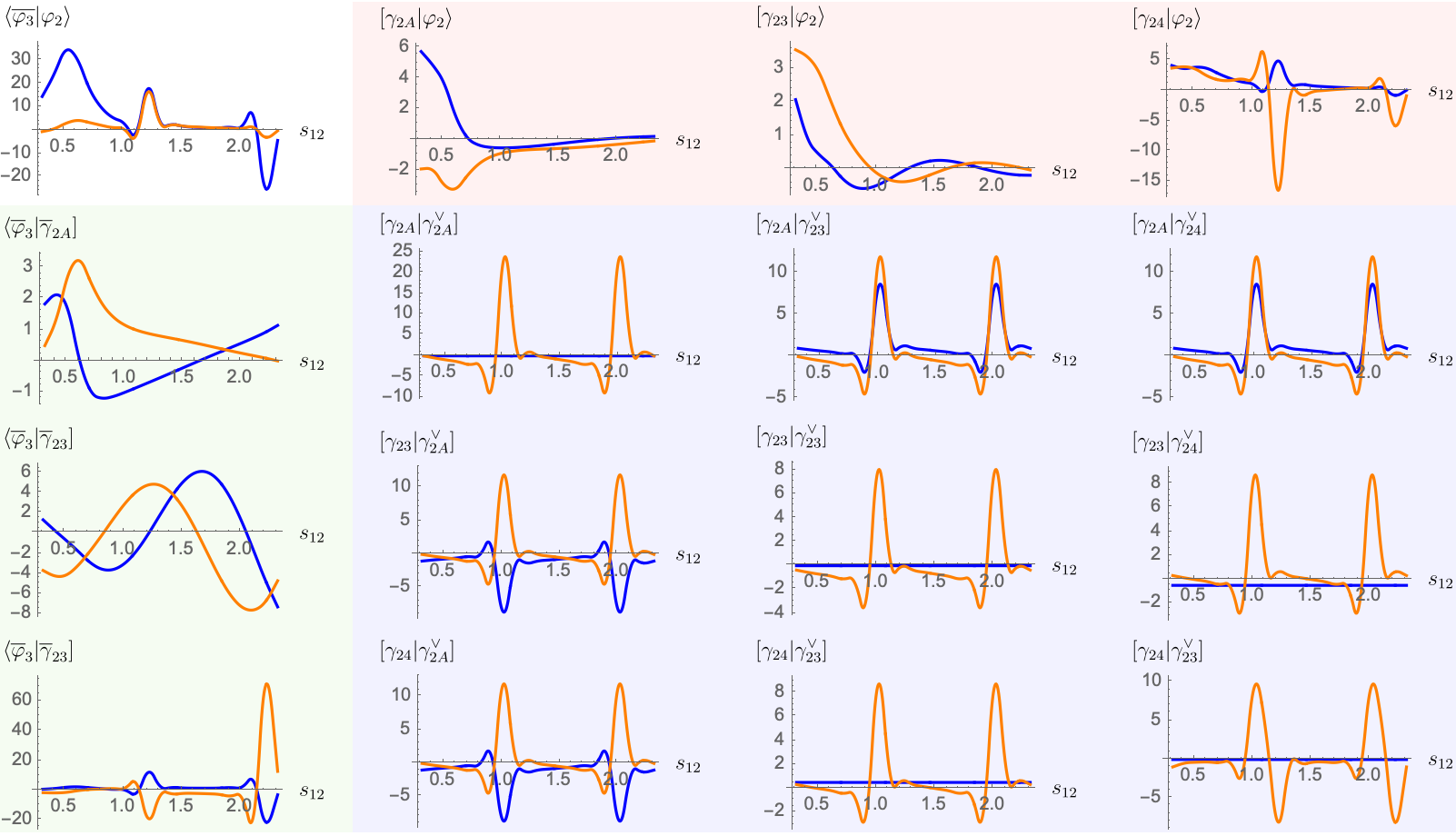}
    \caption{Numerical samples of every component that enters in the computation of $\langle \overline{\varphi}_3 \vert \varphi_2 \rangle$ via the double copy \eqref{eq:complex-RW-integral-klt} for a symmetric choice of homology bases for $n=4$ (i.e. 3-dimensional twisted homology basis). In each subplot, the real part of the quantity is in blue and its imaginary part is in orange. We use $(s_{13},s_{1A},s_{1B},z_3,z_4,\tau)=(-\frac{2}{10},\frac{1}{\pi},-\frac{\pi}{3},\frac{1}{7}+\frac{i}{3},\frac{2}{5}+\frac{i}{2},i+\frac{1}{11})$ and vary $s_{12}$. Note that because we use the symmetric choice in homology bases, the homology intersection numbers are related by complex conjugation.}
    \label{fig:real-mans-homo3}
\end{figure}
\begin{figure}[h]
    \centering
    \includegraphics[scale=0.52]{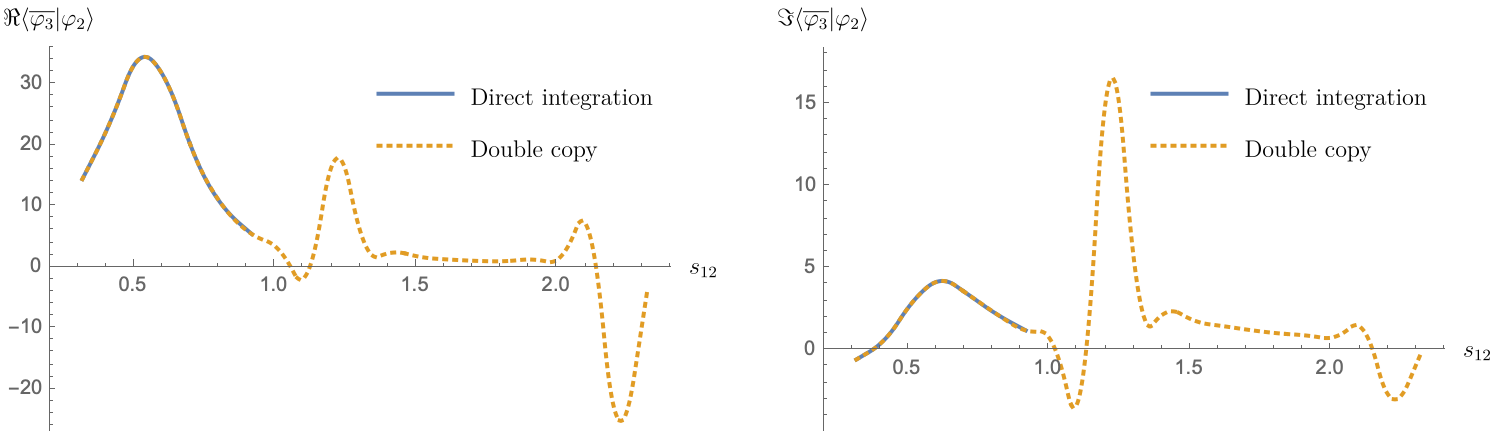}
    \caption{The quantity $\langle \overline{\varphi}_3 \vert \varphi_2 \rangle$ according to the complex integral \eqref{eq:complex-RW-def} (in blue) and the double copy formula \eqref{eq:complex-RW-integral-klt} (in orange) as a function of $s_{12}$. We use the same values of Mandelstam variables, punctures and moduli as in figure \ref{fig:real-mans-homo3}. We note that the definition by \eqref{eq:complex-RW-def}  diverges for $s_{12}>0$. The orange line analytically continues this integral  beyond this region.} 
    \label{fig:real-mans-homo3-together}
\end{figure}

In figure \ref{fig:real-mans-homo3}, we tabulate the building blocks that enter into 
the computation of the integral $\langle \overline{\varphi}_3 \vert \varphi_2 \rangle$
via the double copy with $n=4$ punctures (i.e. a three-dimensional twisted homology). 
In this case, the integral $\langle \overline{\varphi}_3 \vert \varphi_2 \rangle$ has both real and imaginary parts of similar magnitude.
The integral $\langle \overline{\varphi}_3 \vert \varphi_2 \rangle$ as defined by \eqref{eq:complex-RW-def} diverges for $s_{12}>1$. Thus, one \textit{defines} the integral $\langle \overline{\varphi}_3 \vert \varphi_2 \rangle$ by the double copy \eqref{eq:complex-RW-integral-klt} whenever $s_{12}>1$, see figure \ref{fig:real-mans-homo3-together}.
The integral $\langle \overline{\varphi}_3 \vert \varphi_2 \rangle$ as defined by the double copy \eqref{eq:complex-RW-integral-klt} for e.g.  $(s_{12},s_{13})=(\frac{4}{10},-\frac{2}{10})$ requires a regularization of the integral $\la \overline{\varphi}_3 \vert \overline{\gamma}_{23} ]$, but also evaluates to a convergent complex integral via its definition \eqref{eq:complex-RW-def}. Thus, we showcase a harmony between regularization via twisted cycles and convergent complex integrals.

% %%%%%%%%%%%%%%%%%%%%%%%%%%%%%%%%%%%%%%%%%%%%%%%%%
% %%%%%%%%%%%%%%%%%%%%%%%%%%%%%%%%%%%%%%%%%%%%%%%%%
\subsection{The double copy for complex Mandelstam variables  \label{sec:complex_s}}

So far we have assumed that the Mandelstam variables $\{s_{1A},s_{1B},s_{12},\ldots,s_
{1n}\}$ are all real. 
In this section, we will describe one way to relax this condition. 

First and foremost, the action of ``complex conjugation'' will still be relevant, but now we will define it (in this section only) to \textit{not} act on the Mandelstam variables:
\begin{align}
    (\overline{s_{1A}},
    \overline{s_{1j}})
    \rightarrow 
    (s_{1A},
    s_{1j}) 
    \, , 
\end{align}
even when they take complex values.
Complex conjugation acts on the puncture positions $z_j, \, \eta$ and the modulus $\tau$ as usual.
Next, we define the complex Riemann-Wirtinger integral as before \eqref{eq:complex-RW-def} except that $|u(z_1)|^2$ in the integrand is \emph{not} the usual absolute value. 
Instead, $|u(z_1)|^2$ is defined by \eqref{eq:complex-u-def} for complex Mandelstam variables.

Then, we need to remember the ``reality'' condition on the Mandelstam $s_{1B}$:
\begin{align}
\label{eq:conjugated-S1B}
\overline{s_{1B}}= s_{1B} \,
\end{align}
which is non-trivial because $s_{1B}$ is constrained (or defined) by \eqref{eq:eta-def}, and our complex conjugation, by definition, does not act on Mandelstam variables $s_{1A}$ or $s_{1j}$.  For example, one can satisfy this $\overline{s_{1B}}=s_{1B}$ condition by solving for $\eta$ and $\overline{\eta}$, if we first pick a complex value of $s_{1B}$. This is what we do for numerical checks. 
We will describe another way to satisfy this condition in \eqref{eq:solving_for_s1A}, in a way that works that uses the regular definition of complex conjugation (section \ref{sec:modular}).

\begin{figure}
    \centering
    \includegraphics[scale=0.56]{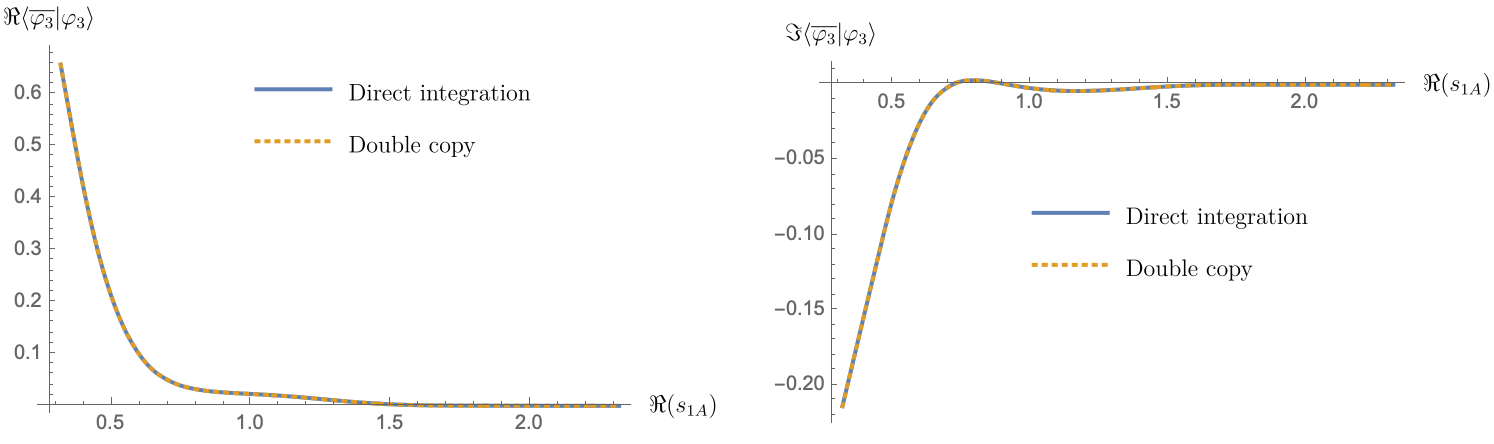}
    \caption{Plots for the quantity $\langle \overline{\varphi}_3 \vert \varphi_3 \rangle$ for $n=3$ and $(s_{12}, \im (s_{1A}),s_{1B},z_3,\tau)=(-\frac{1}{3}+i,\frac{1}{3},\frac{1}{3}+\frac{2i}{3},i+\frac{1}{11})$ as $\re (s_{1A})$  varies. 
    We compute the same quantity via the direct integration \eqref{eq:complex-RW-def} and via the double copy \eqref{eq:complex-RW-integral-klt}. Note that $\langle \overline{\varphi}_3 \vert \varphi_3 \rangle$ has both real and imaginary parts, because the integrand of \eqref{eq:complex-RW-def}
    is not strictly positive for generic complex Mandelstam variables. }
    \label{fig:complex-mans-double copy-Re-Im}
\end{figure}

\begin{figure}
    \centering
    \includegraphics[scale=0.60]{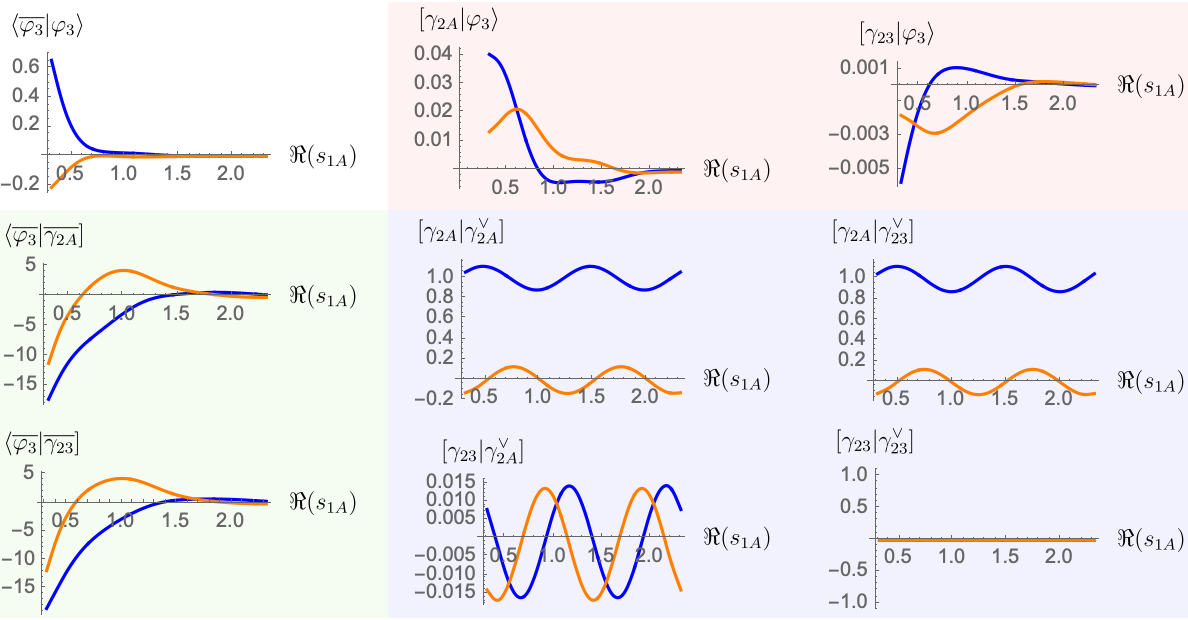}
    \caption{Plots of every component entering \eqref{eq:complex-RW-integral-klt} for $n=3$ and $(s_{12}, \im (s_{1A}),s_{1B},z_3,\tau)=(-\frac{1}{3}+i,\frac{1}{3},\frac{1}{3}+\frac{2i}{3},i+\frac{1}{11})$ as $\re (s_{1A})$  varies. Note that we use the same basis for homology and dual homology, but the quantities above, as complex numbers, are not  related by complex conjugation. Compare this case to the top subplot in figure \ref{fig:real-mans-symmetric}.}
    \label{fig:complex-mans-double copy}
\end{figure}

Then, the twisted Riemann bilinear relations are satisfied as in \eqref{eq:complex-RW-integral-klt}, except that we need to explicitly use the conjugated integrals $[ \overline{\gamma}_{2j}\vert \overline{\vphi}_a\ra$ defined in \eqref{eq:cc-RW-def-integ-real-mandelstams}. 
We note that even in the most symmetric case, the complex Riemann-Wirtinger integrals evaluate to complex numbers, due to the complex powers in $|u(z_1)|^2$. 
The double copy for complex Mandelstams is plotted in figures \ref{fig:complex-mans-double copy-Re-Im} and \ref{fig:complex-mans-double copy}.
In figure \ref{fig:complex-mans-double copy-Re-Im}, we compare the LHS and RHS of the double copy \eqref{eq:complex-RW-integral-klt} and find good numerical agreement. 
Each component entering this double copy is presented in figure \ref{fig:complex-mans-double copy}.

%%%%%%%%%%%%%%%%%%%%%%%%%%%%%%%%%%%
%%%%%%%%%%%%%%%%%%%%%%%%%%%%%%%%%%%
\subsection{The double copy and modularity \label{sec:modular}}

In this section, we relax the reality condition on the Mandelstams (excluding $s_{1B}$) in way that does not change how complex conjugation works. 
We keep $s_{1B}$ real and solve the condition \eqref{eq:realityOfSB} for $s_{1A}$ instead of $\eta$. 
Here, we can think of fixing $s_{1A}$ as a particular limit of the loop momentum. 
Moreover, $\eta$ is now a free variable and can take its usual interpretation as a formal expansion variable.
Interestingly, this approach is naturally associated to the doubly periodic completions of the Kronecker-Eisenstein functions that one sees after performing the integral over the loop momentum in string integrals and connects to the so-called elliptic modular graph forms (eMGFs) \cite{DHoker:2020hlp}. 

Equation \eqref{eq:realityOfSB} ensures that the integrand of the complex Riemann-Wirtinger integral, is doubly-periodic in $z_1$. However, this integral 
\begin{align}
\tilde{N}_{ab}(z_2,z_3,\ldots,z_n,\tau,s_{1A},s_{1j},\eta,\overline{\eta}) 
:=
\langle \bar{\vphi}_a \vert \vphi_b \rangle
\end{align}
is still a function of the  punctures $\{z_2,z_3,\ldots, z_n\}$ and the modulus $\tau$ but with no clear modular properties.
Now, consider solving for $s_{1A}$ in equation \eqref{eq:realityOfSB}:\footnote{
This is strikingly similar to how one integrates out the loop momentum in string theory, see (7.12) in \cite{Mafra:2018pll}.
This generalizes a formula of  Ghazouani and Pirio, see equation (63) of \cite{ghazouani2016moduli}.}
\begin{align}
\label{eq:solving_for_s1A}
s_{1A}
=
\frac{\im \eta - \sum_{j=2}^n s_{1j} \im z_j}{\im \tau} \, .
\end{align}
Substituting this value of $s_{1A}$ into the complex Riemann-Wirtinger integral
$
\tilde{N}_{ab}
$ and adding some ($z_1$-independent) multiplicative factors, we obtain an ``improved'' version of the complex Riemann-Wirtinger integral:
\begin{align}
\label{eq:def-modular-complex-RW}
    N_{kl}(z_2,z_3,\ldots,z_n,\tau,s_{1j},\eta,\overline{\eta}) 
    &=
    \tilde{N}_{kl}(z_2,z_3,\ldots,z_n,s_{1A},\tau,s_{1j},\eta,\overline{\eta})\big{|}_\eqref{eq:solving_for_s1A}
    \nn\\
    &\phantom{=} \notag
    \times \exp \left(-2 \pi i \eta \frac{\im z_l}{\im \tau}+2 \pi i \overline{\eta} \frac{\im z_k}{\im \tau}
    -\sum_{j=2}^n 2\pi s_{1j}  \frac{(\im z_j)^2}{\im \tau} \right)
    \nn\\
    &=\int_{M} v(z_1) \
    \Omega(z_1-z_l,\eta\vert\tau)\ 
    \overline{\Omega(z_1-z_k,\eta\vert\tau)}\ 
    \d^2 z_1 \, ,
\end{align}
where $\Omega$ are the doubly periodic completions of the Kronecker-Eisenstein functions $F$ and 
\begin{align} \label{eq:vdef}
    v(z_1)
    &=\exp\bigg[ 
        \sum_{j=2}^{n}s_{1j} \bigg(\log|\vth(z_{1j}|\tau)|^2  -2\pi \frac{(\im z_{1j})^2}{\im \tau} \bigg)  
    \bigg] 
    \, .
\end{align}
A derivation of equation \eqref{eq:def-modular-complex-RW} is included in appendix \ref{app:solving-for-s1A}. 

We remark that this integral $N_{kl}(z_2,z_3,\ldots,z_n,\tau,s_{1j},\eta,\overline{\eta})$ is doubly periodic in the punctures $\{z_2,z_3,\ldots,z_n\}$. Moreover, this version of the complex Riemann-Wirtinger integral inherits the modular properties of the Kronecker-Eisenstein series $\Omega(z,\eta|\tau)$. In particular, they are Jacobi forms of weight (1,1) and vanishing index:
\begin{align}
N_{kl}(z_2',z_3',\ldots,z_n',\tau',s_{1j},\eta',\overline{\eta}')= |\gamma \tau + \delta|^2 N_{kl}(z_2,z_3,\ldots,z_n,\tau,s_{1j},\eta,\overline{\eta}) \, ,
\end{align}
for 
\begin{align}
\label{eq:SL2-action-def}
\tau'=\frac{\alpha \tau + \beta}{\gamma \tau + \delta} \, ,
\hspace{12 mm}
z_j'=\frac{z_j}{\gamma \tau + \delta} \, ,
\hspace{12 mm}
\eta'=\frac{\eta}{\gamma \tau + \delta} \, ,
\hspace{12 mm}
\begin{pmatrix}
\alpha & \beta \\
\gamma & \delta 
\end{pmatrix}
\in
\textrm{SL}(2,\mathbb{Z}) \, .
\end{align}
Additionally, we can consider the Laurent series of $N_{kl}$ in $\eta$ and $\overline{\eta}$:
\begin{align}
N_{kl}=\sum_{a=0}^{\infty} \sum_{b=0}^{\infty} N^{(a,b)}_{kl} \overline{\eta}^{a-1} \eta^{b-1} \, ,
\end{align}
then every such $N^{(a,b)}_{kl}=N^{(a,b)}_{kl}(z_2,z_3,\ldots,z_n,\tau,s_{1j})$ is a Jacobi form of weight $(b,a)$ and vanishing index:
\begin{align}
N^{(a,b)}_{kl}(z_2',z_3',\ldots,z_n',\tau',s_{1j})= (\gamma \tau + \delta)^b \overline{(\gamma \tau + \delta)}^a N^{(a,b)}_{kl}(z_2,z_3,\ldots,z_n,\tau,s_{1j}) \, ,
\label{eq:modular-coeff-complex-RW}
\end{align}
for an $\text{SL}(2,\mathbb{Z})$ action as in \eqref{eq:SL2-action-def}. 
In fact, the component functions $N^{(a,b)}_{kl}$ are given by the integrals:
\begin{align}\label{eq:modular-complex-RW-component-by-integral}
N^{(a,b)}_{kl}(z_2,z_3,\ldots,z_n,\tau,s_{1j},\eta,\overline{\eta}) 
&=\int_{M} v(z_1) \
   f^{(b)}(z_1-z_l\vert\tau)\ 
   \overline{f^{(a)}(z_1-z_k\vert\tau)}\ 
   \d^2 z_1 \, .
\end{align}
Note that via the twisted period relations \eqref{eq:complex-RW-integral-klt}, we can write these as a bilinear combinations or as the  ``double copy'' of Riemann-Wirtinger integrals up to some $z_1$-independent factors.

The simplest example of \eqref{eq:modular-coeff-complex-RW} is given by $N^{(0,0)}_{2,2}(z_2,z_3,s_{12})$ for the $(n=3)$ Riemann-Wirtinger integral. This is the residue in $\eta$ and $\overline{\eta}$. On the RHS of the double copy \eqref{eq:complex-RW-integral-klt}, $\eta$ and $\overline{\eta}$ poles can only appear from the Kronecker-Eisenstein series, and no subleading contribution in $\eta$ or $\overline{\eta}$ contribute. 
Thus, when imposing \eqref{eq:solving_for_s1A} we only require a leading (in $\eta$, $\overline{\eta}$) version:
\begin{align}\label{eq:solving-s1A-leading-2dim-homology}
s_{1A} \rightarrow\frac{-s_{12}\im z_2-s_{13} \im z_3}{\im \tau} \, .
\end{align}
Choosing a symmetric choice of homology bases $\{\gamma_{2A},\gamma_{23}\}$, the intersection matrix is
\begin{align}
\begin{pmatrix}
[\gamma_{2A} \vert \gamma^{\vee}_{2A}] &
[\gamma_{2A} \vert \gamma^{\vee}_{23}] \\
[\gamma_{23} \vert \gamma^{\vee}_{2A}] &
[\gamma_{23} \vert \gamma^{\vee}_{23}]
\end{pmatrix} \bigg|_{\text{\eqref{eq:solving-s1A-leading-2dim-homology}}}
&=\frac{\sin (\pi s_{1A})}{\sin(\pi s_{12})}
\begin{pmatrix}
2 i \sin(\pi(s_{1A}-s_{12}) )&
e^{i \pi (s_{1A}-s_{12})} \\
-e^{-i \pi (s_{1A}-s_{12})} &
0
\end{pmatrix} \bigg|_{\text{\eqref{eq:solving-s1A-leading-2dim-homology}}}
\notag{}
\\
\label{eq:intMatrix-2dim-leading}
&=
\frac{\sin (\pi s_{12} u_3)}{\sin(\pi s_{12} )}
\begin{pmatrix}
2 i \sin(\pi s_{12}(u_3-1) )&
e^{i \pi s_{12}(u_3-1)} \\
-e^{-i \pi s_{12}(u_3-1)} &
0
\end{pmatrix}\, ,
\end{align}
where we have fixed $z_2=0$, used momentum conservation, and introduced co-moving coordinates $z_3 = v_3 + u_3 \tau$ and $v_3,u_3\in\mathbb{R}$ (note that only $u_3$ appears at this order). 
Now, what used to be a matrix that only depended on the exponents $s_\bullet$ depends on the punctures!

We also need the residue (in $\eta$) of the Riemann-Wirtinger integrals, under the condition \eqref{eq:solving-s1A-leading-2dim-homology}. Interestingly, this lands precisely on the Riemann-Wirtinger integrals with $\eta=0$:
\begin{align}
 [ \gamma_{j} \vert \vphi_a\ra \bigg|\bigg|_{\eta^{-1}} \bigg|_{\text{\eqref{eq:solving-s1A-leading-2dim-homology}}}
 &= \int_{\gamma_{j}}e^{2 \pi i s_{1A} z_1}\ u(z_1)\ \d z_1 \bigg|_{\text{\eqref{eq:solving-s1A-leading-2dim-homology}}}
 \notag
 \\
 &=\int_{\gamma_{j}
 }
 e^{2 \pi i s_{12} u_3 z_1} \bigg[\frac{\vth(z_1|\tau)}{\vth(z_1-z_3|\tau)}\bigg]^{s_{12}} \d z_1 
 =
 [ \gamma_{j }\vert \zeta_1\ra \bigg|_{\text{\eqref{eq:solving-s1A-leading-2dim-homology}}}
\end{align}
where $f(z,\eta)\big|\big|_{\eta^{-1}}$ denotes the coefficient of $\eta^{-1}$ of the Laurent expansion of $f(z,\eta)$. Note, that the twisted cycle $[\gamma_j|$ changes with \eqref{eq:solving-s1A-leading-2dim-homology}\footnote{Here, $[\gamma_j|$ changes in the sense that the $\LSdual$-valued coefficients evaluated with the replacement \eqref{eq:solving-s1A-leading-2dim-homology}, but the topological cycle remains the same.}
and 
\begin{align}
|\varphi_a \rangle \bigg|\bigg|_{\eta^{-1}} = F(z-z_a,\eta|\tau)\ \d z_1 \bigg|\bigg|_{\eta^{-1}} = \d z_1 = |\zeta_1\rangle \, ,
\end{align}
which are twisted 1-forms of different local system (see appendix \ref{sec:RWco-app-0}).

Putting everything together, according to \eqref{eq:complex-RW-integral-klt} and \eqref{eq:def-modular-complex-RW} the leading term of the modular double copy is given by:
\begin{align}
N^{(0,0) }_{ab}&=\exp\left( 2 \pi s_{12} \frac{(\im z_3)^2}{\im \tau}\right)
\frac{\sin (\pi s_{12})}{\sin(\pi s_{12} u_3)} 
\frac{i}{2}
\notag
\\
&\phantom{=}\times
\begin{pmatrix}
 [\gamma_{2A}| \zeta_1\rangle
 &
 [\gamma_{23}| \zeta_1\rangle
\end{pmatrix}
\cdot
\begin{pmatrix}
 0 & e^{- i \pi s_{12}(u_3-1)}
 \\
 -e^{ i \pi s_{12}(u_3-1)} & 2 i \sin(\pi s_{12}(u_3-1))
\end{pmatrix}
\cdot
\begin{pmatrix}
 \overline{[\gamma_{2A}| \zeta_1\rangle}
 \\
 \overline{[\gamma_{23}| \zeta_1\rangle}
\end{pmatrix}
\notag
\\
&=
\exp( 2 \pi s_{12} u^2_3 \im \tau)
\frac{\sin (\pi s_{12})}{\sin(\pi s_{12} u_3)} 
\bigg[\im \big(e^{i \pi s_{12}(u_3-1)} [\gamma_{23}|\zeta_1\rangle \overline{[\gamma_{2A}|\zeta_1 \rangle} \big)
\notag
\\
&\phantom{=\bigg[}
-\sin(\pi s_{12} (u_3-1)) [\gamma_{23}|\zeta_1\rangle \overline{[\gamma_{23}|\zeta_1\rangle}
\bigg] \, .
\label{eq:simplest-modular-double copy}
\end{align}
Note that in \eqref{eq:simplest-modular-double copy} we have used momentum conservation, the comoving coordinate $u_3=\frac{\im z_3}{\im \tau}$, fixed $z_2=0$, and every instance of $[\gamma_{23}|\zeta_1\rangle$ should be read as $[\gamma_{23}|\zeta_1\rangle \big|_{\eqref{eq:solving-s1A-leading-2dim-homology}}$. In the RHS of \eqref{eq:simplest-modular-double copy} the reality of the integral $N^{(0,0)}_{ab}$ is manifest. Also note that the dependence on the punctures $z_a$ and $z_b$ drops out at leading order in $(\eta,\bar{\eta})$.
This equation (without the first exponential factor) was first written by Ghazouani and Pirio in the context of Veech volumes
\cite{ghazouani2016moduli}. 
Formulas for the subleading contributions $N^{(1,0)}_{ab}$ and $N^{(1,1)}_{ab}$ are provided in appendix \ref{sec:modular-subleading}.

We remark that the improved complex integrals $N_{kl}$ in \eqref{eq:def-modular-complex-RW} coincide with generating series of eMGFs, when one imposes momentum conservation on the latter, and rescales the $\eta$ in  $\Omega(z_1-z_j,\eta,\vert \tau)$ in the integrand of $N_{kl}$; see e.g. Equation (4.1) of \cite{DHoker:2020hlp}. 
The $\alpha'$-expansion of \eqref{eq:simplest-modular-double copy} generates double copy representations of eMGFs by sums of products of meromorphic quantities and complex conjugates. Examples of such representation can be found in  \cite{Broedel:2019tlz} at depth one in terms of eMPLs and \cite{Hidding:2022vjf} in terms of iterated $\tau$-integrals. 
% We expect the double copy of Riemann-Wirtinger integrals to be in the image of the single-valued map along the lines of \cite{ Brown:2018omk}.

%%%%%%%%%%%%%%%%%%%%%%%%%%%%%%%%%%%%%%%%%%%%%%%%%
%%%%%%%%%%%%%%%%%%%%%%%%%%%%%%%%%%%%%%%%%%%%%%%%%
%%%%%%%%%%%%%%%%%%%%%%%%%%%%%%%%%%%%%%%%%%%%%%%%%
\section{Towards integrating two punctures 
\label{sec:doubleint}}

In this section, we speculate on how to generalize the Riemann-Writinger integrals to the case where two punctures are integrated. 
Understanding this is an important step towards treating real one-loop string integrals within the formalism of twisted (co)homology.

To make the discussion concrete consider the following natural generalization of the Riemann-Wirtinger integral
\begin{align}
\label{eq:def-z1z2-RW-try}
    \int \d z_1\ \int \d z_2\ 
    u\ F(z_{1a},\eta_1) F(z_{2b},\eta_2) 
    \,,
\end{align}
where 
\begin{align}
\label{eq:def-integrand-uofz1z2}
    u = e^{2\pi i (s_{1A} z_1 + s_{2A} z_2)}
    \prod_{j=2}^n \vth^{s_{1j}}(z_{1j})
    \prod_{k=3}^n \vth^{s_{2k}}(z_{2k})
    \,,
\end{align}
is the $z_1$- and $z_2$-dependent parts of the $n$-point Koba-Nielsen. 
To keep the discussion simple, we assume that $a,b\geq3$ and $a\neq b$.
This twist has the following $A$- and $B$-cycle monodromies 
\begin{align}
\label{eq:def-monodromies-of-twist-z1z2}
    \frac{u_{z_k\to z_k+1}}{u}
    = e^{2\pi i s_{kA}}
    \,,
    \qquad
    \frac{u_{z_k\to z_k+\tau}}{u}
    = e^{2 \pi i ( s_{kA} \tau + \sum_{j\neq k}^n s_{1j} z_j )}
    \qquad
    \text{for}
    \qquad k = 1,2
    \,.
\end{align}
A natural guess for the corresponding local system is the tensor product of the local system considered in the previous sections of this paper
\begin{align}
    \L_2 (s_{1A},s_{1i\neq1},s_{1B}, s_{2A},s_{2i\neq2},s_{2B})
    \sim 
    \L_{\omega_1,\eta_1}(s_{1A},s_{1i\neq1},s_{1B})
    \otimes 
    \L_{\omega_2,\eta_2}(s_{2A},s_{2i\neq2},s_{2B})
    \,,
\end{align}
where 
\begin{align} \label{eq:eta1}
    \eta_1 &= s_{1A} \tau
    + s_{12} z_2
    + \sum_{j = 3}^n s_{1j} z_j 
    - s_{1B}
    \,,
    \\ \label{eq:eta2}
    \eta_2 &= s_{2A} \tau
    + s_{12} z_1
    + \sum_{j=3}^n s_{1j} z_j 
    - s_{2B}
    \,.
\end{align}

To see if this is sensible, we compute the monodromy of the $z_1$ $A$-cycle
\begin{align}
        u\ F(z_{1a},\eta_1) F(z_{2b},\eta_2)
        \vert_{z_1\to z_1 + 1}
    &=
        [u_1\ F(z_{1a},\eta_1) ]
        \vert_{z_1\to z_1 + 1}
        \times
        [\widehat{u}_1\ F(z_{2b},\eta_2)]
        \vert_{z_1\to z_1 + 1}
\end{align}
where $u_1$ is the $z_1$-dependent part of the twist and $\widehat{u}_1$ is $z_1$-independent part of the twist. 
We know what happens to the first term -- it is an overall phase  $e^{2 \pi i s_{1A}}$.
In the second term, substituting $z_1 \to z_1 +1$ into \eqref{eq:eta2} leads to the shift $\eta_2\to \eta_2 + s_{12}$ indicating a change in the second local system in the tensor product $\L_2$
\begin{align}
        u\ F(z_{1a},\eta_1) F(z_{2b},\eta_2)
        \vert_{z_1\to z_1 + 1}
    &=
        [u_1\ F(z_{1a},\eta_1)
        e^{2 \pi i s_{1A}}]
        \times 
        [\widehat{u}_1\ F(z_{2b},\eta_2+s_{12})]
    \nn\\
    &= e^{2 \pi i s_{1A}}
    \left[ 
        u\ F(z_{1a},\eta_1)
        F(z_{2b},\eta_2)
    \right]_{s_{2B} \to s_{2B}^\prime}
\end{align}
where $s_{2B}^\prime = s_{2B} - s_{12}$.
So in the end, we get the phase factor $e^{2 \pi i s_{1A}}$ and a shift of the local system 
\begin{align}
    \L_2 \to \L_2^\prime 
    = \L_2(s_{1A},s_{1i\neq1},s_{1B}, s_{2A},s_{2i\neq2},s_{2B}-s_{12})
    \,.
\end{align}
There is an analogous change of local system for the $z_1$ $B$-cycle. The only difference is that we get phase factor $e^{2\pi i s_{2B}}$ and move to a local system with $s_{2A}^\prime = s_{2A} + s_{12}$. 
Clearly, analogous statements hold for the $z_2$ $A$- and $B$-cycles. 
\begin{figure}[h!]
    \centering
    \includegraphics[scale=0.6]{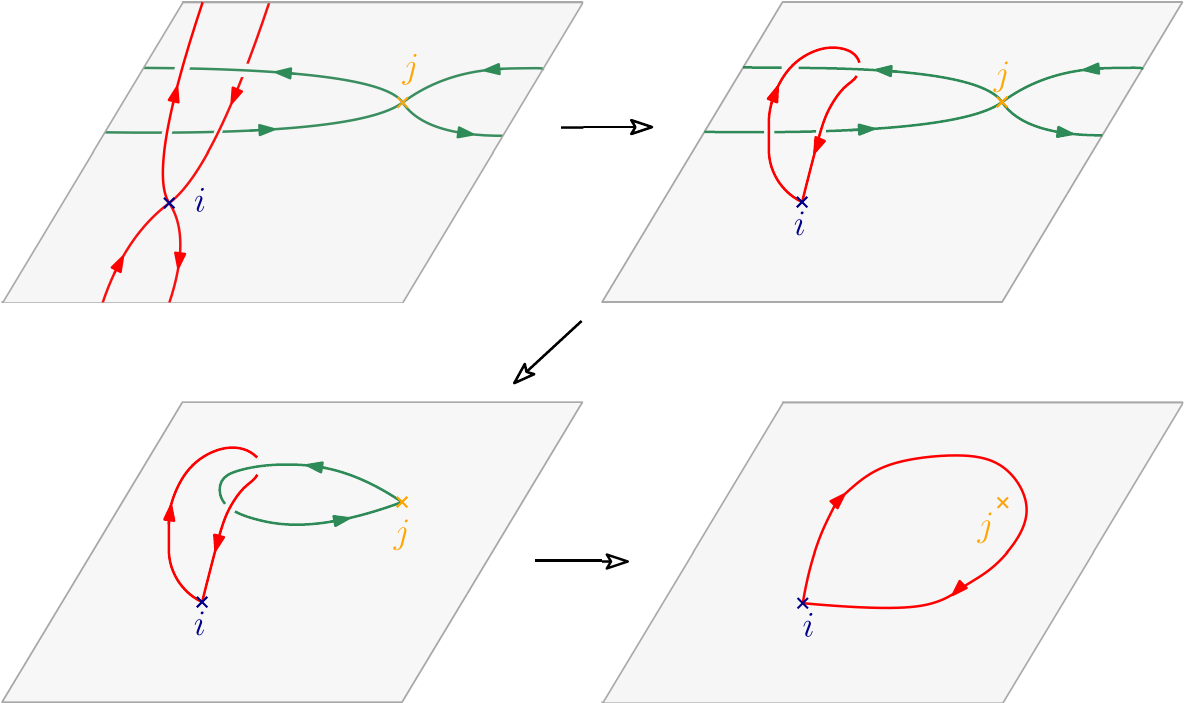}
    \caption{Pictorial proof of the identity $B_i A^{-1}_j B^{-1}_i A_j = M_{ij}$ on the fundamental group of $n\geq3$ punctures on the torus (topologically: a paralellogram with opposite sides identified). The black arrows denote homotopies. The first picture shows, in red, $B$-cycle monodromies for the puncture $z_i$ and, in green, $A$-cycle monodromies for the puncture $z_j$. Our conventions for multiplication are that the product $AB$ is first doing $A$ and then doing $B$, and if they intersect we would write $B$ underneath $A$. Note that the other punctures are not pictured.}
    \label{fig:Monodromy_Rel_braid}
\end{figure}
In fact, one can argue topologically for the need of more structure (e.g. ``shifts of local systems''). 
Recall that the local system is a representation of the fundamental group of the underlying manifold. 
In most situations, the local system that enters the twisted (co)homology is of rank 1 and comes from the \emph{constant} multiplicative monodromies of the twist.
Next, consider the twist $u=u(z_1,z_2;s_{1A},s_{2A},s_{12})$ as defined in \eqref{eq:def-integrand-uofz1z2} but where we have highlighted the dependence of $u$ on $z_1,z_2;s_{1A},s_{2A},s_{12}$. 
More abstractly, the monodromies \eqref{eq:def-monodromies-of-twist-z1z2} arise from the action of the monodromy operators $A^{\pm1}_j$, $B^{\pm1}_j$ that
perform the displacements of a function along the $A$-cycle or $B$-cycle of the corresponding puncture $z_j$:
\begin{align}
    A^{\pm 1}_j : f(z_j) \mapsto f(z_j\pm 1) 
    \qquad \text{and} \qquad
    B^{\pm 1}_j : f(z_j) \mapsto f(z_j \pm \tau) \, .
\end{align}
Summarizing, the monodromies of the twist $u(z_1,z_2;s_{1A},s_{2A},s_{12})$ are:
\begin{align}
A^{\pm 1}_j :&\ u(z_1,z_2;s_{1A},s_{2A},s_{12}) \mapsto e^{\pm 2 \pi i s_{jA}}u(z_1,z_2;s_{1A},s_{2A},s_{12})
\; \;
j=1,2
\\
B^{\pm 1}_1 :&\ u(z_1,z_2;s_{1A},s_{2A},s_{12}) \mapsto e^{\pm 2 \pi i \tilde{s}_{1B}}u(z_1,z_2;s_{1A},s_{2A}\pm s_{12},s_{12})  \, ,
\\
B^{\pm 1}_2 :&\ u(z_1,z_2;s_{1A},s_{2A},s_{12}) \mapsto e^{\pm 2 \pi i \tilde{s}_{2B}}u(z_1,z_2;s_{1A}\pm s_{12},s_{2A},s_{12})  \,  \, 
\end{align}
where the factors $e^{\pm 2 \pi i \tilde{s}_{jB}}$ are independent of $z_1$ and $z_2$. 
Explicitly, 
\begin{align}
\tilde{s}_{1B} &= s_{1A}\tau +\sum^{n}_{j\neq 1,2} s_{1j}z_j  
\\
\tilde{s}_{2B} &= s_{2A}\tau +\sum^{n}_{j\neq 1,2} s_{2j}z_j  \, , 
\end{align}
and are not to be confused with the $s_{kB}$.
Since these phases are independent of the integration variables, they can be
pulled outside of the integral \eqref{eq:def-z1z2-RW-try}. We comment that the shift in $s_{1A}$ in $s_{12}$ caused by the monodromy $B_2$ above physically corresponds to the shift in loop-momentum of $\ell \to \ell + k_2$, in massless kinematics.

Now, in choosing the factors of the twist $u(z_1,z_2;s_{1A},s_{2A},s_{12})$ above, the action of the monodromy that takes the puncture $z_1$ clockwise around $z_2$, $M_{12}$ on the twist is $e^{-2 \pi i s_{12}}$. 
It turns out that the generator $M_{12}$ is not independent from the generators $A_j$, $B_j$.
% in the fundamental group of a punctured torus,
Instead, these generators actually satisfy the identity\footnote{See corollary 5.1 and figure 2a of \cite{birman1969braid}. We identify the $A_i$ here with $\rho_i$ there, and $B_j$ here with $\tau_i$ there. To be precise, different combinations give different ``'kinds'' of $M_{ij}$: going over or under the other fixed punctures.}:
\begin{align}
M_{12}=A^{-1}_1 B_2 A_1 B^{-1}_2 \, .
\end{align}
We showcase an identity of this type in figure \ref{fig:Monodromy_Rel_braid}.
This can be explicitly verified by acting with the RHS of the above identity on the twist
% We can see an incarnation of this relation in the twist $u(z_1,z_2;s_{1A},s_{1B},s_{12}):$
\begin{align}
A^{-1}_1 \circ B_2 \circ A_1 \circ B^{-1}_2  : u(z_1,z_2;s_{1A},s_{1B},s_{12}) \mapsto e^{-2 \pi i s_{12} }u(z_1,z_2;s_{1A},s_{1B},s_{12}) \, , 
\end{align}
which aligns with the topological expectation.

The crucial point above is the following: if it is possible to construct a 1-dimensional representation (or rank-1 local system) of the fundamental group of the configuration space of the punctured torus, any such representation $\rho$ would be commutative and satisfy 
\begin{align}
\rho (A^{-1}_1 B_2 A_1 B^{-1}_1 ) = 1\, .
\end{align}
This, in turn, means that the $e^{2 \pi i s_{12}}=1$. 
That is, $s_{12} \in \mathbb{Z}$ instead of being generic $s_{12}$. 
This is potentially an interesting starting point for studying Riemann-Wirtinger integrals with two integrations. Otherwise, perhaps one should work with higher-rank local systems.

While the shifting of local systems is uncomfortable at first, it is actually known how to map a basis in one local system to a basis in the other using the monodromy relations of \cite{Goto2022}.
This perspective seems promising and could potentially be an important step towards treating actual one-loop string integrals! 
We leave the investigation of such integrals and their potential double copy to future work. 

A similar difficulty arises when considering the following, even simpler integral:
\begin{align}
\label{eq:simplest-RW-def}
\int_{\gamma} \vth^{s_{12}}(z_1\vert \tau)\ e^{2 \pi i s_{1A} z_1}\ F(z_1,\eta \vert \tau)\ \d z_1 \, ,
\end{align}
which looks like a Riemann-Wirtinger integral with $n=2$ punctures. 
However, we need to relax the momentum conservation condition so that $s_{12}$ is non-zero in order to get a non-trivial integral.
Then the following equation becomes relevant:
\begin{align}
\label{eq:problem-simplest-RW}
A_1 B_1 A^{-1}_1 B^{-1}_1 : [\vth(z_{12}\vert \tau)]^{s_{12}} \mapsto e^{- 2 \pi i s_{12}}  [\vth(z_{12}\vert \tau)]^{s_{12}}\, .
\end{align}
Thus, any rank-1 local system that describes the integral \eqref{eq:problem-simplest-RW} must have $s_{12}\in \mathbb{Z}-\{0\}$. 
\new{Interestingly, Stieberger \cite{Stieberger:2023nol} recently found two instances of a double copy formula that work for integrals of the form \eqref{eq:simplest-RW-def} for $s_{12}=1$ at generic complex structure ($\tau$).}
It would be interesting to see if the constructions therein extend for other $s_{12}\in \mathbb{Z}_{\geq 1}$.

%%%%%%%%%%%%%%%%%%%%%%%%%%%%%%%%%%%%%%%%%%%%%%%%%
%%%%%%%%%%%%%%%%%%%%%%%%%%%%%%%%%%%%%%%%%%%%%%%%%
%%%%%%%%%%%%%%%%%%%%%%%%%%%%%%%%%%%%%%%%%%%%%%%%%
\section{Future directions and discussion \label{sec:conclusion}}

Riemann-Wirtinger integrals serve as a simple toy model for one-loop string integrands (see section \ref{sec:1L+RW}).
In particular, we have full mathematical control over their twisted (co)homology. 
This facilitates the computation and study of their differential equations, monodromy relations and, most importantly, leads to a double copy formula for a complex variant of the Riemann-Wirtinger integrals.

\new{In this work, we reviewed the construction of the twisted cohomology and homology of the Riemann-Wirtinger family of integrals in sections \ref{sec:RWco-eta} and \ref{sec:RWhom}. 
In particular, we provide several convenient bases of cohomology and homology. 
We then introduce the intersection number and intersection index -- inner products on cohomology and homology -- since they underpin our double copy construction. 
As a simple example for the utility of the intersection number and intersection index we verify distinguished linear relations between elements of (co)homology in the main text.
}

Then, in section \ref{sec:DC-RW}, we combine the ingredients of sections \ref{sec:RWco-eta} and \ref{sec:RWhom} to construct the double copy of complex Riemann-Wirtinger integrals.
We provide numerical checks of this double copy for real Mandelstams (section \ref{sec:real_s}) and complex Mandelstams (section \ref{sec:complex_s}). 
Then, in section \ref{sec:modular}, we introduce a single-valued and modular Riemann-Wirtinger integral for which we can verify the double copy order by order in the now free parameters $\eta$ and $\bar{\eta}$. 
In particular, this provides a double copy representation for some generating functions of eMGFs \cite{DHoker:2020hlp}. 
Finally, in section \ref{sec:doubleint}, we speculate on how the analysis of this work could be extended to Riemann-Wirtinger integrals where more than one puncture is integrated. 

While Riemann-Wirtinger integrals capture many of the gross features of one-loop string integrands, the constraint \eqref{eq:eta-def} entangling the $\eta$-variable with the moduli is difficult to interpret.
In alternative approaches in mathematics and string theory \cite{Felder1995, Mafra2019}, $\eta$ is an independent parameter organizing different integrands of one-loop string amplitudes into a generating series. 
However, the formalism of twisted (co)homology seems to require the constraint \eqref{eq:eta-def} hinting that perhaps a more robust theory of (co)homology is needed to describe the usual generating series of one-loop string integrands. 
% {\color{red}Even if this is indeed the case, there are single-valued string integrands (four-point open and closed superstring amplitudes \cite{GREEN1982474}, $n$-point tachyonic amplitudes, and four-point open string amplitudes in heterotic string theory \cite{Gerken:2018jrq}) that lie inside the $\eta=0$ twisted (co)homology of the analogous Riemann-Wirtinger integrals.  
% In particular, these string integrands correspond either to the constant function or a single-valued combination of Kronecker-Eisenstein coefficients.}
Within the (co)homology framework at $\eta=0$, the double copy of the constant function, seen in four-point open- and closed- superstring amplitudes \cite{GREEN1982474} and $n$-point tachyonic amplitudes \cite{Gerken:2018jrq}, was tested both numerically and analytically in sections \ref{sec:real_s}, \ref{sec:complex_s} and \ref{sec:modular}. 
Moreover, we have also numerically tested the double copy of some single-valued combination of $\gk{k}$'s for $\eta=0$.
Given that the gap between Riemann-Wirtinger integrals and string amplitudes is easier to bridge at $\eta=0$, this case is a particularly promising starting point to deduce double copy formulae of string amplitudes from those of the Riemann-Wirtinger integral studied in this work.

Perhaps the most direct connection to the usual string integral picture is the modular double copy of section \ref{sec:modular} and its relation to generating functions of eMGFs. 
Here, Riemann-Wirtinger integrals are augmented with additional factors independent of the integration variable, converting the  Kronecker-Eisenstein functions in the integrand into their doubly periodic completion. 
The resulting double copy corresponds to certain generating functions of eMGFs. 
It would be interesting to find an expansion scheme of Riemann-Wirtinger integrals that generates expansions for eMGFs in terms of eMPLs and their complex conjugates. Additionally, the form of the modular double copy in \ref{sec:modular} is interesting because we have solved \eqref{eq:eta-def} for the variable  $s_{1A}$ corresponding to the loop momentum in chiral splitting. 
Thus, to give a string-integral interpretation for this, the closed-string integral \textit{after loop-momentum integration} can be written as a double copy of open-string integrals \textit{where the loop momentum, $s_{1A}$, has a specific value}.

Of course, it would also be interesting to extend the analysis here to the case where more than one puncture is integrated as discussed in section \ref{sec:doubleint}. 
In particular, the $(n-1)$-dimensional Riemann-Wirtinger integral should be a straightforward generalization of the two-dimensional Riemann-Wirtinger integral. 
This generalization would be a huge step towards understanding string integrands in the formalism of twisted (co)homology: placing homology -- i.e. monodromy relations and counting numbers of independent open-string amplitudes \cite{Tourkine:2016bak,Hohenegger:2017kqy,Ochirov:2017jby,Casali:2019ihm} -- and cohomology -- i.e. finding bases of integrands for differential equations \cite{Felder1995,Mafra2019,Broedel:2019gba,Broedel2020,Kaderli:2022qeu} -- on the same ground. This is particularly interesting because in the string-theory setup, the homology counting is most readily done when the integrands are meromorphic (i.e. in the chiral splitting formalism), while the systematic counting of cohomology representatives is more readily done in the doubly-periodic setting~\cite{Rodriguez:2023qir}.

Generalizations of Riemann Wirtinger integrals to higher genus have been pioneered in \cite{watanabe2016twisted,watanabe2016twistedAbelian} with a discussion of cohomology bases. It would be interesting to extend the homology analysis of \cite{Mano2012,Goto2022,ghazouani2016moduli} and the double-copy formulae of this work to the higher-genus setting of \cite{watanabe2016twisted,watanabe2016twistedAbelian}. Twisted (co)homology methods and double-copy approaches at higher genus will have a wealth of implications of string amplitudes, Feynman integrals and mathematics.

%%%
%%%

Finally, it will be  rewarding  to formulate the genus-one KLT relations of Stieberger \cite{Stieberger:2022lss,Stieberger:2023nol} in the language of twisted (co)homology investigate alternative derivations from intersection indices. A possible starting point for this would be to give a twisted-(co)homology understanding of the relations for mixed open- and closed-string amplitudes at genus zero  \cite{Stieberger:2009hq,Stieberger:2015vya} and at genus one \cite{Stieberger:2021daa}, which play a key role in the KLT relations of \cite{Stieberger:2022lss,Stieberger:2023nol}. 
Since the genus-one KLT relations therein work for multiple integrated punctures, these works must process key insights for Riemann-Wirtinger integral with multiple integrated punctures.

\new{It would also be interesting to obtain analytic expressions for RW integrals in terms of eMPLs. 
This would facilitate an analytic double copy and further our understanding of single-valued eMPLs.
While a complete story is missing, progress towards this goal has been presented in appendix \ref{sec:RWco-app}. 
Here, we derive the differential equations satisfied by Riemann-Wirtinger integrals for $\eta \neq 0$ (section \ref{sec:RWco-app-deqs}) and $\eta=0$ (appendix \ref{sec:RWco-app-0}).  
In particular, we verify the Gauss-Manin connection of \cite{Mano2012} ($\eta\neq0$) and \cite{ghazouani2016moduli} ($\eta=0$ and $n=3$) using independent methods.
In section \ref{sec:RWco-app-alphaprime}, we solve the $\eta\neq0$ differential equations to $\mathcal{O}(\alpha')$ explicitly in terms of elliptic polylogarithms. 
Boundary values for the bases of twisted contours presented in section \ref{sec:RWhom-basis} are also provided in section \ref{sec:RWco-app-bdval}.
In particular, the form and solution of the $\eta\neq0$ DEQ is strikingly  similar to that of actual string integrals \cite{Kaderli:2022qeu}.
The $\eta=0$ differential equations and basis are in some sense more complicated due to the presence of unavoidable doubles poles. 
However, while unfamiliar to most iterated integral practitioners, it is possible to make sense of iterated integrals whose kernels have higher order poles \cite{enriquez2023analogues, enriquez2023elliptic} and is in some sense expected at higher genus. 
}
% After providing two distinguished bases of twisted differential forms in section \ref{sec:RWco-basis}, we use the intersection number --- an inner product on cohomology --- to derive the differential equations satisfied by these integrals for $\eta \neq 0$ in section \ref{sec:RWco-app-alphaprime} and for $\eta=0$ in section \ref{sec:RWco-0}.  
% In particular, we verify the Gauss-Manin connection of \cite{Mano2012} ($\eta\neq0$) and \cite{ghazouani2016moduli} ($\eta=0$ and $n=3$) using independent methods.
% In section \ref{sec:alphaprime}, we solve the $\eta\neq0$ differential equations to $\mathcal{O}(\alpha')$ explicitly in terms of elliptic polylogarithms. 
% The $\eta=0$ differential equations and basis are in some sense more complicated due to the presence of unavoidable doubles poles. 
% While unfamiliar to most iterated integral practitioners, it is possible to make sense of iterated integrals whose kernels have higher order poles \cite{enriquez2023analogues, enriquez2023elliptic} and is in some sense expected at higher genus. 

% In section \ref{sec:RWhom}, we construct the twisted homology of Riemann-Wirtinger integrals. 
% The intersection indices necessary for the double copy are computed in section \ref{sec:RWhom-int} and boundary values for the differential equation of section \ref{sec:RWco-app-alphaprime} are provided in section \ref{sec:RWco-app-bdval}.

%%%%%%%%%%%%%%%%%%%%%%%%%%%%%%%%%%%%%%%%%%%%%%%%%
%%%%%%%%%%%%%%%%%%%%%%%%%%%%%%%%%%%%%%%%%%%%%%%%%
%%%%%%%%%%%%%%%%%%%%%%%%%%%%%%%%%%%%%%%%%%%%%%%%%
\acknowledgments 

The authors would like to thank Oliver Schlotterer for many interesting discussions and participation in the early stages of this work. 
The authors would also like to thank 
Francis Brown, Eduardo Casali,
Lorenz Eberhardt, Yoshiaki Goto, 
Saiei-Jaeyeong Matsubara-Heo, Sebastian Mizera,
Franziska Porkert, 
Giulio Salvatori, Marcus Spradlin, Piotr Tourkine, 
Anastasia Volovich, and Federico Zerbini  
for useful discussions. 
AP would like to thank the institute for advanced study for its hospitality and the organizers of the \emph{String Amplitudes at Finite $\alpha^\prime$} workshop where this work was initiated. 
This work was supported in part by the US Department of Energy under contract DESC0010010 Task F (RB, AP, LR) and by Galkin Foundation Fellowship (LR).
The research of CR is supported by the European Research Council under ERC-STG-804286 UNISCAMP.

%%%%%%%%%%%%%%%%%%%%%%%%%%%%%%%%%%%%%%%%%%%%%%%%%
%%%%%%%%%%%%%%%%%%%%%%%%%%%%%%%%%%%%%%%%%%%%%%%%%
%%%%%%%%%%%%%%%%%%%%%%%%%%%%%%%%%%%%%%%%%%%%%%%%%
\appendix

%%%%%%%%%%%%%%%%%%%%%%%%%%%%%%%%%%%%%%%%%%%%%%%%%
%%%%%%%%%%%%%%%%%%%%%%%%%%%%%%%%%%%%%%%%%%%%%%%%%
%%%%%%%%%%%%%%%%%%%%%%%%%%%%%%%%%%%%%%%%%%%%%%%%%
\section{Local systems \label{app:locSys}}

In this appendix, we give a perhaps more geometric description of a local system.
This is included for completeness only and the interested reader can find more details in  \cite{maat-thesis}.

First, we define the monodromy of $\omega$ on $\gamma$ where $\omega$ is a closed 1-form on $M$ and $[\gamma] \in \pi_1(M)$ is a loop in the first homotopy group.
Fixing $\omega$ provides a map $\pi_1 \to \mathbb{C}$ by $[\gamma] \mapsto \int_\gamma \omega$. In fact, this is a homomorphism between groups. 
Our goal is to use this homomorphism to define a representation of the fundamental group.
To get a representation of the fundamental group, we compose the above homomorphism with the exponential map $\text{exp}: \mathbb{C} \to \mathbb{C}^* = \text{GL}(1,\mathbb{C})$ by $[\gamma] \mapsto \exp \int_\gamma \omega$.
Next, we use the above to define a complex line bundle known as a local system. 

Next, let $\tilde{M}$ be the universal cover of $M$. 
By the definition of a universal cover, the fundamental group $\pi_1(M)$ has a free action on $\tilde{M}$ (all stabalizers of $\pi_1$ are trivial) such that the quotient of the action is $M$. 
This means that $\tilde{M}$ is a principal $\pi_1(M)$-bundle over $M$. 
Since we have a principal $\pi_1(M)$-bundle and a representation $\pi_1 \to \mathbb{C}^*$, we can construct the associated vector bundle $(\tilde{M}\times \mathbb{C})/\pi_1(M) \to M$.
In this example, $\mathbb{C}^* = \text{GL}(1;\mathbb{C})$ is the group of $1\times1$ matrices so the associated vector bundle is a line bundle. 
This line bundle $\mathcal{L}_\omega = (\tilde{M}\times \mathbb{C})/\pi_1(M) \to M$ is called the \emph{local system}. 
Note that subscript emphasises that the construction depends on the choice of $\omega$.

While this line bundle is actually trivial, there are new natural smooth structures, such as flat connections, that can be defined. 
In particular, there exists a unique flat connection $\nabla_\omega = \d + \d\log(u)$ on $\mathcal{L}_\omega$ that pulls back to $\d$ on the trivial line bundle $\tilde{M}\times\mathbb{C}$. 
The fact that $\mathcal{L}_\omega$ is trivial means that there exists a global trivialization and implies that we can work with complex valued forms rather than $\mathcal{L}_\omega$-valued forms. 
This only works for cohomology and we must use $\mathcal{L}_\omega$-valued cycles for homology.

%%%%%%%%%%%%%%%%%%%%%%%%%%%%%%%%%%%%%%%%%%%%%%%%%
%%%%%%%%%%%%%%%%%%%%%%%%%%%%%%%%%%%%%%%%%%%%%%%%%
%%%%%%%%%%%%%%%%%%%%%%%%%%%%%%%%%%%%%%%%%%%%%%%%%
\section{Computing intersection indices \label{app:intInd}}

We will now demonstrate with the aid of a few simple examples, the calculation of the intersection indices mentioned in section \ref{sec:RWhom-int}, for a more detailed derivation of \eqref{eq:all_homology_intersection_number} the reader is advised to revisit the original works on this topic \cite{Goto2022,ghazouani2016moduli,ghazouani2017moduli}. Note that for the rest of this appendix, in the figures, the twisted cycles will be denoted by dark blue and the dual will be in green, consistent with the convention stated in section \ref{sec:RWhom-int} for topological intersection indices. Moreover, the intersection points will be colored red and the branch cut $B$ on the torus will be denoted by a red dashed line.

\begin{figure}[H]
    \centering
    \includegraphics[width=80mm]{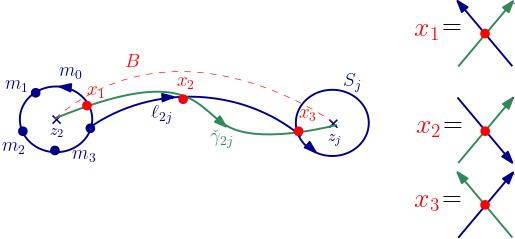}
    \caption{ Self intersection of the $\gamma_{2j}$ cycles. The twisted-cycle intersection is relevant for the $[\gamma_{2j}|\c{\gamma}_{2j}]$ intersection index. The intersection occurs at three points $(x_1,x_2,x_3)$ as depicted above in red. The topological intersection indices at these points are given on the right for reference. The branch cut $B$ is denoted as a  red dashed curve starting at $z_2$ and ending at $z_j$. }
    \label{fig:1j1j_component}
\end{figure}

As a simple example of calculating an intersection index, we start by evaluating the intersection index of $\gamma_{2j}$ with itself, that is $[\gamma_{2j}|\c{\gamma}_{2j}]$. As seen from figure \ref{fig:1j1j_component}, the twisted cycle intersects its dual exactly at three points $(x_1,x_2,x_3)$. Further note that, the way one defines the intersection of the two cycles, is such that in doing so the dual twisted cycle does not cross the branch cut $B$. Using the conventions defined in section \ref{sec:RWhom} for the topological intersection indices and referring to figure \ref{fig:1j1j_component} we can easily read these off at the three points as $\lbrack m_0|\ell_{2j}\rbrack^{\text{top}}_{x_1}=-1$, $\lbrack \ell_{2j}|\ell_{2j}\rbrack^{\text{top}}_{x_2}=1$ and $\lbrack S_{j}|\ell_{2j}\rbrack^{\text{top}}_{x_3}=1$. Then using the master formula for the intersection index of twisted cycles and their duals \eqref{eq:Hint2}, and the expressions for regularised cycle $\text{Reg}[\gamma_{2j}]$, we derive the desired intersection index as follows:
\begin{align}
    [\gamma_{2j}|\c{\gamma}_{2j}] &= \lbrack m_0|\ell_{2j}\rbrack^{\text{top}}_{x_1}\frac{1}{e^{2\pi i s_{12}}-1} +\lbrack \ell_{2j}|\c{\ell}_{2j}\rbrack^{\text{top}}_{x_2}+\lbrack S_{j}|\c{\ell}_{2j}\rbrack^{\text{top}}_{x_3}\frac{1}{e^{2\pi i s_{1j}}-1} \label{eq:1j1j_component}\\
    &= \frac{1-e^{2\pi i(s_{12}+s_{1j})}}{(1-e^{2\pi is_{12}})(1-e^{2\pi is_{1j}})}\nonumber,
\end{align}
which agrees with \eqref{eq:all_homology_intersection_number}.
 Notice in the last example, we did not pick up any phase due to the discontinuity around the branch cut $B$, this is because we chose to intersect the twisted cycles in such a way that we avoid crossing the branch cut. However, we could have very well considered self-intersecting the twisted cycle such that one ends up with fewer intersection points, but the last term in \eqref{eq:1j1j_component} picks up an overall phase due to the monodromy for the Koba-Nielsen twist \eqref{eq:KN} around $B$. The next example that we are going to consider will demonstrate how to account for the monodromy properties of the twist while performing such calculations. 
 \begin{figure}[H]
    \centering 
    \includegraphics[scale=0.4]{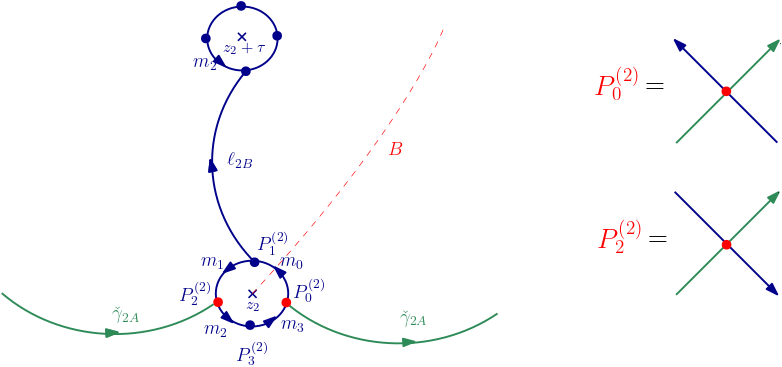}
    \caption{ Intersection of the $A$-cycle and $B$-cycle. The twisted-cycle intersection is relevant for the $[\gamma_{2B}|\c{\gamma}_{2A}]$ intersection index. The intersection occurs at two points $(P^{(2)}_0,P^{(2)}_1)$ as depicted above in red.}
    \label{fig:101infty_component}
\end{figure}

 Let us now evaluate the $[\gamma_{2B}|\c{\gamma}_{2A}]$ intersection index. Figure \ref{fig:101infty_component} describes the topological intersection of the two twisted cycles, in particular we see that the twisted cycles intersect exactly at two points $(P^{(2)}_0,P^{(2)}_2)$. Any point on the line segment $l_{2A}$ that lies just outside the fundamental domain, is related to a point on the circular arc $m_2$ that lies just inside up to a phase $e^{2\pi i s_{1A}}$, corresponding to the $A$-cycle monodromy. As a result this time one picks up a phase at $P^{(2)}_2$, as for the twists in equation \eqref{eq:Hint2} we have
\begin{equation}
    u_{m_1}(P^{(2)}_2)u_{\c{\gamma}_{2A}}^{-1}(P^{(2)}_2) =  e^{2\pi i s_{1A} z}e^{-2\pi i s_{1A}(z+1)} =  e^{-2\pi i s_{1A}}
\end{equation}
note that in the above equation the Kronecker theta functions cancel each other out due to the $\text{SL}(2,\mathbb{Z})$ modular transformation property $\vth(z+1)=-\vth(z)$ and the momentum conservation for the Mandelstam variables $\sum_{k=1}^ns_{1k}=0$.
Then referring to figure \ref{fig:1j1j_component}, we have that the topological intersection indices at these points are $\lbrack m_0|\c{\ell}_{2A}\rbrack^{\text{top}}_{P^{(2)}_0}=-1$ and $\lbrack m_1|\c{\ell}_{2A}\rbrack^{\text{top}}_{P^{(2)}_2}=1$, thus again using equations \eqref{eq:reg_gammaA} and \eqref{eq:reg_gammaB} and plugging them into the formula \eqref{eq:Hint2} one gets 
\begin{align}
    [\gamma_{2B}|\c{\gamma}_{2A}] &= \lbrack m_0|\c{\ell}_{2A}\rbrack^{\text{top}}_{P^{(2)}_0}\frac{1-e^{-2\pi i s_{1B}}}{e^{2\pi i s_{12}}-1}+\lbrack m_1|\c{\ell}_{2A}\rbrack^{\text{top}}_{P^{(2)}_2}e^{-2\pi is_{1A}}\frac{1-e^{2\pi i (-s_{1B}+s_{12})}}{e^{2\pi i s_{12}}-1}\nonumber\\
    &= \frac{1-e^{-2\pi i s_{1B}}-e^{-2\pi i s_{1A}}+e^{2\pi i (s_{12}-s_{1B}-s_{1A})}}{1-e^{2\pi i s_{12}}}
\end{align}
in agreement with \eqref{eq:all_homology_intersection_number}.

\begin{figure}[H]
    \centering
    \includegraphics[width=65mm]{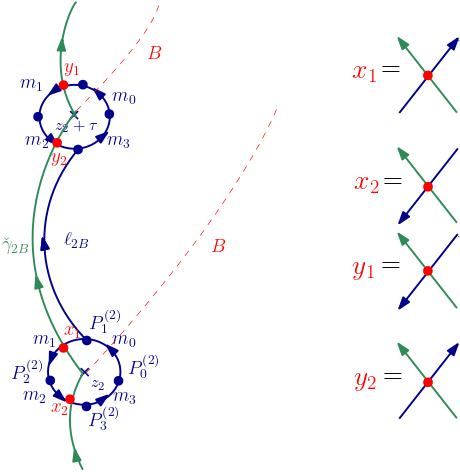}
    \caption{Self intersection of the $B$-cycles. The twisted-cycle intersection is relevant for the $[\gamma_{2A}|\c{\gamma}_{2B}]$ intersection index. The intersection occurs at four points $(x_1,x_2,y_1,y_2)$ as depicted above in red.}
    \label{fig:1infty1infty_component}
\end{figure}

We further explore this notion of picking up non-trivial phase factors due to different branch choices at a given simplex and its dual, with a highly non-trivial example of intersection index of the twisted $B$-cycle with itself i.e. $[\gamma_{2B}|\c{\gamma}_{2B}]$ (see figure \ref{fig:1infty1infty_component} for reference). Inferring from the figure \ref{fig:1infty1infty_component}, we see that the dual cycle intersects twisted cycle exactly at four points $(x_1,x_2,y_1,y_2)$. At each of these points, using the convention introduced earlier in this section, the topological intersection indices are 
\begin{equation}
    \lbrack m_1^i|\c{\ell}_{2B}\rbrack^{\text{top}}_{x_1} = -1 = \lbrack m_1^f|\c{\ell}_{2B}\rbrack^{\text{top}}_{y_1}~~~\text{and}~~~ \lbrack m_2^i|\c{\ell}_{2B}\rbrack^{\text{top}}_{x_2} = 1 = \lbrack m_2^f|\c{\ell}_{2B}\rbrack^{\text{top}}_{y_2}\label{eq: topological_intersections_for_ABcycles}
\end{equation}
where we have that $m^i$ refers to the initial arc circling $z_2$ and $m^f$ refers to the final arc circling $z_2+\tau$.
 Unlike before, there is now a need to distinguish between the two arcs $\{m_2^i,m_2^f\}$ on $\mathbb{C}/\Lambda_{\tau}$, the branch choice $u\cdot F$ for each of these simplices is different, as they lie on different domains of the torus. However, much like the previous example, the effect that this branch choice has is that it leads to non-trivial monodromy phase factors. The first non-trivial contribution comes from the intersection between the arc $m_2^i$ on the twisted cycle $\text{Reg}[\gamma_{2B}]$ and the dual cycle $\c{\gamma}_{2B}$ at $x_2$. This leads to
\begin{align}
    &u_{m_2^i}(x_2)F(x_2)(u_{\c{\gamma}_{2B}}(x_2)F(x_2))^{-1}=\nonumber\\
   &~~~~~ u(x_2;s_{1A},\{s_{1k}\})F(x_2;s_{1A},\{s_{1k}\})\left(u(x_2-\tau;-s_{1A},\{-s_{1k}\})F(x_2-\tau;-s_{1A},\{-s_{1k}\})\right)^{-1}\nonumber\\ 
    &~~~~~~~~~~~= e^{2\pi i s_{1B}}~,\label{eq: monodromy_at_x2}
\end{align}
where in the last line we used the $B$-cycle monodromy relation for the twist given by \eqref{eq:KN-Bmon} and the same for the Kronecker--Eisenstein generating function given by \eqref{eq:F-Bmon}. 
Another non-trivial contribution comes from the branch choice at the intersection between $m_2^{f} \subset \text{Reg}[\gamma_{2B}]$ and $\c{\gamma}_{2B}$ at $y_2$. Similar arguments hold as before, and one again picks up the same phase factor
\begin{align}
    &u_{m_2^f}(y_2)F(y_2)(u_{\c{\gamma}_{2B}}(y_2)F(y_2))^{-1}=\nonumber\\
   &~~~~~ u(y_2+\tau;s_{1A},\{s_{1k}\})F(y_2+\tau;s_{1A},\{s_{1k}\})\left(u(y_2;-s_{1A},\{-s_{1k}\})F(y_2;-s_{1A},\{-s_{1k}\})\right)^{-1}\nonumber\\ 
    &~~~~~~~~~~~= e^{2\pi i s_{1B}}~,\label{eq: monodromy_at_y1}
\end{align}
however, this time it is due to the monodromy around the
twisted cycle as opposed to the case at $x_2$, where one picks up a phase around the dual-twisted cycle.
Finally, using the expressions for the regularised cycles \eqref{eq:reg_gammaA} and \eqref{eq:reg_gammaB}, the topological intersection indices \eqref{eq: topological_intersections_for_ABcycles}, and the phases \eqref{eq: monodromy_at_x2} and \eqref{eq: monodromy_at_y1} one concludes after a few lines of algebra 
\begin{align}
    [\gamma_{2B}|\c{\gamma}_{2B}] &=  \lbrack m_1^i|\c{\ell}_{2B}\rbrack^{\text{top}}_{x_1}\frac{1}{e^{2\pi i s_{12}}-1}+ \lbrack m_2^i|\c{\ell}_{2B}\rbrack^{\text{top}}_{x_2}\frac{e^{2\pi is_{1B}}}{e^{2\pi i s_{12}}-1}\nonumber\\
    &~~~- \lbrack m_1^f|\c{\ell}_{2B}\rbrack^{\text{top}}_{y_1}\frac{e^{2\pi is_{12}}}{e^{2\pi i s_{12}}-1}e^{-2\pi is_{1B}}-\lbrack m_2^f|\c{\ell}_{2B}\rbrack^{\text{top}}_{y_2}\frac{e^{2\pi i s_{12}}}{e^{2\pi i s_{12}}-1}\nonumber\\
    &= -\frac{(e^{2\pi is_{1B}}-1)(e^{2\pi is_{12}}-e^{2\pi is_{1B}})}{e^{2\pi is_{1B}}(e^{2\pi is_{12}}-1)},
\end{align}
as required by \eqref{eq:all_homology_intersection_number}.

%%%%%%%%%%%%%%%%%%%%%%%%%%%%%%%%%%%%%%%%%%%%%%%%%
%%%%%%%%%%%%%%%%%%%%%%%%%%%%%%%%%%%%%%%%%%%%%%%%%
%%%%%%%%%%%%%%%%%%%%%%%%%%%%%%%%%%%%%%%%%%%%%%%%%
\section{Solving for $s_{1A}$}\label{app:solving-for-s1A}

In this section, we provide a derivation of \eqref{eq:def-modular-complex-RW}. 

We start by inserting \eqref{eq:solving_for_s1A} into  $e^{s_{1A}(z_1-\overline{z_1})}$:
\begin{align}
e^{2 \pi i s_{1A} (z_1-\overline{z_1})} 
&= 
e^{2 \pi i \frac{\im \eta}{\im \tau}(z_1-\overline{z_1})} \exp\bigg[-\frac{2 \pi i}{\im \tau}\big(\sum_{j=2}^{n} s_{1j} \im z_{j}\big)(z_1-\overline{z_1})\bigg]
\notag
\\
&= e^{2 \pi i \frac{\im z_1}{\im \tau} (\eta-\overline{\eta})} \exp \bigg[4 \pi \im z_1 \sum_{j=2}^{n} s_{1j} \im z_j\bigg] \, .
\label{eq:deriving-modular-double copy-s1A}
\end{align}
Now, let's look at the first factor in \eqref{eq:deriving-modular-double copy-s1A} with the Kronecker-Eisenstein series in the integrand of \eqref{eq:complex-RW-def}:
\begin{align}
&F(z_1-z_l,\eta|\tau) \overline{F(z_1-z_k,\eta|\tau)}
e^{2 \pi i \frac{\im z_1}{\im \tau} (\eta-\overline{\eta})} 
\notag
\\
&=
F(z_1-z_l,\eta|\tau)e^{2 \pi i \eta \frac  {\im (z_1 - z_l+z_l)}{\im \tau}}
\overline{F(z_1-z_k,\eta|\tau)}e^{-2 \pi i \overline{\eta} \frac  {\im (z_1 - z_k+z_k)}{\im \tau}}
\notag
\\
&=
\Omega(z_1-z_l,\eta|\tau) \overline{\Omega(z_1-z_k,\eta|\tau)} \exp\big[2 \pi i \eta \frac{\im z_l}{\im \tau} - 2 \pi i \overline{\eta} \frac{\im z_k}{\im \tau}\big]
\, ,
\label{eq:modular-rw-derivation-nonholo-Eisenstein}
\end{align}
where we have obtained the definition of the doubly-periodic non-holomorphic Eisenstein series.

Next, look at the second factor in \eqref{eq:deriving-modular-double copy-s1A} and the theta functions inside the complex Riemann-Wirtinger integral \eqref{eq:complex-RW-def}:
\begin{align}
&\bigg[\prod_{j=2}^{n} |\vth(z_1-z_j|\tau)|^{2 s_{1j}} \bigg] \exp\bigg[ 4 \pi \frac{\im z_1}{\im \tau}\sum_{j=2}^n s_{1j} \im z_j\bigg] 
\notag
\\
&=\exp \bigg[ \sum_{j=2}^{n} s_{1j} \log |\vth(z_1-z_j|\tau)|^2\bigg] 
\notag
\exp \bigg\{
2\pi \sum_{j=2}^{n}\frac{s_{1j}}{\im \tau}
 \bigg[
  - (\im z_1)^2 + 2 \im z_1 \im z_j - (\im z_j)^2
 \bigg]
\notag
\\
&\phantom{=\exp \bigg\{}
+2 \pi \sum_{j=2}^{n}\frac{s_{1j}}{\im \tau}(\im z_1)^2 
+ 2 \pi \sum_{j=2}^{n}\frac{s_{1j}}{\im \tau}(\im z_j)^2
\bigg\}
\notag
\\
&=\exp \bigg[ \sum_{j=2}^{n} s_{1j} \log |\vth(z_1-z_j|\tau)|^2\bigg] 
\notag
\exp \bigg[ -2 \pi \sum_{j=2}^{n}s_{1j}\frac{(\im (z_1-z_j))^2}{\im \tau}  + 2 \pi \sum_{j=2}^{n} s_{1j} \frac{(\im z_j)^2}{\im \tau}\bigg]
\notag
\\
&=\exp \bigg\{ \sum_{j=2}^{n} s_{1j}\bigg[ \log |\vth(z_1-z_j|\tau)|^2 -2 \pi \frac{(\im(z_1-z_j))^2}{\im \tau}\bigg]\bigg\} 
\exp \bigg[ 2 \pi \sum_{j=2}^{n} s_{1j} \frac{(\im z_j)^2}{\im \tau} \bigg] \, ,
\label{eq:modular-rw-derivation-theta-funcs}
\end{align}
where we note that we completed the square to form $(\im(z_1-z_j))^2$ and that the term $\sum_{j=2}^{n}\frac{s_{1j}}{\im \tau}(\im z_1)^2 $ vanishes due to momentum conservation.

Gathering the results of \eqref{eq:modular-rw-derivation-nonholo-Eisenstein} and \eqref{eq:modular-rw-derivation-theta-funcs}, we obtain the following formula for the complex Riemann-Wirtinger integral:
\begin{align}
\label{eq:modular-RW-derivation-last-step}
\langle \bar{\vphi}_k \vert \vphi_l \rangle &=
\exp\big[2 \pi i \eta \frac{\im z_l}{\im \tau} - 2 \pi i \overline{\eta} \frac{\im z_k}{\im \tau} 
   +
2 \pi \sum_{j=2}^{n} s_{1j} \frac{(\im z_j)^2}{\im \tau}
\big]
\\
    \notag
   &\phantom{=} \times\int_{M} \exp \bigg\{ \sum_{j=2}^{n} s_{1j}\bigg[ \log |\vth(z_1-z_j|\tau)|^2 -2 \pi \frac{(\im(z_1-z_j))^2}{\im \tau}\bigg]\bigg\}  
   \\
   \notag
   &\phantom{= \times\int_{M}}
   \times
   \Omega(z_1-z_l,\eta\vert\tau)\ 
   \overline{\Omega(z_1-z_k,\eta\vert\tau)}\ 
    \d^2 z_1  \, .
\end{align}
Equation \eqref{eq:def-modular-complex-RW} follows immediately from \eqref{eq:modular-RW-derivation-last-step}.

\section{The modular double copy: subleading terms} \label{sec:modular-subleading}

We will now write the modular forms $N_{2,2}^{(0,1)}$ and $N_{2,2}^{(1,1)}$ that an be obtained from Riemann-Wirtinger integrals when $(n=3)$. For concreteness, we will focus on the case of real Mandelstam variable $s_{12}$ and fixing $z_2=0$. We will also use the comoving coordinate $u_3=\frac{\im z_3}{\im \tau}$.

We will need to introduce some notation for the expansion of the Riemann-Wirtinger integral $[ \gamma \vert \varphi_j \rangle $ and its complex-conjugate as follows:
\begin{align}
\label{eq:app-eta=expand-RW}
[ \gamma \vert \varphi_j \rangle \big{|}_\eqref{eq:solving-s1A-leading-2dim-homology}  & = \sum_{n=-1}^{\infty} \sum_{m=0}^{\infty}\eta^n \overline{\eta}^m [ \gamma \vert \varphi_j \rangle^{(m,n)} \, ,
\\
\overline{[ \gamma \vert \varphi_j \rangle}\big{|}_\eqref{eq:solving-s1A-leading-2dim-homology}  &= \sum_{n=0}^{\infty} \sum_{m=-1}^{\infty}\eta^m \overline{\eta}^n \overline{[ \gamma \vert \varphi_j \rangle}^{(m,n)}\, , 
\end{align}
where we need to include powers of both $\eta$ and $\overline{\eta}$ because of \eqref{eq:solving-s1A-leading-2dim-homology}. Because of the reality of $s_{12}$ we have 
\begin{align}
\overline{[ \gamma \vert \varphi_j \rangle^{(m,n)} }
=
[ \overline{\gamma \vert \varphi_j \rangle}^{(n,m)}\, .
\end{align}
Thus, we will be able to write $N_{2,2}^{(0,1)}$ and $N_{2,2}^{(1,1)}$ in terms of only the Riemann-Wirtinger integrals $[\gamma | \varphi_2 \rangle$ . From this point onward, we omit any explicit reference to \eqref{eq:solving-s1A-leading-2dim-homology}, which is understood to hold for the rests of this appendix.

The component of the modular double copy, $N_{22}^{(0,1)}$, is given by:\begin{align}
N_{22}^{(0,1)}&= \frac{i}{2}\csc(\pi s_{12} u_3 )\sin(\pi s_{12}) e^{2 \pi s_{12} u^2_3 \im \tau}
\\
&\phantom{=} \times \bigg\{
\notag
\frac{1}{\im \tau} \bigg[ \bigg( \frac{\pi \, e^{i \pi s_{12}(u_3-1)}}{e^{2\pi i s_{12} u_3}-1}\bigg) \bigg( e^{2 \pi i s_{12}}
\big(
[\gamma_{23},\varphi_2\rangle^{(0,-1)}-[\gamma_{2A},\varphi_2\rangle^{(0,-1)}\big) \overline{[\gamma_{23},\varphi_2\rangle^{(0,-1)}} 
\\
\notag
&\hspace{6em}
+ [\gamma_{23},\varphi_2\rangle^{(0,-1)}
\big(
\overline{[\gamma_{2A},\varphi_2\rangle^{(0,-1)}}
-
\overline{[\gamma_{23},\varphi_2\rangle^{(0,-1)}}
\big)
\bigg)
\bigg]
\\
\notag
&
+(\im \tau)^0 \bigg[ e^{- i \pi s_{12} (u_3-1)} \bigg(
-[\gamma_{23},\varphi_2\rangle^{(0,0)}
\overline{[\gamma_{23},\varphi_2\rangle^{(0,-1)}}
+[\gamma_{2A},\varphi_2\rangle^{(0,0)}
\overline{[\gamma_{23},\varphi_2\rangle^{(0,-1)}}
\\
\notag
&\hspace{6em}
-[\gamma_{23},\varphi_2\rangle^{(0,-1)}
\overline{[\gamma_{23},\varphi_2\rangle^{(1,-1)}}
+[\gamma_{2A},\varphi_2\rangle^{(0,-1)}
\overline{[\gamma_{23},\varphi_2\rangle^{(1,-1)}}
\\
\notag
&\hspace{6em}
+e^{2 \pi i s_{12}(u_3-1)}
[\gamma_{23},\varphi_2\rangle^{(0,0)}
\big(
\overline{[\gamma_{23},\varphi_2\rangle^{(-1,0)}}
-
\overline{[\gamma_{2A},\varphi_2\rangle^{(-1,0)}}
\big)
\\
\notag
&\hspace{6em}  
+e^{2 \pi i s_{12}(u_3-1)}
[\gamma_{23},\varphi_2\rangle^{(0,-1)}
\big(
\overline{[\gamma_{23},\varphi_2\rangle^{(1,-1)}}
-
\overline{[\gamma_{2A},\varphi_2\rangle^{(1,-1)}}
\big) \bigg] \bigg\} \, .
\end{align}
Note that in the formula for $N_{22}^{(0,1)}$  above we have collected the powers of $\im \tau$ that appear from the double copy, other than the power in the exponential in the first line. 
Like $N_{22}^{(0,1)}$ above, we  write $N_{22}^{(1,1)}$ as a sum of three terms, by isolating negative powers of $\im \tau$:
\begin{align}
N_{22}^{(1,1)} = \frac{i}{2}\csc(\pi s_{12} u_3 )\sin(\pi s_{12}) e^{2 \pi s_{12} u^2_3 \im \tau}\bigg( \frac{1}{(\im \tau)^2}N^{(2)}
+\frac{1}{\im \tau}N^{(1)}
+(\im \tau)^0 N^{(0)} \bigg) \, ,
\end{align}
and where $N^{(2)}$, $N^{(1)}$, $N^{(0)}$ are given by:
\begin{align}
N^{(2)}&
= - \frac{i \pi^2 e^{i \pi s_{12}(u_3-1)}}{e^{2\pi i s_{12} u_3}-1} \cot (\pi s_{12 u_3})
\bigg[
e^{2 \pi i s_{12}} 
\big(
[\gamma_{23},\varphi_2\rangle^{(0,-1)}
-
[\gamma_{2A},\varphi_2\rangle^{(0,-1)}
\big)
\overline{[\gamma_{23},\varphi_2\rangle^{(0,-1)}}
\nn\\
&
\phantom{= - \frac{i \pi^2 e^{i \pi s_{12}(u_3-1)}}{e^{2\pi i s_{12} u_3}-1} )
\bigg[}
[\gamma_{23},\varphi_2\rangle^{(0,-1)}
\big(
-\overline{[\gamma_{23},\varphi_2\rangle^{(0,-1)}}
+\overline{[\gamma_{2A},\varphi_2\rangle^{(0,-1)}}
\big)
\bigg] \, ,
\end{align}
%%%%%%%%%%%%%%%%%%%%%%%%%%%%%%%%%%%%%%%%%%%%%%%%%
\begin{align}
N^{(1)}&=\frac{\pi e^{\pi i s_{12}(u_3-1)}}{e^{2 \pi i s_{12}u_3}-1} \bigg[
e^{2 \pi i s_{12}}
\bigg(
\big(
-[\gamma_{2A},\varphi_2\rangle^{(1,-1)}
+
[\gamma_{2A},\varphi_2\rangle^{(0,0)}
\big)
\overline{[\gamma_{23},\varphi_2\rangle^{(0,-1)}}
\\
\notag
&\hspace{6em}
-\big(
[\gamma_{23},\varphi_2\rangle^{(0,-1)}
-[\gamma_{2A},\varphi_2\rangle^{(0,-1)}
\big)
\big(
\overline{[\gamma_{23},\varphi_2\rangle^{(1,-1)}}-
-\overline{[\gamma_{23},\varphi_2\rangle^{(0,0)}}
\big)
\bigg)
\\
\notag
&\hspace{6em}
+[\gamma_{23},\varphi_2\rangle^{(1,-1)}
\big(
(e^{2 \pi i s_{12}}-1)
\overline{[\gamma_{23},\varphi_2\rangle^{(0,-1)}}
+
\overline{[\gamma_{2A},\varphi_2\rangle^{(0,-1)}}
\big)
\\
\notag
&\hspace{6em}
-[\gamma_{23},\varphi_2\rangle^{(0,0)}
\big(
(e^{2 \pi i s_{12}}-1)
\overline{[\gamma_{23},\varphi_2\rangle^{(0,-1)}}
+
\overline{[\gamma_{2A},\varphi_2\rangle^{(0,-1)}}
\big)
\\
\notag
&\hspace{6em}
+[\gamma_{23},\varphi_2\rangle^{(0,-1)}
\big(
\overline{[\gamma_{23},\varphi_2\rangle^{(1,-1)}}
-\overline{[\gamma_{23},\varphi_2\rangle^{(0,0)}}
\\
\notag
&\hspace{6em}
-\overline{[\gamma_{2A},\varphi_2\rangle^{(1,-1)}}
+\overline{[\gamma_{2A},\varphi_2\rangle^{(0,0)}}
\big)
\bigg)
\bigg] \, ,
\end{align}
\begin{align}
N^{(0)}&= e^{- \pi i s_{12} (u_3-1)} \bigg[
-[\gamma_{23},\varphi_2\rangle^{(1,0)}
\overline{[\gamma_{23},\varphi_2\rangle^{(0,-1)}}
+[\gamma_{2A},\varphi_2\rangle^{(1,0)}
\overline{[\gamma_{23},\varphi_2\rangle^{(0,-1)}}
\\
\notag
&\phantom{= e^{- \pi i s_{12} (u_3-1)} \bigg[}
-[\gamma_{23},\varphi_2\rangle^{(1,-1)}
\overline{[\gamma_{23},\varphi_2\rangle^{(1,-1)}}
+[\gamma_{2A},\varphi_2\rangle^{(1,-1)}
\overline{[\gamma_{23},\varphi_2\rangle^{(1,-1)}}
\\
\notag
&\phantom{= e^{- \pi i s_{12} (u_3-1)} \bigg[}
-[\gamma_{23},\varphi_2\rangle^{(0,0)}
\overline{[\gamma_{23},\varphi_2\rangle^{(0,0)}}
+[\gamma_{2A},\varphi_2\rangle^{(0,0)}
\overline{[\gamma_{23},\varphi_2\rangle^{(0,-1)}}
\\
\notag
&\phantom{= e^{- \pi i s_{12} (u_3-1)} \bigg[}
-[\gamma_{23},\varphi_2\rangle^{(0,-1)}
\overline{[\gamma_{23},\varphi_2\rangle^{(1,0)}}
+[\gamma_{2A},\varphi_2\rangle^{(0,-1)}
\overline{[\gamma_{23},\varphi_2\rangle^{(1,0)}}
\\
\notag
&\phantom{= e^{- \pi i s_{12} (u_3-1)} \bigg[}
+e^{2 \pi i s_{12}(u_3-1)}\bigg(
[\gamma_{23},\varphi_2\rangle^{(1,0)}
\big(
\overline{[\gamma_{23},\varphi_2\rangle^{(0,-1)}}
-
\overline{[\gamma_{2A},\varphi_2\rangle^{(0,-1)}}
\big)
\\
\notag
&\phantom{= e^{- \pi i s_{12} (u_3-1)} \bigg[
+e^{2 \pi i s_{12}(u_3-1)}\bigg(
}
+[\gamma_{23},\varphi_2\rangle^{(1,-1)}
\big(
\overline{[\gamma_{23},\varphi_2\rangle^{(1,-1)}}
-
\overline{[\gamma_{2A},\varphi_2\rangle^{(1,-1)}}
\big)
\\
\notag
&\phantom{= e^{- \pi i s_{12} (u_3-1)} \bigg[
+e^{2 \pi i s_{12}(u_3-1)}\bigg(
}
+
[\gamma_{23},\varphi_2\rangle^{(0,0)}
\big(
\overline{[\gamma_{23},\varphi_2\rangle^{(0,0)}}
-
\overline{[\gamma_{2A},\varphi_2\rangle^{(0,0)}}
\big)
\\
\notag
&\phantom{= e^{- \pi i s_{12} (u_3-1)} \bigg[
+e^{2 \pi i s_{12}(u_3-1)}\bigg(
}
+
[\gamma_{23},\varphi_2\rangle^{(0,-1)}
\big(
\overline{[\gamma_{23},\varphi_2\rangle^{(1,0)}}
-
\overline{[\gamma_{2A},\varphi_2\rangle^{(1,0)}}
\big)
\bigg)\bigg] \, .
\end{align}
As a supplementary material, we attach a \texttt{Mathematica} notebook with the formulas for $N_{22}^{(1,0)}$ and $N_{22}^{(1,1)}$.

%%%%%%%%%%%%%%%%%%%%%%%%%%%%%%%%%%%%%%%%%%%%%%%%%
%%%%%%%%%%%%%%%%%%%%%%%%%%%%%%%%%%%%%%%%%%%%%%%%%
%%%%%%%%%%%%%%%%%%%%%%%%%%%%%%%%%%%%%%%%%%%%%%%%%
\section{Towards the analytic Riemann-Wirtinger integral in terms of eMPLs \label{sec:RWco-app}}

\new{In this appendix, we compute the differential equations satisfied by the Riemann-Wirtinger integrals in section \ref{sec:RWco-app-deqs}.
In section \ref{sec:RWco-app-alphaprime}, we solve the differential equations to order $\mathcal{O}(\alpha')$ in terms of eMPLs for generic boundary values.
Boundary values for the contours introduced in section \ref{sec:RWhom-basis} are derived in \ref{sec:RWco-app-bdval}. 
Lastly, in appendix \ref{sec:RWco-app-0}, we describe the cohomology in the $\eta\to0$ limit and discuss some of new features of the corresponding differential equations.
}

\subsection{Differential equations for the $\vphi$-basis \label{sec:RWco-app-deqs}}

Using intersection numbers, we compute the Gauss-Manin differential equation satisfied by the Riemann-Wirtinger integrals where we differentiate with respect to the external un-integrated punctures $z_{i\geq2}$. 

The differential operators in the external punctures must preserve the monodromy structure of the local system in order for the differential equations to close. 
For $f\ \d z_1 \in \Omega^1_\eta$, the Gauss-Manin differential operators that preserve monodromies are \cite{Mano2012}:
\begin{align}
    \label{eq:nablaKinz}
    \nabla_{\text{ext},a} f\ \d z_1 
    &:= (\partial_{a} f)\ \d z_1  
        + (\omega_a f)\ \d z_1 
        + s_{1a} (\partial_\eta f)\ \d z_1
    \qquad \text{for} \qquad a=2,3,\dots,n
    \,,
    \\
    \label{eq:nablaKintau}
    \nabla_{\text{ext},\tau} f\ \d z_1
    &:= (\partial_\tau f)\ \d z_1 
        + (\omega_\tau f)\ \d z_1
        + s_{1A} (\partial_\eta f)\ \d z_1
        - \frac{1}{2 \pi i} \nabla_{\mathcal{M}/\mathcal{B}} (\partial_\eta f)
    \,.
\end{align}
Here, $f$ is a meromorphic function on $M$ with $B$-cycle monodromy $e^{-2 \pi i \eta}$.
Understanding the precise form of the above differential operators requires understanding how varying the punctures and moduli on the total space $M \to \mathcal{M} \overset{\pi}{\to} \mathcal{B}=\{(z_2,\dots,z_n,\tau)\in\mathbb{C}^{n-1}\times\mathbb{H} \vert z_i \neq z_j \}$ descends to $M$. 
For Riemann-Wirtinger integrals, the differential operator $\nabla_{\M/\mathcal{B}}$ is simply $\nabla$ (see \cite{ghazouani2016moduli} appendix \ref{app:intInd} for the derivation in the $\eta=0$ case)\footnote{See also \cite{deligne2006equations} or \cite{KatzOda} for the general theory.}. 
It is also important to note that we cannot throw away $\nabla_{\M/\mathcal{B}}(\partial_\eta f)$ even though it is a total covariant derivative because $\partial_\eta f$ does not have the allowed multiplicative monodromy.

To motivate the form of the Gauss-Manin differentials above, we examine how the connection on $\mathcal{M}$ transforms under the $B$-cycle. 
We let $\tilde{\d}$ be the exterior derivative on $\mathcal{M}$ and set $\tilde{\omega} = \tilde{\d}\log(u)$. 
This is the obvious uplift of the connection on $M$ to $\mathcal{M}$.
The $B$-cycle monodromy of $\tilde{\omega}$ is 
\begin{align}
    \tilde{\omega} = \tilde{\d}\log(u) & \underset{z_1\to z_1+\tau}{\longrightarrow}
    \tilde{\omega} + \tilde{\d}\log\left(
        e^{
            2\pi i s_{1A} \tau 
            + 2\pi i \sum_{i=2}^n s_{1i} z_i 
        } 
    \right)
    \nn\\
    &= \tilde{\omega} 
    + 2\pi i\ \tilde{\d} \left( 
             s_{1A} \tau 
            + \sum_{i=2}^n s_{1i} z_i
        \right)
    = \tilde{\omega} + 2\pi i\ \tilde{\d} \eta,
\end{align}
When $\eta=0$, the above connection is doubly periodic.
Next, we compute the $B$-cycle monodromy of the image of the covariant derivative on $\mathcal{M}$: $\tilde{\nabla}=\tilde{\d}+\tilde{\omega}\wedge$. 
Explicitly, 
\begin{align}
    &\tilde{\nabla} \left( 
        F(z_{1a},\eta)\ \tilde{\d} z_1 
    \right)
    \underset{z_1\to z_1+\tau}{\longrightarrow}
    \left( \tilde{\nabla} + 2\pi i\ \tilde{\d}\eta\wedge \right) 
    \left( e^{-2\pi i\eta} F(z_{1a},\eta)\ (\tilde{\d} z_1 + \tilde{\d}\tau)  \right)
    ,
    \nn\\
    &= e^{-2\pi i\eta}  \bigg[ 
        (\nabla_\tau - \nabla_1) F(z_{1a},\eta)\ \tilde{\d} \tau \wedge \tilde{\d} z_1 
        + \sum_{i=2}^n \nabla_i F(z_{1a},\eta)\ \tilde{\d} z_i \wedge \tilde{\d} z_1 
        + \cdots
    \bigg]
    ,
    \label{eq:nablaTotMonodromy}
\end{align}
where the dots above indicate differential forms that are independent of $\tilde{\d}z_1$.
By adding $-\frac{1}{2 \pi i} \nabla_1 \partial_\eta$ to the $\d\tau$-part of the covariant derivative $\tilde{\nabla}$ one obtains a differential form whose $\tilde{\d}z_1$ part has the allowed $B$-cycle monodromy. 
In fact, the components of this differential are precisely \eqref{eq:nablaKinz} and \eqref{eq:nablaKintau}. 
Note that we have made the action of the partial derivatives on $\eta$ explicit in \eqref{eq:nablaKinz} and \eqref{eq:nablaKintau} but left this implicit in \eqref{eq:nablaTotMonodromy}. 

The differential equation for the family of Riemann-Wirtinger integrals or Gauss-Manin connection on the base space $\mathcal{B}$ is 
\begin{align}
    \d_\text{ext} [\gamma\vert \vphi_a \ra = A_{ab} [\gamma\vert \vphi_b \ra 
    \iff 
    \nabla_{\text{ext}} \vert \vphi_a \ra
    = A_{ab} \wedge \vert \vphi_b \ra
\end{align}
where 
\begin{align}
    A_{ab}\vert_{\d z_i}
    = C_{bc}^{-1} \la \c{\vphi}_c \vert \nabla_{\text{ext},i}\ \vphi_a \ra
    = -\frac{1}{s_{1b}} \la \c{\vphi}_b \vert \nabla_{\text{ext},i}\ \vphi_a \ra
    . 
\end{align}
Breaking this up into the off-diagonal and diagonal parts, the general formula for the $\d z_c$-components of the differential equation are
 \begin{align}
    \label{eq:Omegaz-off-diag}
    \left(A^{\text{off-diag}}_{c}\right)_{ab}
        =&
            \sum_{d \neq c} 
            \delta_{bd}\ \delta_{ac}\ 
            s_{1b}\ F(z_{bc},\eta)
            % \nn\\&\qquad
        -
            \sum_{d \neq c} 
            \delta_{ad}\ \delta_{bc}\ 
            s_{1c}\ F(z_{ca},\eta)
        \,,
        \\
    \label{eq:Omegaz-diag}
    \left(A^{\text{diag}}_c\right)_{ab}
        = \delta_{ab} \Bigg[&
            -s_{1c} (1-\delta_{ca})\ 
                g^{(1)}(z_{ac})
            % \nn\\&
            + \delta_{ac}
            \bigg(
                2 \pi i s_{1A} 
                + \sum_{d \neq c} s_{1d}\ 
                    g^{(1)}(z_{cd})
            \bigg)
        \Bigg]
        \,,
\end{align}
where $a,b = 2,\dots n$.
Similarly, the general formula for the $\d\tau$-component is 
\begin{align}
    \label{eq:Omegatau-off-diag}
    \left(A^\text{off-diag}_\tau\right)_{ab} &= 
    -\frac{(1-\delta_{ab})}{2\pi i}
    \bigg[
        s_{1b}\ \partial_{\eta} F(z_{ba},\eta)
    \bigg] \d\tau
    \,,
    \\
    \label{eq:Omegatau-diag}
    \left(A^\text{diag}_\tau\right)_{ab} &= 
    \frac{\delta_{ab}}{2\pi i}
    \bigg[
        \sum_{c=2}^n s_{1c}\ \gk{2}(z_{ac})
        - s_{1a}\ \partial_\eta F(0,\eta)
    \bigg] \d\tau
    \,,
\end{align}
where 
\begin{align}
    \partial_\eta F(0,\eta)
    = \frac{\vth^{\prime\prime}(\eta)}{\vth(\eta)}
    -\left(\frac{\vth^\prime(\eta)}{\vth(\eta)}\right)^2
    = - \left( \wp(\eta) + G_2(\tau) \right)
    \,,
\end{align}
and $G_2(\tau) = -\frac{\vth^{\prime\prime\prime}(0)}{3\vth^\prime(0)}$ is the second Eisenstein series. 
For example, the $\d z_2$- and $\d\tau$-components of $\mat{A}$ for $n=4$ are 
\begin{align} \label{eq:4ptManoDEQ2}
    \mat{A}_{z_{2}}
    &= \begin{pmatrix}
        2\pi i s_{1A} 
        {+} s_{13}\ \gk{1}_{23}
        {+} s_{14}\ \gk{1}_{24}
        & s_{13}\ F(z_{32},\eta) 
        & s_{14}\ F(z_{42},\eta) 
        \\
        -s_{12}\ F(z_{23}, \eta) 
        & s_{12}\ g^{(1)}_{23}
        & 0 
        \\
        -s_{12}\ F(z_{24},\eta)
        & 0
        & s_{12}\ g^{(1)}_{24}
    \end{pmatrix}
    ,
    \\
    \mat{A}_{\tau}
    &= \frac{1}{2\pi i}
    \begin{pmatrix} \label{eq:4ptManoDEQ2_tau}
        \sum_{c=2}^4 \frac{s_{1c}}{2}\ \gk{2}(z_{2c})
        & 0
        & 0 
        \\
        0
        & \sum_{c=2}^4 \frac{s_{1c}}{2}\ \gk{2}(z_{3c})
        & 0 
        \\
        0
        & 0
        & \sum_{c=2}^4 \frac{s_{1c}}{2}\ \gk{2}(z_{4c})
    \end{pmatrix}
    \nn\\&\qquad
    - \frac{1}{2\pi i}
    \begin{pmatrix}
        s_{12}\ \partial_\eta F(0,\eta)
        & s_{13}\ \partial_\eta F(z_{32},\eta)
        & s_{14}\ \partial_\eta F(z_{42},\eta)
        \\
         s_{12}\ \partial_\eta F(z_{23},\eta)
        & s_{13}\ \partial_\eta F(0,\eta)
        & s_{14}\ \partial_\eta F(z_{43},\eta)
        \\
        s_{12}\ \partial_\eta F(z_{24},\eta)
        & s_{13}\ \partial_\eta F(z_{34},\eta)
        & s_{14}\ \partial_\eta F(0,\eta)
    \end{pmatrix}
    .
\end{align}
We have also checked that equations \eqref{eq:Omegaz-off-diag}-\eqref{eq:Omegatau-diag} match the differential equations obtained by  different methods in \cite{Mano2012}.
Moreover, our results agree with those obtained using standard integration-by-parts and the Fay identity \cite{Kaderli:2022qeu} once the condition \eqref{eq:eta-def} is imposed and expand on this below. 

It is interesting that these equations differ from those previously obtained in the math and physics literature, especially in the context of KZB equations \cite{Felder1995,Mafra2019,Broedel2020,Kaderli:2022qeu}. Since the authors of \cite{Mano2012} have already commented on how their differential equations relate to those in \cite{Felder1995}, we focus on the differences in the $z_j$-differential equation of this work and section (3.3.1) of \cite{Kaderli:2022qeu}. 
Imposing momentum conservation on the differential equations of \cite{Kaderli:2022qeu} yields \eqref{eq:4ptManoDEQ2} and \eqref{eq:4ptManoDEQ2_tau} up to one key difference: 
what we mean by the derivative $\partial_{z_i}$ in our work is actually a partial derivative at constant $s_{1B}$, $(\partial_{z_i})_{s_{1B}}$, while the partial derivative $\partial_{z_i}$ of \cite{Kaderli:2022qeu} makes no such distinction. 
Thus, these two derivatives are related by:
\begin{align}
\label{eq:eta-derivative}
(\partial_{z_i})_{s_{1B}}
&=
\partial_{z_i}\big|_{[{\color{magenta}117}]} +  s_{1i} \partial_{\eta} \, ,
\end{align}
where we one has to  exchange the punctures $z_1$ and $z_2$ in \cite{Kaderli:2022qeu} to convert into the conventions used here.
% (so that the Mandelstam variable relating the integrated puncture and $z_3$ is $s_{13}$). 
Once we take equation \eqref{eq:eta-derivative} into account, we can readily recover the $z_i-$differential equation in our work from \cite{Kaderli:2022qeu}.

As an important cross-check, we show that the above differential equation is integrable. 
Writing the connection on kinematic space as $\mat{A} = \sum_{i=2}^n \mat{A}_i \d z_i\ + \mat{A}_\tau \d \tau$, we want to show
\begin{align}
    \label{eq:zz-integrability}
    \partial_{a} \mat{A}_b
    -\partial_{b} \mat{A}_a
    + [\mat{A}_a, \mat{A}_b]
    &= 0
    \,,
    \\
    \label{eq:ztau-integrability}
    \partial_\tau \mat{A}_a
    -\partial_{a} \mat{A}_\tau
    + [\mat{A}_\tau, \mat{A}_a]
    &=0
    \,,
\end{align}
for all $a,b=2,\dots,n$ and $a \neq b$.
To show integrability, we will also need the help from the following identity
\begin{align}
    \partial_\eta F_{jk} 
    + (\gk{1}_{ij} - \gk{1}_{ik}) F_{jk}  
    + F_{ji}\ F_{ik} 
    = 0   
    \,,
\end{align}
which follows from the $\mathcal{O}(\xi^0)$ term of \eqref{eq:fay} with $(i,j,k)=(2,1,3)$, $\zeta_i=\xi$ and $\zeta_j=\eta-\xi$.
Concentrating on the integrability of the $\d z_a$-components, one finds that
\begin{equation}
    \begin{aligned}
        0 &= \bigg( 
            \pa_{c} \mat{A}_e 
            - \pa_{e} \mat{A}_c 
            + [\mat{A}_c, \mat{A}_e]
        \bigg)_{ab}
        \\&= \left(1 - \delta_{ab} \right) \sum_{d\neq c,e} \Bigg\lbrack 
        -\left( 
            \delta_{ac} \delta_{be} 
            - \delta_{ae} \delta_{bc} 
        \right) 
        s_{1b} s_{1d} 
        \left( 
            \partial_\eta F_{ba} 
            + (g_{db}^{(1)} - g_{da}^{(1)}) F_{ba}
            + F_{da} F_{bd} 
        \right) 
        \\& \hspace{8em}
        + \left( 
            \delta_{ac} \delta_{bd} 
            - \delta_{bc} \delta_{ad} 
        \right) 
        s_{1b} s_{1e} 
        \left( 
            \partial_\eta F_{ba}
            + ( g_{eb}^{(1)} - g_{ea}^{(1)}) F_{ba}
            + F_{be}F_{ea} 
        \right) 
        \\& \hspace{8em}
        - \left( 
            \delta_{ae} \delta_{bd} 
            - \delta_{be} \delta_{ad} 
        \right) 
        s_{1b} s_{1c} 
        \left( 
            \partial_\eta F_{ba}
            + ( g_{cb}^{(1)} - g_{ca}^{(1)} ) F_{ba}
            + F_{bc}F_{ca} 
        \right) 
        \Bigg\rbrack 
        \, ,
    \end{aligned}
\end{equation}
where we have introduced the shorthand $F_{ab} = F(z_{ab},\eta)$. 
The $\tau$-integrability follows similarly, but requires more complicated Fay identities.

%%%%%%%%%%%%%%%%%%%%%%%%%%%%%%%%%%%%%%%%%%%%%%%%%
%%%%%%%%%%%%%%%%%%%%%%%%%%%%%%%%%%%%%%%%%%%%%%%%%
\subsection{The $\mathcal{O}(\alpha^\prime)$ solution to the differential equations \label{sec:RWco-app-alphaprime}}

In this section, we study the $\alpha^\prime$ expansions of the differential equations \eqref{eq:Omegaz-off-diag}-\eqref{eq:Omegatau-diag} with $n=4$ punctures and the resulting solutions.
The purpose of this section is to obtain some zero-th order information about the $\alpha'$-expansion of Riemann-Wirtinger integrals; 
to identify the kind of objects that appear and how they relate to known functions in the elliptic Feynman integral and loop string amplitude literature. 
At leading order in the $\alpha'$-expansion, the Riemann-Wirtinger integrals have an analytic structure very similar to the leading $\alpha'$-expansion of the analogous string integrals \cite{Kaderli:2022qeu}. 
However, at higher orders, the two solutions start to diverge.

We start by making the $\alpha'$ dependence apparent in the Mandelstams by setting $s_{ij} = \alpha' \tilde{s}_{ij}$, where $\tilde{s}_{ij} = k_i\cdot k_j$ are the new $\alpha'$ independent Mandelstams. 
Next, we expand the Kronecker-Eisenstein functions  $F(z_{ij},\eta)$ in the cohomology basis elements in orders of $\alpha'$ remembering that $\eta = -s_{1B}+\alpha'\left(\tilde{s}_{1A}\tau+\sum_{i=2}^n\tilde{s}_{1i}z_i\right)$. 
While all of the Mandelstam variables have a true factor of $\alpha'$, it is unclear if $s_{1B}$ should also be assigned a power of $\alpha'$.
In the following we take the conservative approach and assume that $s_{1B}$ does not scale with $\alpha'$ to ensure that $\eta\neq0$ since the structure of the cohomology changes in the $\eta\to0$ limit.

To sub-leading order in $\alpha'$, 
\begin{equation}\begin{aligned}
    F(z_{ij},\eta) 
    &\approx F(z_{ij},{-}s_{1B})
    % \\&\times 
    \left[
        1        
        {+} \alpha'\bigg(\tilde{s}_{1A}\tau {+} \sum_{i=2}^n\tilde{s}_{1i}z_i\bigg)
        \left[ 
            \gk{1}(z_{ij}{-}s_{1B}) 
            {-} \gk{1}({-}s_{1B})
        \right]
        % {+} \mathcal{O}\big(\alpha'^2\big)
    \right]. 
\end{aligned}\end{equation}
Similarly, the kinematic connections \eqref{eq:Omegaz-off-diag}-\eqref{eq:Omegatau-diag}
have an $\alpha^\prime$ expansion 
\begin{equation}
    \mat{A}_a =  \sum_{\ell=1}^{\infty}
    (\alpha')^{\ell}\
    \tilde{\mat{A}}^{(\ell)}_a
    \,,
\end{equation}
that does not truncate due to the non-uniform dependence of $\eta$ on $\alpha'$.
This means that the differential equations are not in $\alpha'$-form complicating the expansion of the path ordered exponential. 
Still, one can obtain the $\alpha'$ expanded solution order-by-order where the order $k$ DEQ has contributions from all previous orders. 
Explicitly, we have 
\begin{align}
    \d_\text{ext}\mathbf{I}_\gamma^{(0)} 
    &= 0
    \,,
    \\
    \d_\text{ext}\mathbf{I}_\gamma^{(1)} 
    &= \sum_{i=2}^n \d z_i\ \tilde{\mat{A}}^{(1)}_i \cdot \mathbf{I}^{(0)}_\gamma
    + \d\tau\ \tilde{\mat{A}}^{(1)}_\tau  \cdot \mathbf{I}^{(0)}_\gamma
    \,,
    \\
    \d_\text{ext}\mathbf{I}_\gamma^{(k\geq1)} 
    &= \sum_{i=2}^n \d z_i\ 
    \sum_{l+m=k} \tilde{\mat{A}}^{(l)}_i \cdot \mathbf{I}^{(m)}_\gamma
    + \d\tau\ \sum_{l+m=k}
    \tilde{\mat{A}}^{(l)}_\tau  \cdot \mathbf{I}^{(m)}_\gamma
    \,,
\end{align}
where we have set $\mathbf{I}_\gamma := \alpha'\ [\gamma\vert\boldsymbol{\vphi}\ra$.
The factor of $\alpha'$ in the definition of $\mathbf{I}_\gamma$ is there to ensure that there are no $\alpha'$-poles in the series expansion of $\mathbf{I}_\gamma$ for the contours introduced in section \ref{sec:RWhom}.

To explore the properties of these functions, we solve the DEQs to the first non-trivial order in $\alpha'$. 
Here, the leading order approximation of the connection matrices is
\begin{align}
    \label{eq:alphaprimeDEQ_tau}
    &\left(\tilde{A}^{(1)}_\tau\right)_{ab}
    = 
    \frac{\delta_{ab}}{2\pi i}
    \bigg[
        \sum_{c=2}^n s_{1c}\ \gk{2}(z_{ac})
        + s_{1a}\ \partial_{s_{1B}} F(0,-s_{1B})
    \bigg] 
    \nn\\&\hspace{6em}
    +\frac{(1-\delta_{ab})}{2\pi i}
    \bigg[
        s_{1b}\ \partial_{s_{1B}} F(z_{ba},-s_{1B})
    \bigg]
    \,,
    \\ \label{eq:alphaprimeDEQ_z}
    &\left(\tilde{A}_{c=2,\dots,4}^{(1)}\right)_{ab}
        = \Bigg[
            \sum_{d \neq c} 
            \delta_{bd}\ \delta_{ac}\ 
            \tilde{s}_{1b}\ F(z_{bc},-s_{1B})
            % \nn\\&\qquad
        -
            \sum_{d \neq c} 
            \delta_{ad}\ \delta_{bc}\ 
            \tilde{s}_{1c}\ F(z_{ca},-s_{1B})
        \Bigg]
        \nn\\
        &\hspace{6em}
        + \delta_{ab} \Bigg[
            -\tilde{s}_{1c} (1-\delta_{ca})\ 
                g^{(1)}(z_{ac})
            % \nn\\&
            + \delta_{ac}
            \bigg(
                2 \pi i \tilde{s}_{1A} 
                + \sum_{d \neq c} \tilde{s}_{1d}\ 
                    g^{(1)}(z_{cd})
            \bigg)
        \Bigg]
    \,.
\end{align}
We also set $\mathbf{I}_\gamma(z_{i=2,\dots,n}^*,\tau^*) = \sum_{\ell=0}^\infty \left(\alpha'\right)^{\ell} \mathbf{c}^{(\ell)}_\gamma$ as the boundary value.
While the boundary values for a basis of contours is given in section \ref{sec:RWco-app-bdval}, we will keep $\mathbf{c}^{(\ell)}_\gamma$ generic in this section to avoid subtleties with the limit $\tau\to i \infty$ where the torus degenerates into a nodal sphere.

Obviously, the first term in the solution is simply the leading term of the boundary value: $\mathbf{I}_\gamma^{(0)}=\mathbf{c}^{(0)}_{\gamma}$. 
Next, we integrate in $\tau$ to find $\mathbf{I}_\gamma^{(1)}(z_{i\geq2}^*,\tau) = \int_{\tau^*}^\tau \d\tau'\ \tilde{\mat{A}}^{(1)}_\tau(z_{i\geq2}^*,\tau') \cdot \mathbf{I}_\gamma^{(0)}(z_{i\geq2}^*,\tau')$.
Explicitly, 
\begin{align}
    {I}_{\gamma,a}^{(1)}(z_{i\geq2}^*,\tau)
    &= c_{\gamma,a}^{(0)} 
    \sum_{b=2}^n s_{1b}  \log \left[
        \frac{\vth^\prime(z_{ab}^*,\tau)}{\vth^\prime(z_{ab}^*,\tau^*)}
        \frac{\eta_{\text{\tiny De}}(\tau^*)}{\eta_{\text{\tiny De}}(\tau)}
    \right]
    + \sum_{b=2}^n c_{\gamma,b}^{(0)} s_{1b}\ \mathcal{F}(z_{ba}^*,\tau)
    + c_{\gamma,a}^{(1)}
    \,,\label{eq:solution_tau_DEQ}
\end{align}
where $\eta_{\text{\tiny De}}(\tau)$ is the Dedekind eta-function\footnote{The subscript on the Dedekind eta-function is to differentiate from the variable $\eta = s_{1A} \tau + \sum_{i=2}^n s_{1i} z_{i}$.}, 
\begin{align} 
    \mathcal{F}(z,\tau) 
    &= \int_{\tau^*}^\tau \d\tau'\ \partial_{s_{1B}} F(z,-s_{1B}\vert\tau')
    % \nn\\&
    = \int_{\tau^*}^\tau \d\tau'\ 
        \left[ 
            \frac{1}{s_{1B}^2}
            % + 0 \times \frac{1}{s_{1B}}
            - \sum_{k=2}^\infty (k-1) \gk{k}(z,\tau') (-s_{1B})^{k-2}
        \right]
    \,,
    \nn\\
    &= \frac{\tau-\tau^*}{s_{1B}^2}
    - \sum_{k=2}^\infty (-s_{1B})^{k-2} (2\pi i)^{k-1} 
    \left[ 
        \Omega^{(k-1)}(z_{ba}^*,\tau) 
        - \Omega^{(k-1)}(z_{ba}^*,\tau^*) 
    \right]
    \,,
\end{align}
and $\Omega^{(k)}(z,\tau)$ are the known primitives of the $g^{(k)}(z,\tau)$ \cite{Wilhelm:2022wow}
\begin{align}
    g^{(k)}(z,\tau) = (2\pi i)^{k-1} \partial_{z}\Omega^{(k)}(z,\tau) 
    = \frac{(2\pi i)^{k-1}}{k-1} \partial_\tau \Omega^{(k-1)}(z,\tau)
    \,.
\end{align}
The primitives $\Omega^{(k)}$ are the symbol letters of elliptic multiple polylogarithms (eMPLs).
In fact, the $\Omega^{(k)}$ are actually depth-1 eMPLs themselves. 
It is also important to note that $\mathcal{F}(0,\tau)$ is well-defined since $\partial_{s_{1B}} F(0,-s_{1B})$ appears in the differential equations. 
To see this, observe that the primitives are finite at the origin since the $\gk{k\geq2}(0,\tau)<\infty$ are finite at the origin. 
Moreover, note that the above expression for $\mathcal{F}$ only makes sense if the difference $\tau-\tau^*$ is finite otherwise one has to apply the tangential base point regularization of \cite{brown2014multiple} and set $\int^{\tau}_{i\infty} \d\tau' = \tau$. 
However, we find boundary values in section \ref{sec:RWco-app-bdval}, which ensure that the $\tau$-integral is finite without applying tangential base point regularization. 

Once a boundary value is known at finite $\tau$ or when the $\tau$ integration converges, the $z_i$-integrals are readily evaluated in terms of familiar eMPLs
\begin{align}
    {I}_{\gamma,a}^{(1)}(z_{i\geq2},\tau)
    &= \left( 
        2\pi i\tilde{s}_{1A}\ \tilde{\Gamma}
            \left(\begin{smallmatrix}0\\0\end{smallmatrix};z_a{-}z_a^*\vert\tau\right)
        + \sum_{b\neq a} \tilde{s}_{1b}\ 
        \left[ 
            \tilde{\Gamma}
            \left(\begin{smallmatrix}1\\0\end{smallmatrix};z_{ab}\vert\tau\right)
            - \tilde{\Gamma}
            \left(\begin{smallmatrix}1\\0\end{smallmatrix};z_{ab}^*\vert\tau\right)
        \right]
    \right) c^{(0)}_{\gamma,a}
    \nonumber\\
    &~~~
    + \sum_{b\neq a} \tilde{s}_{1b}\ c^{(0)}_{\gamma,b}\
    \left[
         \Phi\left(z_{ab};s_{1B}\vert\tau\right)
        - \Phi\left(z_{ab}^*;s_{1B}\vert\tau\right)
    \right]
    % + c^{(1)}_{\gamma,a}(\tau)
    + {I}_{\gamma,a}^{(1)}(z_{i\geq2}^*,\tau)
    \,,
\end{align}
where we define the $z$-integral of the Kronecker-Eisenstein function to be
\begin{equation}
    \Phi\left(z_{ij};s_{1B}|\tau\right) := \sum_{k\geq 0}s_{1B}^{k-1}\tilde{\Gamma}\left(\begin{smallmatrix}
    k\\
    0
    \end{smallmatrix};z_{ij}|\tau\right)~, 
\end{equation}
and $\tilde{\Gamma}\left(\begin{smallmatrix}0\\0\end{smallmatrix};z_a{-}z_a^*\vert\tau\right) = z_a{-}z_a^*$.
Note that this expression may only be well defined after shuffle regularization \cite{Broedel:2019tlz}. 

Unsurprisingly, the first $\alpha'$ correction, $\mathbf{I}^{(1)}_\gamma$, has a functional form very similar to that found in \cite{Kaderli:2022qeu} where there was no constraint on $\eta$. 
However, at order $\mathcal{O}(\alpha'^2)$ we start to see integrals with powers of $\tilde{s}_{1A}\tau {+} \sum_{i=2}^n\tilde{s}_{1i}z_i$ in the integrand and terms with different transcendentality start to mix. 
Thus, we expect that only the leading term $\mathbf{I}^{(1)}_\gamma$ is actually representative of the analogous string integral.

Unfortunately, twisted (co)homology seems to force the condition \eqref{eq:eta-def} on us. 
To get closer to real string integrals, we must find a way to understand the (co)homology for unconstrained $\eta$ \eqref{eq:eta-def}. 
However, at the level of the differential equation, we can always shift the $\eta$-derivatives similarly to \eqref{eq:eta-derivative} to define differential operators that are unconstrained by \eqref{eq:eta-def}.

%%%%%%%%%%%%%%%%%%%%%%%%%%%%%%%%%%%%%%%%%%%%%%%%%
\subsection{Boundary values}\label{sec:RWco-app-bdval}

In this section, we show that the $\tau\to i\infty$ limit of the Riemann-Wirtinger family corresponds to generalized Lauricella-D hypergeometric functions. 
This degeneration to a nodal sphere yields one method for computing the boundary values for the differential equations derived in section \ref{sec:RWco-app-alphaprime}. 
For $n=3$, this degeneration can be found in \cite{mano2009}. 
However, we fix a typo there and provide an $n$-point generalization that has been numerically verified.
Finally, using a neat trick, we also provide boundary conditions at finite $\tau$ for all but the $A$- and $B$-cycles.

\begin{figure}
    \centering
    \includegraphics[scale=.8]{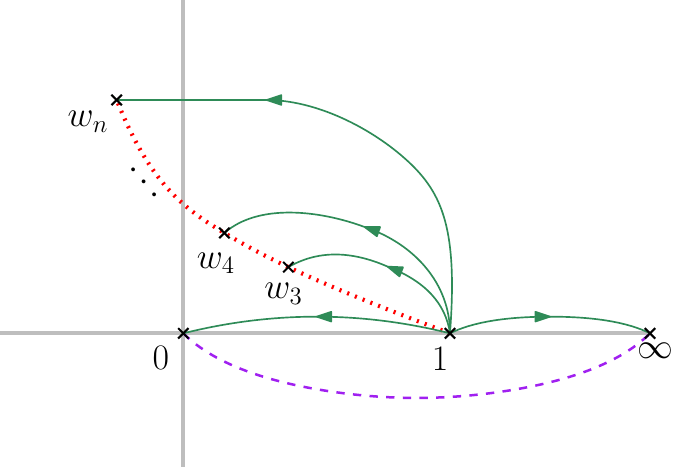} 
    \caption{The branch choice for the tree-level twist (Koba-Nielsen) factor given by \eqref{eq:RW-degen-twist}. 
    The branch choice in the fundamental domain of the torus maps to the {\color{BrickRed}red dotted line} connecting the marked points on the rational nodal curve parameterized by $w=e^{2\pi i z}$. 
    The $A$- and $B$-cycle discontinuities are captured by the branch cut from $0$ to $\infty$ denoted by the {\color{Plum}purple dashed line}.
    }
    \label{fig:branch_choice_rational_curve}
\end{figure}

To understand how to compute this degenerate limit, we need to know the form of the integrand in the $\tau\to i\infty$ limit and how the cycles on the torus map to cycles on the nodal sphere. 
Using
\begin{align}
    \frac{\vth(z_{1a}+\eta_0 + s_{1A} \tau)}{\vth(\eta_0 + s_{1A} \tau)}
    &\ \underset{\tau\to i\infty}{\longrightarrow} e^{\pi i z_{1a}}
    \quad\text{for}\quad 
    0 < \re (s_{1A}) < 1
    \,,
    \\
    \frac{\vth^\prime(0)}{\vth(z_{1a})}
    &\ \underset{\tau\to i\infty}{\longrightarrow} 
    \frac{1}{\sin(\pi z_{1a})}
    \,,
\end{align}
we find that the $\tau\to i \infty$ limit of the Riemann-Wirtinger integrand becomes
\begin{align}
    \label{eq:RW-degen-integrand}
    u\
    F(z_{1a},\eta)\
    \d z_1 
    &\ \underset{\tau\to i\infty}{\longrightarrow}
    u_\mathrm{tree}\
    \frac{\d w_1}{w_{1a}}
    \,,
\end{align}
where $w_i = e^{2\pi i z_i}$ are the new coordinates on the sphere and 
\begin{align}
    \label{eq:RW-degen-twist}
    u_\mathrm{tree} = 
    w_1^{s_{1A}}
    \left(
        \prod_{i=2}^n 
        w_i^{-\frac{s_{1i}}{2}} 
        w_{1i}^{s_{1i}}
    \right)
    \,,
\end{align}
is the analogous twist for tree-level string integrals. 
Above, $\eta_0 = \sum_{i=2}^n s_{1i} z_i - s_{1B}$ is the $\tau$-independent part of $\eta$ and drops out of all quantities in the limit.
Equations \eqref{eq:RW-degen-integrand} and \eqref{eq:RW-degen-twist} are the $n$-point generalization of the analogous formulas in \cite{mano2009} where the factors of $w_i^{-\frac{s_{1A}}{2}}$ are missing.
Like \cite{mano2009}, we assume that $0<\re (s_{1A}) <1$ since this leads to convergent integrals. 
Just from the integrand, it is easy to see that the $\tau\to i\infty$ limit results in a linear combination of generalized Lauricella-D hypergeometric functions (see figure \ref{fig:branch_choice_rational_curve}).

The contours $\gamma_{2j}$ are the easiest to describe since the path from $z_2$ to $z_j$ on the torus maps to an analogous path from $w_2$ to $w_j$ with no phase factors. 
Gauge fixing $z_2=0$, we find
\begin{align} \label{eq:Iinf-2j}
    [\gamma_{2j} \vert \vphi_k\ra_{i\infty}
    := \lim_{\tau\to i\infty} [\gamma_{2j} \vert \vphi_a\ra
    = \int_{w_2=1}^{w_j} u_\mathrm{tree}\ \frac{\d w_1}{w_{1a}}
    \,.
\end{align}
On the other hand, the $A$- and $B$-cycle integrals become 
\begin{align}
    \label{eq:Iinf-2A}
    [\gamma_{2A} \vert \vphi_k\ra_{i\infty}
    &:= \lim_{\tau\to i\infty} [\gamma_{2j} \vert \vphi_a\ra
    = (1-e^{2\pi i s_{1A}}) \int_{w_2=1}^{\infty} u_\mathrm{tree}\ \frac{\d w_1}{w_{1a}}
    \,,
    \\
    \label{eq:Iinf-2B}
    [\gamma_{2B} \vert \vphi_k\ra_{i\infty}
    &:= \lim_{\tau\to i\infty} [\gamma_{2j} \vert \vphi_a\ra
    = \int_{w_2=1}^{0} u_\mathrm{tree}\ \frac{\d w_1}{w_{1a}}
    - e^{2\pi i s_{1B}} \int_{w_2=1}^{\infty} u_\mathrm{tree}\ \frac{\d w_1}{w_{1a}}
    \,.
\end{align}
Note that the $A$- and $B$-cycle degenerations come with explicit factors of the $A$- and $B$-cycle monodromies even though the twist $u_\text{tree}$ knows nothing about $s_{1B}$.

The phase factor and contour on the nodal sphere for the $A$-cycle integral \eqref{eq:Iinf-2A} is fairly straightforward to derive. 
The path from $z_1=0$ to $z_1=1$ on the torus maps to the contour from $\text{arg}(w_1) = 0$ to $\text{arg}(w_1)=2\pi$ with $|w_1|=1$ on the nodal sphere. 
Moreover, the corresponding punctures are all inside the unit circle since $\im z_i \geq0$ 
\begin{equation}\label{eq:limitA1}
    \begin{aligned}
        \lbrack \gamma_{2A} \vert =& 
        \includegraphics[align=c,scale=0.4]{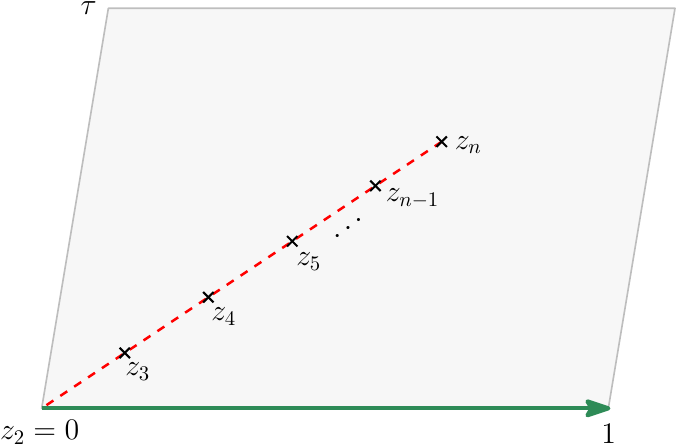} 
        \ \underset{\tau\to i \infty}{\longrightarrow} \  
        \includegraphics[align=c,scale=0.4]{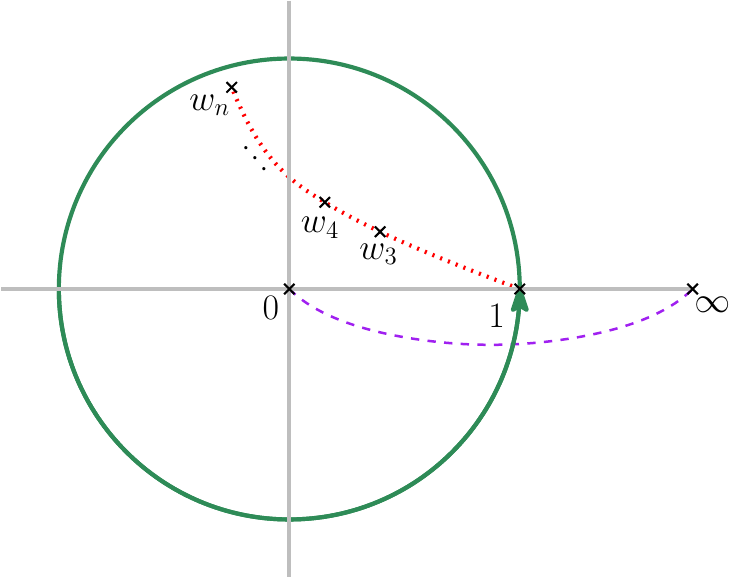} 
        \,.
    \end{aligned}
\end{equation}
We can then deform the contour to be: $\infty + i 0^- \to 1 \to \infty + i 0^+$
\begin{equation}\label{eq:limitA}
    \begin{aligned}
        \lbrack \gamma_{2A} \vert &
        \underset{\tau\to i \infty}{\longrightarrow}\ 
        \includegraphics[align=c,scale=0.39]{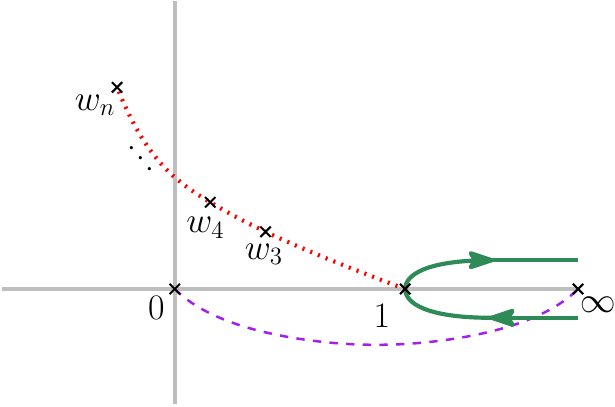} 
        = 
        (1 - e^{2\pi i s_{1A}}) \left( 
        \includegraphics[align=c,scale=0.39]{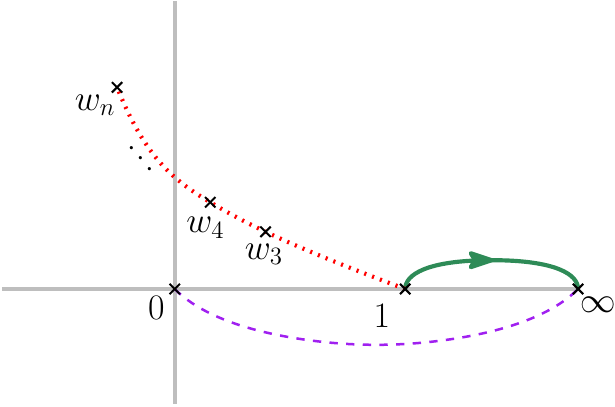} 
        \right) 
        \,,
    \end{aligned}
\end{equation}
where the phase comes from crossing the purple branch cut.
Equation \eqref{eq:limitA} is precisely the contour in the boundary value \eqref{eq:Iinf-2A}!

On the other hand, the degeneration of the $B$-cycle is more subtle
since the $\tau \to i\infty$ limit of a cycle only makes sense for paths whose imaginary part is much smaller than $|\tau|$. 
Thus, we cannot directly map the $B$-cycle to the nodal sphere.  
Instead, we define a new path that includes the $B$-cycle and can be deformed to one with imaginary part much smaller than $|\tau|$. 
To this end, we consider the following linear combination:
\begin{equation}\label{eq:deformBAB}
\begin{aligned}
    &\lbrack \gamma_{2B} \vert + \lbrack \gamma_{2A} \vert_{z_1 \to z_1 + \tau} - \lbrack \gamma_{2B} \vert_{z_{1} \to z_{1}+1} 
    &= (1 - e^{2\pi i s_{1A}}) \lbrack \gamma_{2B} \vert + e^{2\pi i s_{1B}} \lbrack \gamma_{2A} \vert 
    \,,
    \\[1em]
    &= \includegraphics[align=c,scale=0.4]{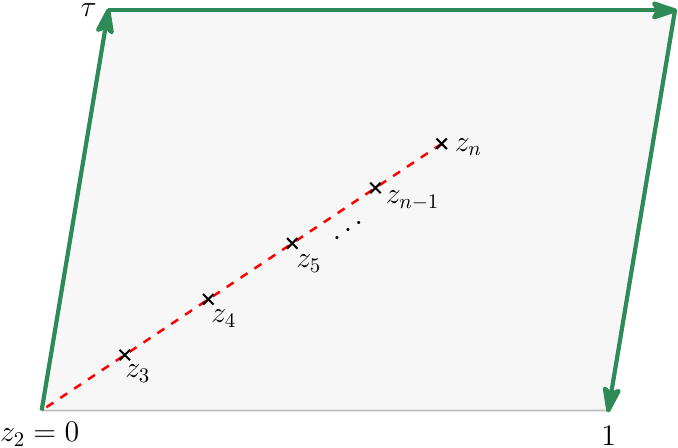} 
    &= \quad  \includegraphics[align=c,scale=0.4]{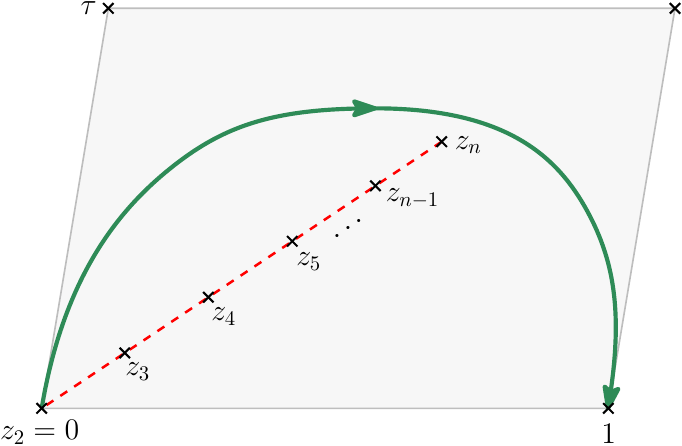} 
    \,.
\end{aligned}
\end{equation}
Then, the above path maps to the nodal sphere as follows
\begin{equation}\label{eq:limitBAB1}
    \begin{aligned}
        \lbrack \gamma_{2B} \vert + \lbrack \gamma_{2A} \vert_{z_1 \to z_1 + \tau} - \lbrack \gamma_{2B} \vert_{z_{1} \to z_{1}+1} 
        \ \underset{\tau\to i \infty}{\longrightarrow} \ 
        \includegraphics[align=c,scale=0.4]{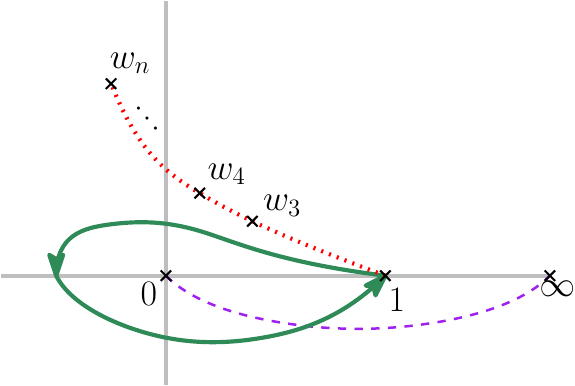}
        \,.
    \end{aligned}
\end{equation}
Accounting for the phase from crossing the branch cut, we find
\begin{equation}\label{eq:limitBAB}
    \begin{aligned}
        & \lbrack \gamma_{2B} \vert + \lbrack \gamma_{2A} \vert_{z_1 \to z_1 + \tau} - \lbrack \gamma_{2B} \vert_{z_{1} \to z_{1}+1} 
        \underset{\tau\to i \infty}{\longrightarrow} \ 
        =& (1 - e^{2\pi i s_{1A}}) \left(
        \includegraphics[align=c,scale=0.35]{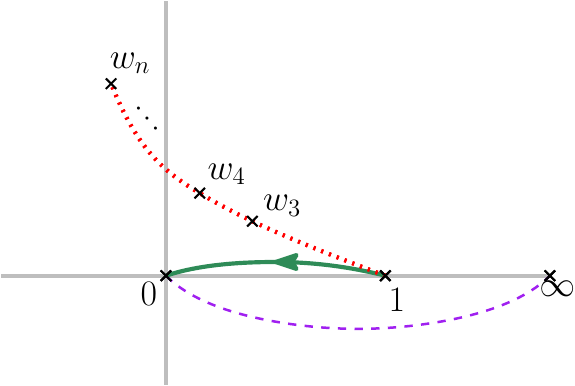}
        \right) 
    \end{aligned}
\end{equation}
Combining \eqref{eq:deformBAB} and \eqref{eq:limitBAB} and inserting \eqref{eq:limitA}, we obtain the contour in \eqref{eq:Iinf-2B}.

\begin{figure}
    \centering
    \includegraphics[scale=0.5]{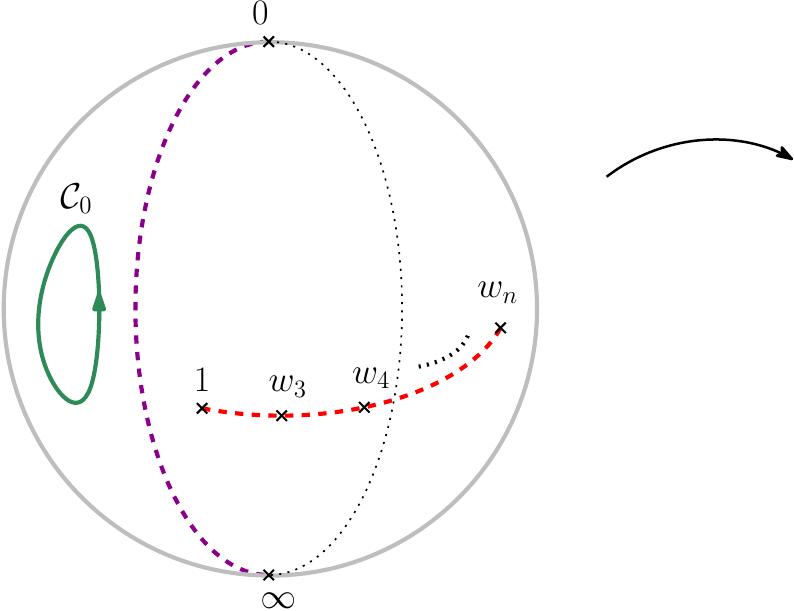}
    \hspace{1em}
    \includegraphics[scale=0.5]{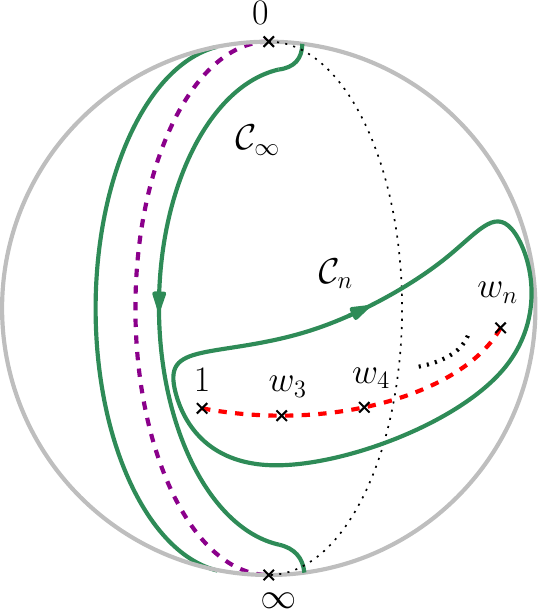}
    \caption{Contour deformation argument for the over-completness of cycles on the sphere. 
    Via Cauchy's theorem, the integral over the contour $\mathcal{C}_0$ vanishes. 
    Moreover, $\mathcal{C}_0$ can be deformed into the contours $\mathcal{C}_\infty + \mathcal{C}_n$, which can further be decomposed into the contours of figure \ref{fig:branch_choice_rational_curve}. 
    Thus, one finds a linear relation between the contours of figure \ref{fig:branch_choice_rational_curve}.
    }
    \label{fig:contour_deform_overcompleteness_g0}
\end{figure}

As a sanity check, we show that these degenerate integrals satisfy the over-completeness relation \eqref{eq:homology_cycle_overcomplete_basis}. 
Taking the $\tau \to i \infty$ limit of \eqref{eq:homology_cycle_overcomplete_basis} yields 
\begin{align}
    \lim_{\tau \to i \infty} \eqref{eq:homology_cycle_overcomplete_basis}
    &= (1-e^{2\pi i s_{1A}}) 
        \int_{0}^{\infty} u_\mathrm{tree}\ \frac{\d w_1}{w_{1a}}
    - \sum_{j=3}^n 
        e^{-2 \pi i (s_{12} + \cdots + s_{1j})}
        (1 - e^{2 \pi i s_{1j}})
        [\gamma_{2j}\vert\vphi_a\ra_{i\infty}
    \,,
   \nn \\
    &=
    \int_{\mathcal{C}_\infty} u_\mathrm{tree}\ \frac{\d w_1}{w_{1a}}
    - \sum_{j=3}^n 
        e^{-2 \pi i (s_{12} + \cdots + s_{1j})}
        (1 - e^{2 \pi i s_{1j}})
        [\gamma_{2j}\vert\vphi_a\ra_{i\infty}
    \,,
\end{align}
where all of the dependence on the $B$-cycle monodromy has dropped out and we have identified the first term with the contour $\mathcal{C}_\infty$ in figure \ref{fig:contour_deform_overcompleteness_g0}. 
Next, we note that 
\begin{align}
    &\sum_{j=3}^n 
        e^{-2 \pi i (s_{12} + \cdots + s_{1j})}
        (1 - e^{2 \pi i s_{1j}})
        [\gamma_{2j}\vert\vphi_a\ra_{i\infty}
    = \sum_{j=3}^n \left( 1-e^{-2\pi i \sum_{k=2}^{j-1} s_{1k}} \right) 
        \int_{w_{j-1}}^{w_j} u_\mathrm{tree}\ \frac{\d w_1}{w_{1a}}
    \,
    \nn\\
    &\qquad\qquad
    = \sum_{j=3}^n \left( 1-e^{2\pi i \sum_{k=j}^{n} s_{1k}} \right) 
        \int_{w_{j-1}}^{w_j} u_\mathrm{tree}\ \frac{\d w_1}{w_{1a}}  
    \,
    = - \int_{\mathcal{C}_n} u_\mathrm{tree}\ \frac{\d w_1}{w_{1a}}  
    \,.
\end{align}
Putting the last two equations together, yields 
\begin{align}
    \lim_{\tau \to i \infty} \eqref{eq:homology_cycle_overcomplete_basis}
    = \int_{\mathcal{C}_\infty + \mathcal{C}_n} u_\mathrm{tree}\ \frac{\d w_1}{w_{1a}}
    = \int_{\mathcal{C}_0} u_\mathrm{tree}\ \frac{\d w_1}{w_{1a}}
    = 0
\end{align}
since the contour $\mathcal{C}_0 = \mathcal{C}_\infty + \mathcal{C}_n$ in figure \ref{fig:contour_deform_overcompleteness_g0} is clearly contractible with no singularities inside and thus the corresponding integral vanishes due to Cauchy. 

The drawback to equations (\ref{eq:Iinf-2j}-\ref{eq:Iinf-2B}) is that one has to be careful when propagating the boundary value at $\tau=i \infty$ to finite $\tau$ since equation \eqref{eq:solution_tau_DEQ} is naively divergent at $\tau=i\infty$.
Fortunately, we can avoid the complication of regulating the $\tau$-integration for the $\gamma_{2j}$ cycles because we can find a boundary value at finite $\tau$ by taking the $z_i$ to a special configuration!

By taking all of the punctures to the origin, the $\tau$-dependence drops out. 
However, this limit must be taken carefully using regularized objects to get a finite answer.
In the end, we find the following prescription for the boundary value where all punctures are sent to the origin
\begin{align}
    \label{eq: general-n_regularised_limit}
    \lbrack \gamma_{2i} \vert \vphi_a\ra_{0} 
    &:= \left( 
        \prod_{k\neq2,i} 
        \lim_{z_k\to0}(- 2 \pi i z_k)^{-s_{1k}}
    \right)
    \lim_{z_i\to0}( - 2\pi i z_i)^{-s_{12}-s_{1i}}
    \int_{\gamma_{2i}} \d z_1\ u\
    F(z_{1a},\eta\vert \tau)
    \,.
\end{align}
In the above, we set $z_1 = x z_i$ and the integration region is $x\in [0,1]$ in $x$-space runs from $0$ to $1$. 
It is also important to note that the limit $z_i\to0$ is taken before all others.

One can think of the prefactors in \eqref{eq: general-n_regularised_limit} as defining a regularized limit consistent with the shuffle regularisation of eMPLs \cite{Broedel2020, Broedel:2019tlz}.
The eMPL of weight and length one makes an appearance in the twist because it is simply related to the log of the Jacobi theta function and its regularization is well understood. 
The unregulated eMPL is 
\begin{equation}
    \tilde{\Gamma}\left(\begin{smallmatrix}1\\0\end{smallmatrix};z|\tau\right) 
    = \lim_{\epsilon \to 0}\int_{\epsilon}^{z}
    \d w~g^{(1)}(w;\tau)
    = \Omega^{(1)}(z,\tau) - \lim_{\epsilon\to0} \Omega^{(1)}(\epsilon,\tau)
    \,,
    \label{eq:divergent_eMPLs}
\end{equation}
where $\Omega^{(1)}(z,\tau) = \log{\left(\vth(z\vert\tau)/\etaDe(\tau)\right)}$ (recall from section \ref{sec:RWco-app-alphaprime} that $\etaDe$ is the Dedekind eta function and $\Omega^{(1)}$ is an elliptic symbol letter \cite{Wilhelm:2022wow}). 
As one can clearly see, this is divergent for all $z_1$ except $z_1=0$ where it vanishes. 
For any $z_1\neq0$, one needs to regularize $\tilde{\Gamma}\left(\begin{smallmatrix}
    1
    \\
    0
\end{smallmatrix};z|\tau\right)$
in order to make sense of the lower integration boundary.
The shuffle regularized eMPL is obtained by subtracting off the logarithmic singularity at the lower boundary \cite{Broedel:2019tlz}
\begin{align}
    \label{eq:GammaReg}
    {\tilde{\Gamma}_{\text{reg}}}\left(\begin{smallmatrix}
        1\\
        0
    \end{smallmatrix};z|\tau\right) 
    &:= \lim_{\epsilon \to 0}
    \left[
        \int_{\epsilon}^{z} \d w~g^{(1)}(w;\tau)
        + \log(1-e^{2\pi i \epsilon})
    \right]
    \nn\\& = 
        \log(1-e^{2\pi iz}) - \pi i z 
        + 4\pi \sum_{k,l>0} \frac{1}{2\pi k} (1-\cos(2\pi kz)) e^{\pi i kl\tau}
    \,.
\end{align}
Now, the regularized eMPL has a logarithmic singularity at $z=0$, 
$
{\tilde{\Gamma}_{\text{reg}}}\left(\begin{smallmatrix}1\\0\end{smallmatrix};z\sim0|\tau\right) \sim \log(-2\pi i z)
$, 
but is finite everywhere else.
Writing the Riemann-Wirtinger integral in terms of the regulated eMPL, we find
\begin{equation}
    [\gamma_{2i} \vert \vphi_a\ra = 
    \int_{\gamma_{2i}} 
    \d z_1\ 
    e^{2\pi i s_{1A}z_1}
    \exp{\bigg(
        \sum_{j=2}^n s_{1j}\
        {\tilde{\Gamma}_{\text{reg}}}
        \left(\begin{smallmatrix}
            1\\
            0
        \end{smallmatrix};z_1-z_j|\tau\right)
    \bigg)}
    F(z_1-z_a,\eta|\tau)
    \,,
\end{equation}
where momentum conservation ensures that additional ``constants'' introduced from the regularization drop out.
The prefactors in \eqref{eq: general-n_regularised_limit} can now be understood as canceling the non-analytic behavior of $
{\tilde{\Gamma}_{\text{reg}}}\left(\begin{smallmatrix}1\\0\end{smallmatrix};z|\tau\right)
$ for $z\sim0$. 
Since $\tilde{\Gamma}_{\text{reg}}\left(\begin{smallmatrix}
        1\\
        0
\end{smallmatrix};z_1|\tau\right)$ 
and 
$\tilde{\Gamma}_{\text{reg}}\left(\begin{smallmatrix}
        1\\
        0
\end{smallmatrix};z_1-z_i|\tau\right)$ 
diverge at both the boundaries of $\gamma_{2i}$ ($z_1=0$ and $z_1=z_i$), we need a factor of $(-2\pi i z_i)^{-s_{12}-s_{1i}}$ to render the $z_i\to0$ limit finite. 
Similarly, $\tilde{\Gamma}_{\text{reg}}\left(\begin{smallmatrix}
        1\\
        0
\end{smallmatrix};z_1-z_k|\tau\right)$  for $k\neq i$ diverges at the integration boundary $z_1=0$. 
Therefore, we need a factor of $(-2\pi i z_k)^{-s_{1k}}$ for all $k \neq 2,i$. 

Another way to take the limit is to use \eqref{eq:divergent_eMPLs} when $z\sim0$ and \eqref{eq:GammaReg} otherwise. 
To see how this works in practice, consider the boundary value $\lbrack \gamma_{24} \vert \vphi_a\ra_{0}$ when $n=4$. 
Taking the limit $ z_4\to 0$ is subtle because the integration contour is being ``squished'' to a neighborhood of 0 and the Kronecker-Eisenstein functions diverges when $a=2$ and $a=4$. 
Making the substitution $z_1=xz_4$, yields
\begin{align}
    \bs{\vphi}
    &= \begin{pmatrix}
        F(z_1,\eta\vert\tau)
        \\
        F(z_1-z_3,\eta\vert\tau)
        \\
        F(z_1-z_4,\eta\vert\tau)
    \end{pmatrix}
    \d z_1 
    = \begin{pmatrix}
        \frac{1}{x}\ 
        \\
        0
        \\
        -\frac{1}{1-x}\ 
    \end{pmatrix}
    \d x\
    + \mathcal{O}(z_4)
    \,.
\end{align}
Next, we expand the twist around $z_4=0$.
Since the regulated eMPLs $\tilde{\Gamma}_{\text{reg}}\left(\begin{smallmatrix}
        1\\
        0
\end{smallmatrix};z_1|\tau\right)$ 
and 
$\tilde{\Gamma}_{\text{reg}}\left(\begin{smallmatrix}
        1\\
        0
\end{smallmatrix};z_1-z_4|\tau\right)$ 
diverge at both the boundaries of $\gamma_{24}$, we use \eqref{eq:divergent_eMPLs} to define the limit
\begin{align}
    \lim_{z_{4}\to 0} \tilde{\Gamma}\left(\begin{smallmatrix}1\\0\end{smallmatrix};z_1|\tau\right)
    &= \lim_{z_{4}\to 0} 
    \lim_{\epsilon\to0}
    \log\frac{\vth(x z_4)}{\vth(\epsilon)} 
    = \log x 
    \,,
    \\
    \lim_{z_{4}\to 0} \tilde{\Gamma}\left(\begin{smallmatrix}1\\0\end{smallmatrix};z_1-z_4|\tau\right)
    &= \lim_{z_{4}\to 0} 
    \lim_{\epsilon\to0}
    \log\frac{\vth((x -1)z_4)}{\vth(\epsilon)} 
    = \log (x-1) 
    \,,
\end{align}
where we have set $\epsilon = z_4$ and
\begin{align}
    \lim_{z_4\to0} \Gtreg{z_1-z_3}
    = \Gtreg{-z_3}
    \,.
\end{align}
Using \eqref{eq:divergent_eMPLs} again, the $z_{3}\to0$ limit of the only remaining eMPL becomes
\begin{align}
    \lim_{z_{3}\to0} \Gt{-z_3} 
    = \lim_{z_{3}\to 0} 
    \lim_{\epsilon\to0}
    \log\frac{\vth(-z_3)}{\vth(\epsilon)} 
    = \log(-1) 
    = i \pi 
    \,,
\end{align}
where we have set $\epsilon = z_3$.
Putting everything together, we find the finite $\tau$ boundary value 
\begin{align}
    [ \gamma_{24} \vert \bs{\vphi} \ra_{0}
    &= \int_0^1 
    \d x\ 
    (-1)^{-s_{12}} x^{s_{12}} (1-x)^{s_{14}} 
    \begin{pmatrix}
        \frac{1}{x}\ 
        \\
        0
        \\
        -\frac{1}{1-x}\ 
    \end{pmatrix}
    \nn\\
    &= (-1)^{-s_{12}}
    \frac{\Gamma(1+s_{12})\Gamma(1+s_{14})}{\Gamma(1+s_{12}+s_{14})}
    \begin{pmatrix}
        \frac{1}{s_{12}}\ 
        \\
        0
        \\
        -\frac{1}{s_{14}}\ 
    \end{pmatrix}
    \,.
\end{align}
Notice that this is the same boundary value for the analogous integral defined with $\eta$ independent of the punctures \cite{Kaderli:2022qeu} (up to signs coming from a different convention for the Mandelstams).  
Generalisation to arbitrary $n$ is straight forward, 
\begin{equation}
        \lbrack \gamma_{2i} \vert \bs{\vphi}\ra_{0} =  (-1)^{-s_{12}}
        \frac{\Gamma(1+s_{12})\Gamma(1+s_{1i})}{\Gamma(1+s_{12}+s_{1i})}\begin{pmatrix}
        \frac{1}{s_{12}}\ 
        \\
        0\\
        \vdots\\
        0\\
        -\frac{1}{s_{1i}}  \\
        0\\
        \vdots\\
        0
        \end{pmatrix},\label{eq:n-pt_initial_value_gamma2i}
\end{equation}
where the second non-trivial entry shows up in the $i$\textsuperscript{th} row.

Notice that these boundary values are independent of $\tau$! In fact, one can verify that the \eqref{eq:n-pt_initial_value_gamma2i} matches \eqref{eq:Iinf-2j}: $[\gamma_{2j}\vert\vphi_a\ra_0 = [\gamma_{2j}\vert\vphi_a\ra_{i\infty}$.
One has to be careful when taking the $z_i\to0$ limit of the hypergeometric functions produced by \eqref{eq:Iinf-2j} and this equivalence was checked for $n=3,4$ using the \texttt{HPL} and \texttt{HypExp} \texttt{Mathematica} packages \cite{Maitre:2005uu, Huber:2005yg}.
Moreover, this boundary value conspires to cancel the naive divergence in \eqref{eq:solution_tau_DEQ} from setting $\tau^* = i\infty$ when integrating the $\tau$-part of differential equation. 
Substituting this boundary value into \eqref{eq:solution_tau_DEQ}, we find 
\begin{align}
    \mathbf{I}^{(1)}_{\gamma_{2j}}(z^*_{i\geq2},\tau)
    = \mathbf{I}^{(1)}_{\gamma_{2j}}(z^*_{i\geq2},\tau^*)
    = \mathbf{c}^{(1)}
\end{align}
where $\mathbf{c}^{(1)}$ is the order $\alpha^\prime$ term in the boundary value.
When all $z_i\to0$, the $\log$ term in \eqref{eq:solution_tau_DEQ} drops out due to momentum conservation and the boundary value \eqref{eq:n-pt_initial_value_gamma2i} ensures that the sum over the $\mathcal{F}$'s vanish.

%%%%%%%%%%%%%%%%%%%%%%%%%%%%%%%%%%%%%%%%%%%%%%%%%
%%%%%%%%%%%%%%%%%%%%%%%%%%%%%%%%%%%%%%%%%%%%%%%%%
\subsection{The $\eta\to0$ limit \label{sec:RWco-app-0}}

The $\eta$-dependent differential forms considered in section \ref{sec:RWco-eta} are useful because they can be thought of as generating integrals where one performs a formal series expansion in $\eta$. 
On the other hand, true string integrals do not depend on the parameter $\eta$. 
For this reason, it is interesting to examine the $\eta\to0$ limit of the Riemann-Wirtinger integrals. 
However, unlike true string integrands, the Riemann-Wirtinger integrand must not have $B$-cycle monodromies.\footnote{The integral over the loop-momentum guarantees the double periodicity of string integrals. Therefore, it is possible to have string integrands that are not doubly periodic.} 
Anything that breaks this condition is not part of our cohomology.

It turns out that the spanning set $\{\xi^{(p)}_a\}$ of $H^1(M,\LS)$ (equation \eqref{eq:xi-basis}) has a smooth $\eta\to0$ limit that also spans the cohomology when $\eta=0$: $H^1(M,\L_{\omega,\eta=0})$ \cite{Goto2022}.
Explicitly,  we let $\zeta^{(p)}=\lim_{\eta\to0} \xi^{(p)}$ and find
\begin{align}
	\zeta^{(p)}_1 &= \d z_1 , 
	\nn \\ \label{eq:zeta-basis}
	\zeta^{(p)}_p &= \partial_{z_1} \gk{1}(z_{1p})\, \d z_1 , 
	\\
	\zeta^{(p)}_{a \geq 2} &= \left[
		\gk{1}(z_{1a}) - \gk{1}(z_{1p})
	\right]\, \d z_1  
	\quad \text{for} \quad a \neq p
	. 
	\nn
\end{align}
Each element of this set has no $B$-cycle monodromy.
Also note that, to have a basis, the element $\zeta^{(p)}$ with a double pole must be included. 
This spanning set is also subject to the relation
\begin{align}
    \label{eq:xirel-0}
    2 \pi i s_{1A}\ \vert \zeta^{(p)}_1\ra
%    - (s_{1,p} - 1) \eta \vert \zeta^{(p)}_p\ra
    + \underset{j\neq p}{\sum_{j=2}^n} s_{1j}\,
    \ \vert \zeta^{(p)}_j\ra \simeq 0
\end{align}
that follows from the $\eta\to0$ limit of \eqref{eq:xirel}.
Moreover, in this limit, the defining equation for $\eta$ places restrictions on the punctures in the problem
\begin{align}
    0 = s_{1A} \tau + \sum_{i=2}^n s_{1i} z_i - s_{1B}.
\end{align}
Importantly, for the differential equation, this means that the differentials satisfy the relation 
\begin{align}
    0 = s_{1A} \d\tau + \sum_{i=2}^n s_{1i} \d z_i \ . 
\end{align}

We can compute the intersection numbers directly in the $\eta\neq0$ limit case using \eqref{eq:intNum} as before. 
Of course, this yields the same result as taking the $\eta\to0$ limit of \eqref{eq:xi-int1}-\eqref{eq:xi-int3}. 
For example, at $n=4$ with $p=4$, the explicit form of the intersection matrix is 
\begin{align}
    \frac{\la \zeta^{(4)}_a \vert \zeta^{(4)}_b \ra}{2\pi i}
    =
    \scalebox{.9}{$
        \begin{pmatrix}
		 0 & 0 & 0 & \frac{1}{s_{14}-1}
	\\
		 0 
		 & 
		 \frac{ s_{13} }{ s_{12} s_{14} } 
		 &
		  -\frac{1}{s_{14}} 
		  &
		  \frac{
		  	2 i \pi  s_1
			+ s_{13} \left[ \gk{1}(z_{24}) - \gk{1}(z_{34}) \right]
		  }{(s_{14}-1) s_{14}} 
	\\
		 0 
		 & 
		 -\frac{1}{s_{14}} 
		 & 
		 \frac{ s_{12} }{s_{13} s_{14}} 
		 & 
		 \frac{ 
		 	2 i \pi  s_1
		 	-s_{12} \left[ \gk{1}(z_{24}) - \gk{1}(z_{34}) \right]
		 }{(s_{14}-1) s_{14}}
	\\
		 \frac{1}{s_{14}+1}  
		 &
 	 	\frac{
			2 i \pi  s_1
			+ s_{13} \left[ \gk{1}(z_{24}) - \gk{1}(z_{34}) \right]
		}{s_{14} (s_{14}+1)}
		& 
		\frac{
			2 i \pi  s_1
			- s_{12} \left[ \gk{1}(z_{24}) - \gk{1}(z_{34}) \right]
		}{s_{14} (s_{14}+1)}
		& 
		\frac{\la \zeta^{(4)}_4 \vert \zeta^{(4)}_4 \ra}{2\pi i}
	\end{pmatrix}
   $}
    \,,
\end{align}
where 
\begin{align}
	\frac{\la \zeta^{(4)}_4 \vert \zeta^{(4)}_4 \ra}{2\pi i}
	&= \frac{1}{ (s_{14}-1) s_{14} (s_{14}+1) } \bigg[
		-4 \pi ^2 s_1^2
		- 2 s_{14} \left( s_{12} - s_{14} \right) G_2(\tau)
	\nn\\&\qquad 
	+ 4 i \pi  s_1 \left( 
		s_{12}\ \gk{1}(z_{24}) 
		+ s_{13}\ \gk{1}(z_{34})
	\right)
	+ s_{12} s_{13} \left( \gk{1}(z_{24}) - \gk{1}(z_{34}) \right)^2
	\nn\\&\qquad	
	+ 2 s_{14} \left( s_{12}\ \gk{2}(z_{24}) + s_{13}\ \gk{2}(z_{34}) \right)
	\bigg]
    \,.
\end{align}

Choosing $\zeta^{(p)}_p$ and any set of $n-2$ other $\zeta^{(p)}_{a \neq p}$ forms a basis of the cohomology $H^1(M,\L_{\omega,\eta=0})$.
After we choosing a basis, the DEQs can be computed in the usual way.
For example, choosing $\zeta^{(p)}_{a>1}$ for our basis and choosing to eliminate $\d z_4$ using the $\eta=0$ condition, we find the following $\d z_2$ component of the differential equation
\begin{align}
    \mat{A}_\zeta \vert_{\d z_2} 
    = \sum_{i=1}^9 \mat{\Xi}_i X_i 
\end{align}
where the $X_i$ are monomials in the $g$-functions (including the constant function)
\begin{align}
    X_i \in \bigg\{&
        1, 
        g^{(1)}\left(z_{23}\right),
        g^{(1)}\left(z_{34}\right),
        g^{(1)}\left(z_{24}\right),
        g^{(2)}\left(z_{34}\right),
        g^{(2)}\left(z_{24}\right),
        \nn \\ & \qquad
        \left(g^{(1)}\left(z_{34}\right)\right)^2,
        g^{(1)}\left(z_{24}\right) g^{\text{(1)}}\left(z_{34}\right),
        \left(g^{(1)}\left(z_{24}\right)\right)^2
        \bigg\}
        \,,
\end{align}
and the coefficient matrices are
\begin{align}
\begin{aligned}
& \mat{\Xi}_1 = \left(
\begin{array}{ccc}
 -\frac{2 i \pi  s_1 s_{13} \left(s_{12}-s_{14}\right)}{s_{12} s_{14}^2} & \frac{2 i \pi  s_1 s_{12}}{s_{14}^2} & \frac{s_{13} G_2(\tau )}{s_{14}}-\frac{4 \pi ^2 s_1^2 s_{12}}{\left(1-s_{14}\right) s_{14}^2} \\
 \frac{2 i \pi  s_1 s_{12}}{s_{14}^2} & \frac{2 i \pi  s_1 s_{12}}{s_{14}^2} & -\frac{s_{12} G_2(\tau )}{s_{14}}-\frac{4 \pi ^2 s_1^2 s_{12}}{\left(1-s_{14}\right) s_{14}^2} \\
 0 & 0 & 0 \\
\end{array}
\right),
\\& 
\mat{\Xi}_2 = \left(
\begin{array}{ccc}
 -\frac{s_{13}}{s_{12}} & 1 & 0 \\
 1 & -\frac{s_{12}}{s_{13}} & 0 \\
 0 & 0 & 0 \\
\end{array}
\right),
\qquad 
\mat{\Xi}_3 = 
\left(
\begin{array}{ccc}
 -\frac{s_{12} s_{13}}{s_{14}^2} & \frac{s_{12}^2}{s_{14}^2} & -\frac{4 i \pi  s_1 s_{12} s_{13}}{\left(1-s_{14}\right) s_{14}^2} \\
 \frac{s_{12}^2}{s_{14}^2} & -\frac{s_{12}^3}{s_{13} s_{14}^2} & \frac{2 i \pi  s_1 s_{12} \left(s_{12}-s_{13}\right)}{\left(1-s_{14}\right) s_{14}^2} \\
 0 & 0 & 0 \\
\end{array}
\right),
\\&
\mat{\Xi}_4 = 
\left(
\begin{array}{ccc}
 \frac{s_{13}^3}{s_{12} s_{14}^2} & -\frac{s_{13}^2}{s_{14}^2} & \frac{4 i \pi  s_1 s_{12} s_{13}}{\left(1-s_{14}\right) s_{14}^2} \\
 -\frac{s_{13}^2}{s_{14}^2} & \frac{s_{12} s_{13}}{s_{14}^2} & -\frac{2 i \pi  s_1 s_{12} \left(s_{12}-s_{13}\right)}{\left(1-s_{14}\right) s_{14}^2} \\
 0 & -\frac{2 i \pi  s_1 s_{12}}{\left(-s_{14}-1\right) s_{14}} & 0 \\
\end{array}
\right),
\qquad
\mat{\Xi}_5 =
\left(
\begin{array}{ccc}
 0 & 0 & \frac{2 s_{12} s_{13}}{s_{14}^2} \\
 0 & 0 & -\frac{2 s_{12}^2}{s_{14}^2} \\
 0 & \frac{s_{12}}{s_{14}+1} & 0 \\
\end{array}
\right),
\\&
\mat{\Xi}_6 = 
\left(
\begin{array}{ccc}
 0 & 0 & \frac{2 s_{13}^2}{s_{14}^2} \\
 0 & 0 & -\frac{2 s_{12} s_{13}}{s_{14}^2} \\
 0 & \frac{2 s_{12}}{s_{14}+1} & 0 \\
\end{array}
\right),
\qquad
\mat{\Xi}_7 = \left(
\begin{array}{ccc}
 0 & 0 & -\frac{s_{12} \left(s_{12}+1\right) s_{13}}{\left(1-s_{14}\right) s_{14}^2} \\
 0 & 0 & \frac{s_{12}^2 \left(s_{12}+1\right)}{\left(1-s_{14}\right) s_{14}^2} \\
 0 & 0 & 0 \\
\end{array}
\right),
\\&
\mat{\Xi}_8 = 
\left(
\begin{array}{ccc}
 0 & 0 & -\frac{2 s_{12} s_{13}^2}{\left(1-s_{14}\right) s_{14}^2} \\
 0 & 0 & \frac{2 s_{12}^2 s_{13}}{\left(1-s_{14}\right) s_{14}^2} \\
 0 & \frac{s_{12}^2}{\left(s_{14}+1\right) s_{14}} & 0 \\
\end{array}
\right),
\qquad
\mat{\Xi}_9 = 
\left(
\begin{array}{ccc}
 0 & 0 & -\frac{s_{13}^2 \left(s_{13}+1\right)}{\left(1-s_{14}\right) s_{14}^2} \\
 0 & 0 & \frac{s_{12} s_{13} \left(s_{13}+1\right)}{\left(1-s_{14}\right) s_{14}^2} \\
 0 & \frac{s_{12} s_{13}}{\left(s_{14}+1\right) s_{14}} & 0 \\
\end{array}
\right).
\end{aligned}
\end{align}
It is interesting to note that the $\eta=0$ differential equations are much more complicated than their $\eta \neq 0$ cousins. 
Most of the complexity comes from having to use a form with a double pole in our basis.
This translates into double poles in the differential equation. 
That is, the double poles coming from the monomials $\left(g^{(1)}\left(z_{34}\right)\right)^2, g^{(1)}\left(z_{24}\right) g^{\text{(1)}}\left(z_{34}\right)$ and $\left(g^{(1)}\left(z_{24}\right)\right)^2$ are not spurious. 
Connections with higher order poles appear at higher genus and it is possible to make sense of iterated integrals with such a connection \cite{enriquez2023analogues, enriquez2023elliptic}.

The $\eta=0$ twisted (co)homology was previously studied in \cite{ghazouani2016moduli}. 
Here, the authors study the moduli space of flat tori and elliptic hypergeometric functions using algebro-geometric techniques similar to those used in this paper. 
In particular, they compute the $n=2$ differential equations (see their appendix \ref{app:intInd}) and construct the double copy in order to obtain an explicit expression for the so-called Veech map.
In both cases, we find agreement.

%%%%%%%%%%%%%%%%%%%%%%%%%%%%%%%%%%%%%%%%%%%%%%%%%
%%%%%%%%%%%%%%%%%%%%%%%%%%%%%%%%%%%%%%%%%%%%%%%%%
%%%%%%%%%%%%%%%%%%%%%%%%%%%%%%%%%%%%%%%%%%%%%%%%%
\bibliographystyle{JHEP}
\bibliography{refs.bib}
\end{document}